\documentclass{aa}  
\usepackage{amsmath}
\usepackage{graphicx}
\usepackage{txfonts}
\usepackage{natbib}

\bibpunct{(}{)}{;}{a}{}{,} 
\def\ie{{\it i.e.,}\,}
\def\eg{{\it e.g.,}\,}

\def\la{\hbox{\raise.5ex\hbox{$<$} 
    \kern-1.1em\lower.5ex\hbox{$\sim$}}} 
\def\ga{\hbox{\raise.5ex\hbox{$>$} 
    \kern-1.1em\lower.5ex\hbox{$\sim$}}}

\newcommand{\dgr}{\mbox{$^\circ$}}           
\newcommand{\Msun}{\mbox{$M_\odot$}}         
\newcommand{\Lsun}{\mbox{$L_\odot$}}         
\newcommand{\ad}{\mbox{d}}                   
\newcommand{\cm}{\mbox{\ cm}}                
\newcommand{\s}{\mbox{\ s}}                  
\newcommand{\K}{\mbox{\ K}}                  
\newcommand{\erg}{\mbox{\ erg }}              
\newcommand{\cms}{\mbox{\ cm s${}^{-1}$}}    
\newcommand{\Ks}{\mbox{\ K s${}^{-1}$}}    
\newcommand{\mes}{\mbox{\ m s${}^{-1}$}}    
\newcommand{\gcm}{\mbox{\ g cm${}^{-3}$}}    
\newcommand{\ergs}{\mbox{$\erg\s^{-1}$}}  
\newcommand{\radcm}{\mbox{\ rad$^2 \cm^{-2}$\,}}  

\begin{document}
\bibliographystyle{aa}
   \title{The core helium flash revisited}

   \subtitle{II. Two and three-dimensional hydrodynamic simulations}

   \author{M. Moc\'ak,
           E. M\"uller,
           A.Weiss,
           \and K.Kifonidis}



   \institute{Max-Planck-Institut f\"ur Astrophysik,
              Postfach 1312, 85741 Garching, Germany\\
              \email{mmocak@mpa-garching.mpg.de}}

   \date{Received  ........................... }


  \abstract
   { We study turbulent convection during the core helium flash close
     to its peak by comparing the results of two and three-dimensional 
     hydrodynamic simulations.}
   { In a previous study we found that the temporal evolution and the
     properties of the convection inferred from two-dimensional
     hydrodynamic studies are similar to those predicted by
     quasi-hydrostatic stellar evolutionary calculations. However, as
     vorticity is conserved in axisymmetric flows, two-dimensional
     simulations of convection are characterized by incorrect
     dominant spatial scales and exaggerated velocities.  Here, we
     present three-dimensional simulations 
     that eliminate the restrictions and flaws of two-dimensional
     models, and that provide a geometrically unbiased insight into
     the hydrodynamics of the core helium flash. In particular, we
     study whether the assumptions and predictions of stellar
     evolutionary calculations based on the mixing-length theory can
     be confirmed by hydrodynamic simulations. }
   { We use a multidimensional Eulerian hydrodynamics code based on
     state-of-the-art numerical techniques to simulate the evolution
     of the helium core of a $1.25 M_{\odot}$ Pop I star.}
   { Our three-dimensional hydrodynamic simulations of the evolution
     of a star during the peak of the core helium flash do not show
     any explosive behavior. The
     convective flow patterns developing in the three-dimensional
     models are structurally different from those of the corresponding
     two-dimensional models, and the typical convective velocities are
     smaller than those found in their two-dimensional counterparts.
     Three-dimensional models also tend to agree better with the
     predictions of mixing length theory. Our hydrodynamic simulations
     show the presence of turbulent entrainment that results in a
     growth of the convection zone on a dynamic time scale. Contrary
     to mixing length theory, the outer part of the convection zone is
     characterized by a sub-adiabatic temperature gradient. }
   {}


   \maketitle
%

\section{Introduction}
\label{sect:1}

The core helium flash is the most violent event in the life of a star
with an initial mass between approximately $0.7 M_{\odot}$ and $2.2
M_{\odot}$ \citep{SweigertGross1978}. Electron-degeneracy in the
helium core at the time of the flash leads to a thermonuclear runaway
producing, at its peak, large amounts of energy ($\sim 10^{10}
L_{\odot}$) within the stellar core over a very short period of time
($\sim$ days).  The temperature rises up to several $10^8\,$Kelvins until
the degeneracy of the electron gas is eventually lifted. Nevertheless,
the event seems not to be catastrophic for the star. It results only
in a slow expansion of the helium core (typically with a few$\mes$), since
energy transport by turbulent convection, heat conduction, and
radiation seems to be able to deliver most of the flash energy
quiescently from the stellar interior to the outer stellar
layers. However, computations of the flash have a confusing history
predicting either severe explosions
\citep{Edwards1969,Zimmermann1970,Villere1976,Wickett1977,ColeDeupree1980,
  ColeDeupree1981,DeupreeCole1983,Deupree1984a,Deupree1984b,ColeDemDeupree1985}
or a quiescent behavior \citep{Deupree1996,Dearborn2006,
  LattanzioDearborn2006}.

Previous two-dimensional hydrodynamic simulations of turbulent
convection within the helium core during the peak of the core helium
flash \citep{Mocak2008} support a quiescent scenario in agreement with
stellar evolutionary calculations. However, they also showed that the
convection zone, powered by helium burning, is characterized by
convective velocities that are roughly four times larger than those
predicted by mixing-length theory, and that the width of the
convection zone grows on a dynamical timescale. This may lead to mixing of
hydrogen into the helium core as it is known from one-dimensional
simulations of extremely metal-poor stars \citep{FujimotoIben1990,
SchlattlCassisiSalaris2001,CassisiSchlattl2003,WeissSchlattl2004,Campbell2008}.

\begin{table*} 
\caption[]{Some properties of the initial model: total mass $M$,
  stellar population, metal content $Z$, mass $M_{He}$ and radius
  $R_{He}$ of the helium core ($X(^{4}He) > 0.98$), nuclear energy
  production in the helium core $L_{He}$, temperature maximum
  $T_{max}$, radius $r_{max}$, and density $\rho_{max}$ at the
  temperature maximum.}
\begin{tabular}{l|lllllllll} 
Model & $M$  & Pop. & $Z$ & $M_{He}$  & $R_{He}$    & 
$L_{He}$      & $T_{max}$  & $r_{max}$   & $\rho_{max}$ \\ 
      & $[\Msun]$ &      &     & $[\Msun]$ & $[10^9\cm]$ &
$[10^9\Lsun]$ & $[10^8\K]$ & $[10^8\cm]$ & $[10^5\gcm]$  \\
\hline 
M  & $1.25$ & I & $0.02$ & $0.47$ & $1.91$  & $1.03$ 
   & $1.70$ & $4.71$ & $3.44$ \\
\end{tabular} 
\label{imodtab} 
\end{table*}

\begin{figure*} 
\includegraphics[width=0.49\hsize]{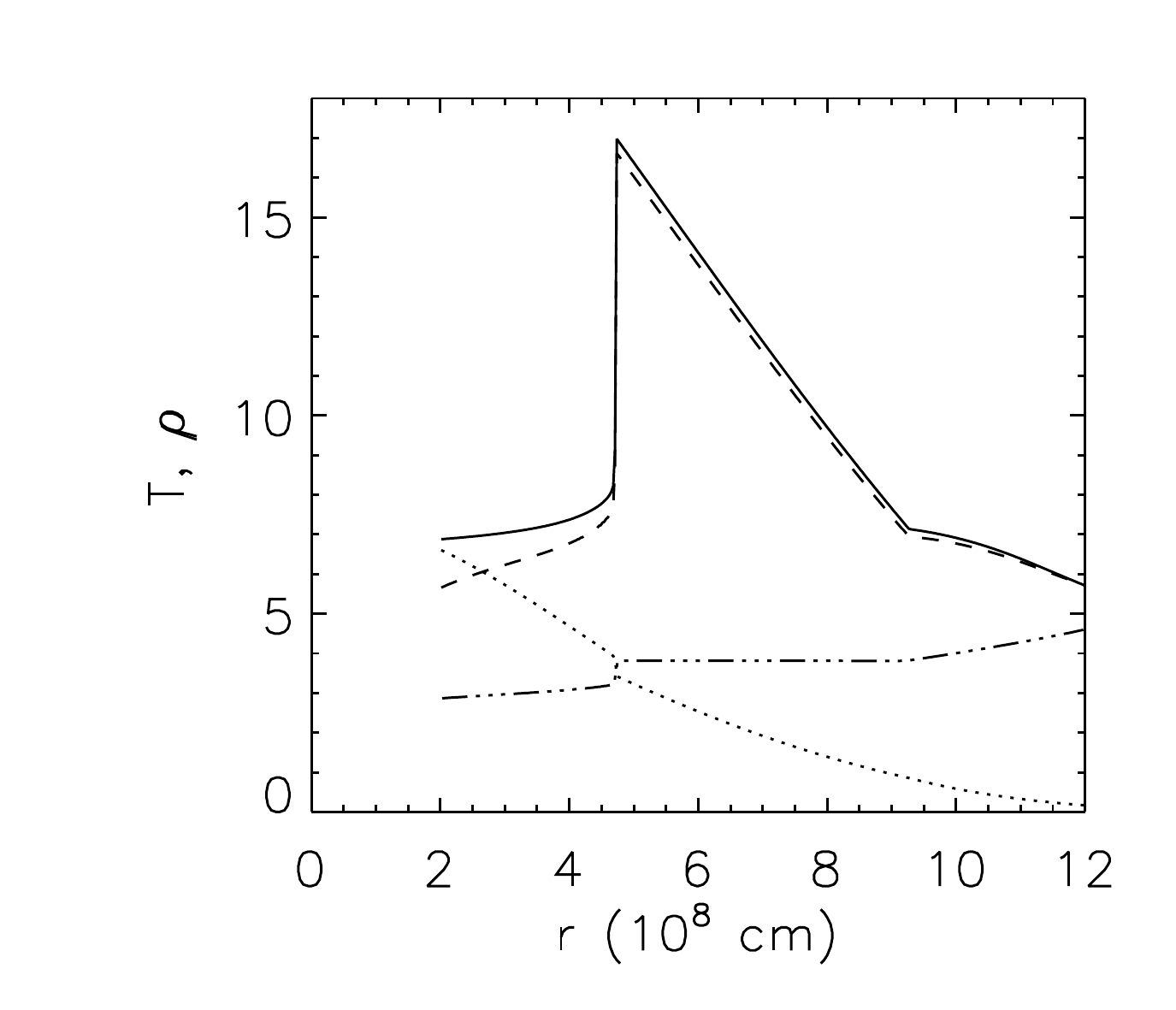} 
\includegraphics[width=0.49\hsize]{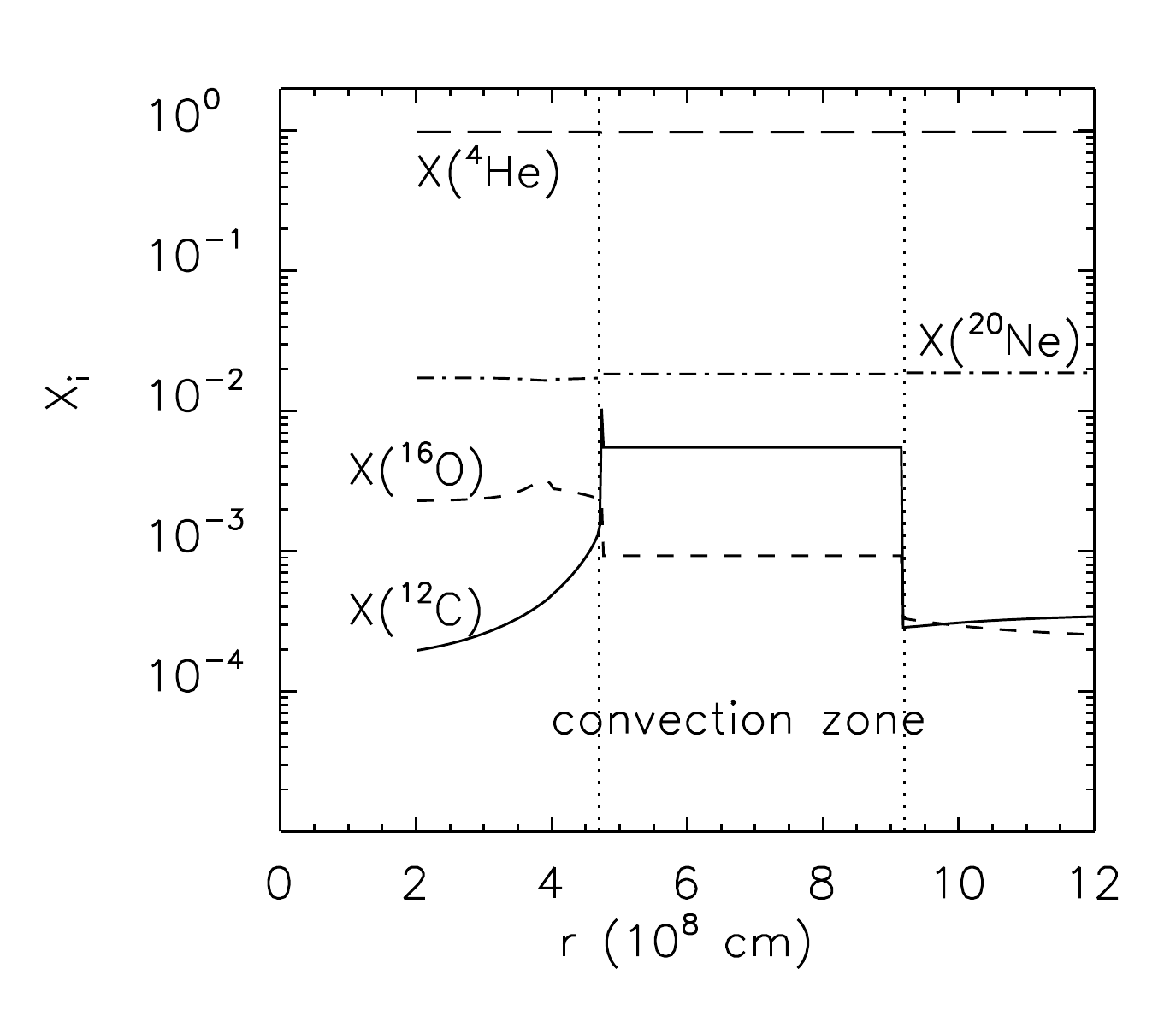} 
\caption{{\it{Left panel:}} Temperature (in units of $10^{7}\K$)
  distribution of the mapped (dashed) and stabilized (solid) initial
  model displayed together with the density (in units of
  $10^{5}\,\gcm$) and entropy (in k$_B$ per baryon) stratification of
  the stellar evolution model (dotted and dash-dotted), respectively.
  {\it{Right panel:}} Chemical composition of the initial model. The
  dotted vertical lines mark the edges of the convection zone.}
\label{fig2.1.2}
\end{figure*}

It is well known that two-dimensional hydrodynamic simulations of
turbulence are seriously biased due to the imposed symmetry
restrictions. Opposite to 3D flows, the turbulent kinetic energy
increases from small to large scales in 2D simulations, i.e., the
energy cascade to smaller length scales characteristic of turbulent
flows is not reproduced \citep{Canuto2000, Bazan1998}. Hence, the mean
convective velocities, the amount of overshooting, and the size of
turbulent structures are too large. Thus, as pointed out already by
\eg \citet{Muthsam1995} and \citet{Bazan1998}, three-dimensional 
simulations are required to validate the predictions of 
two-dimensional simulations.

With increasing computational capabilities, the importance of
multidimensional hydrodynamic simulations in stellar evolution studies
grows rapidly because they are essentially ``parameter free''. In
contrast, canonical one-dimensional stellar evolutionary calculations involve
free parameters like the mixing length or the overshooting distance
\citep{BohmVitense1958,CanutoMazzitelli1991}, which are chosen in an
appropriate manner to fit observational data \citep{Montalban2004}.
Comparison of the results obtained with both approaches is crucial,
because it allows one to constrain, validate or disprove the free
parameters used in stellar evolutionary calculations, and because it
can provide a clue to the applicability limits of the canonical (1D)
treatment
\citep{Bazan1998,Kercek1998,Kercek1999,Asida2000,Herwig2006,Meakin2006,
  Dearborn2006,Arnett2007,MeakinArnett2007}.
 
In the following we present an investigation of the core helium flash
by means of two and three-dimensional hydrodynamic simulation using
state-of-the-art numerical techniques, a detailed equation of state,
and a time-dependent gravitational potential. 
Our hydrodynamic models are characterized by a convectively unstable
layer (the convection zone) embedded in between two stable layers
composed of several chemical nuclear species and of a partially
degenerate electron gas. Similar systems were studied in the past by
many authors \eg \citet{HurlburtToomre1986, HurlburtToomre1994,
  Muthsam1995, Singh1995, Singh1998} and \citet{Brummell2002} assuming, however,
that the stellar matter is composed of a single ideal Boltzmann
gas. This gives extra relevance to our simulations because they allow
us to study, e.g., the dependence of turbulent entrainment and the
structure of convective boundary layers on the composition of the
stellar gas, and on the composition gradients present in the stellar
model.

 We introduce the
stellar model used as input for our multidimensional simulations in
the next section. We then briefly describe our hydrodynamics code in
Sect.\,3, and discuss and compare the results of the two and
three-dimensional simulations in Sect.\,4. In particular, we compare
the temporal evolution, the kinetic energy density, the power spectra,
the thermodynamic fluctuation amplitudes, the stability of the flow
structures, and the turbulent entrainment at the convective boundaries
of our models. We discuss our results in the context of the
predictions of mixing-length theory in Sect.\,5. Finally, we present a
high resolution 2D simulation covering almost 1.5 days of core helium
flash evolution in Sect.\,6. We end with a summary of our findings and
some conclusions in Sect.\,7.

%
\section{Initial model}
\label{sect:2}
The initial model was obtained with the stellar evolution code Garstec
\citep{WeissSchlattl2000,WeissSchlattl2007}. Some of its properties
are listed in Table\,\ref{imodtab}, and the distributions of
temperature, density and composition are displayed in
Fig.\,\ref{fig2.1.2}. The model possesses a white dwarf like
degenerate structure with an off-center temperature maximum resulting
from plasma and photo-neutrino cooling. Its central density is
$7\times 10^5$\gcm. At the outer edge of the isothermal region in the
center of the helium core the temperature jumps up to $T_{max} \sim
1.7\times 10^{8}\,$K, and the adjacent convection zone has a
super-adiabatic temperature gradient. The core is
predominantly composed of $^{4}$He (mass fraction X($^{4}$He)$ >$
0.98), and some small amounts of $^{1}$H, $^{3}$He, $^{12}$C,
$^{13}$C, $^{14}$N, $^{15}$N and $^{16}$O, respectively . For our
hydrodynamic simulations we adopt the abundances of $^{4}$He, $^{12}$C
and $^{16}$O as the triple-$\alpha$
reaction dominates the nuclear energy production during the flash. 
The remaining composition is assumed to be adequately represented by 
a gas with a mean molecular weight equal to that of $^{20}$Ne.

As our multidimensional hydrodynamics code Herakles (see next section)
utilizes an Eulerian computational grid, our initial data for the 
hydrodynamic simulations are 
obtained by polynomial interpolation from the stellar model which
was computed on a Lagrangian grid. The initial
hydrodynamics model obtained in this way is not in hydrostatic 
equilibrium, because the
equation of state of \citet{TimmesSwesty2000} implemented in our
hydrodynamics code differs from that of \citet{Rogers1996} implemented
in the Garstec code (for a given density and temperature in the initial
stellar model, the pressure
differs by about 1\%). Stabilization of this initial hydrodynamics
model resulted in a small decrease of the temperature
compared to the stellar model (Fig.\ref{fig2.1.2}).

%
\section{Hydrodynamic code}
\label{sect:3}
The hydrodynamic simulations were performed with an enhanced version
of the grid-based code Herakles \citep{Mocak2008}.  The adopted
mathematical model implemented in the code consists of the Euler
equations in spherical coordinates ($r, \theta$, and $\phi$; see
Appendix \ref{app:euler_eqs}) coupled to the source terms describing
thermal transport, self-gravity, and nuclear burning. The hydrodynamic
equations are integrated using the PPM reconstruction scheme
\citep{ColellaWoodward1984}, and a Riemann solver for real gases
according to \citet{ColellaGlaz1984}. The chemical species are evolved
by a set of additional coupled continuity equations
\citep{PlewaMueller1999}. Source terms in the momentum and energy
equations due to self-gravity and nuclear burning are treated by means
of operator splitting at the end of every integration step.  The
gravitational potential is approximated by a one-dimensional Newtonian
potential which is obtained from the spherically averaged mass
distribution. The stiff nuclear reaction network is integrated using
the semi-implicit Bader-Deuflhard method,
\citep{BaderDeuflhard1983,Press1992} which allows for large time
integration steps.

In Herakles, a program cycle consists of two hydrodynamic time steps and
proceeds as follows (in 2D simulations Step (3) is omitted): 
\begin{enumerate}
\item The hydrodynamic equations are integrated in $r$-direction
  (r-sweep) for one time step including the effects of heat conduction. The time
  averaged gravitational forces are computed, and the momentum and the
  total energy are updated to account for the respective source terms.
  Subsequently, the equation of state is called to update the
  thermodynamic state due to the change of the total energy.
\item Step (1) is repeated in $\theta$-direction ($\theta$-sweep).
\item Step (1) is repeated in $\phi$-direction ($\phi$-sweep).
\item The nuclear reaction network is solved in all zones with
  significant nuclear burning (i.e, where $T > 10^{8}\,$K), and then
  the equation of state is called to update the pressure and the
  temperature.
\item In the subsequent time step, the order of Step (1), (2), and (3)
  is reversed to guarantee second-order accuracy of the time
  integration, and Step (4) is repeated with the updated state
  quantities.
\item The size of the time step for the next cycle is determined.
\end{enumerate} 

The velocities in the convection zone, even at the peak of the core
helium flash, are subsonic corresponding to Mach numbers $M \sim 0.01$
\citep{Mocak2008}.  In this regime, the applicability of Riemann
solver based methods, like PPM, has been questioned 
\citep{Schneider1999,Turkel1999,Almgren2006}. However, a
recent study by \citet{Meakin2007ane} based on a direct comparison of
anelastic \citep{Kuhlen2003} and fully compressible simulations
computed with PPM shows that at Mach numbers around 10$^{-2}$, PPM 
can capture the properties of convective flows well.


\begin{table*} 
\caption{Some properties of the three and two-dimensional simulations:
  number of grid points in $r$ ($N_{r}$), $\theta$ ($N_{\theta}$) and
  , $\phi$ ($N_{\phi}$) direction, spatial resolution in r ($\Delta r$
  in 10$^{6}$ cm), $\theta$ ($\Delta \theta$), and $\phi$ ($\Delta
  \phi$) direction, characteristic velocity $v_{c}$ (in
  10$^{6}\,\cms$) of the flow during the first 6000\,s, expansion
  velocity at temperature maximum $v_{exp}$ (in $\mes$), typical
  convective turnover time $t_{o} = 2\,R/v_{c}$ (in s) where $R$ is
  the height of the convection zone, and maximum
  evolutionary time $t_{max}$ (in s), respectively.}

\begin{center}
\begin{tabular}{p{1.cm}|p{2.cm}p{0.7cm}p{0.7cm}p{0.7cm}
  p{0.7cm}p{0.7cm}p{0.7cm}p{1.1cm}} 
\hline
\hline
run & grid & $\Delta r$ & $\Delta\theta$ & $\Delta\phi$ & 
$v_{c}$ & $v_{exp}$ & $t_{o}$ & $t_{max}$ 
\\
\hline 
hefl.2d.a & $180\times90$ & 5.55 & 1.5\degr & -  &
$1.44$  & 24. & 650  & 6000\\
hefl.2d.b & $360\times240$ & 2.77 & 0.75\degr & -  &
$1.84$  & 92. & 510 & 130000\\
hefl.3d & $180\times90\times90$ & 5.55 & 1.5\degr & 1.5\degr &
$0.85$ & 10. & 1105 & 6000\\
\hline
\end{tabular} 
\end{center}
\label{modtab} 
\end{table*}

%
\section{Hydrodynamic simulations}
\label{sect:4}
We performed two two-dimensional and one three-dimensional simulation
whose properties are summarized in Table\,\ref{modtab}. 
\footnote{We simulated another 3D model at lower resolution than model
  hefl.3d (using a wedge of $90 \times 80 \times 80$ zones centered at
  the equator with $\Delta r = 11.1\times 10^{6}\,$cm and $\Delta
  \theta = \Delta \phi = 1.5\dgr$), but we do not discuss this model
  here any further, because of its larger numerical viscosity and its
  almost 50\% smaller convective velocities.}
All simulations, except model hefl.2d.b, cover 6000\,s of evolutionary
time during the peak of the core helium flash, and were computed on a
computational grid spanning a wedge of $120\dgr$ in
angular directions centered at the equator. The grid had a radial
resolution of $\Delta r = 5.55\times 10^{6}\,$cm, and an angular
resolution of $\Delta \theta = \Delta \phi = 1.5\dgr$).  The rather
wide angular grid appeared to be necessary for the three-dimensional
simulations, due to the size of the largest vortices ($\sim$\,40\dgr)
found in previous two-dimensional simulations \citep{Mocak2008}.

The 2D model hefl.2d.b was evolved for almost 34 hrs (130\,000 s) on a
grid covering the full $180\dgr$ angle in $\theta$ direction. It was
simulated at almost twice the resolution as the other 
models. The properties of model hefl.2d.b and its evolution are
discussed in Section\,\ref{sect:6}, and to some extent also in
Sections\,\ref{sect:4}\,, and \ref{sect:5}.

All models posses a convection zone that spans 1.5 density scale
heights and that is enclosed by convectively stable layers extending
out to a radius of $1.2\times 10^{9}$\,cm. We used reflective boundary
conditions in every coordinate direction and for all models, except
model hefl.2d.a. For this model, it turned out to be necessary
to impose periodic boundary conditions in angular direction, because
reflective boundaries together with the 120 \dgr wedge size affect the
large scale convective flow adversely leading to higher convection
velocities.
 
\begin{figure} 
\includegraphics[width=0.99\hsize]{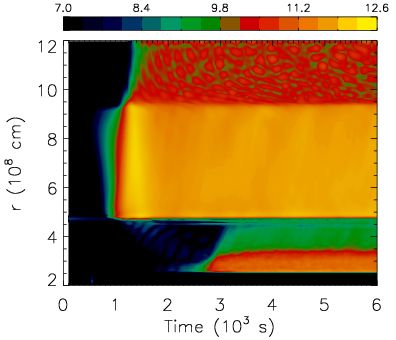}
\includegraphics[width=0.99\hsize]{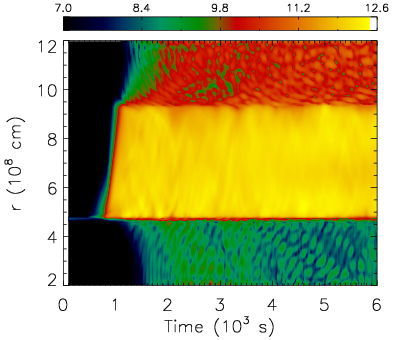}
\caption{Temporal evolution of the angular averaged kinetic energy
  density (in \ergs) of models hefl.3d (upper) and hefl.2d.a (lower),
  respectively.}
\label{fig4.1.2}
\end{figure}

\begin{figure} 
\includegraphics[width=0.99\hsize]{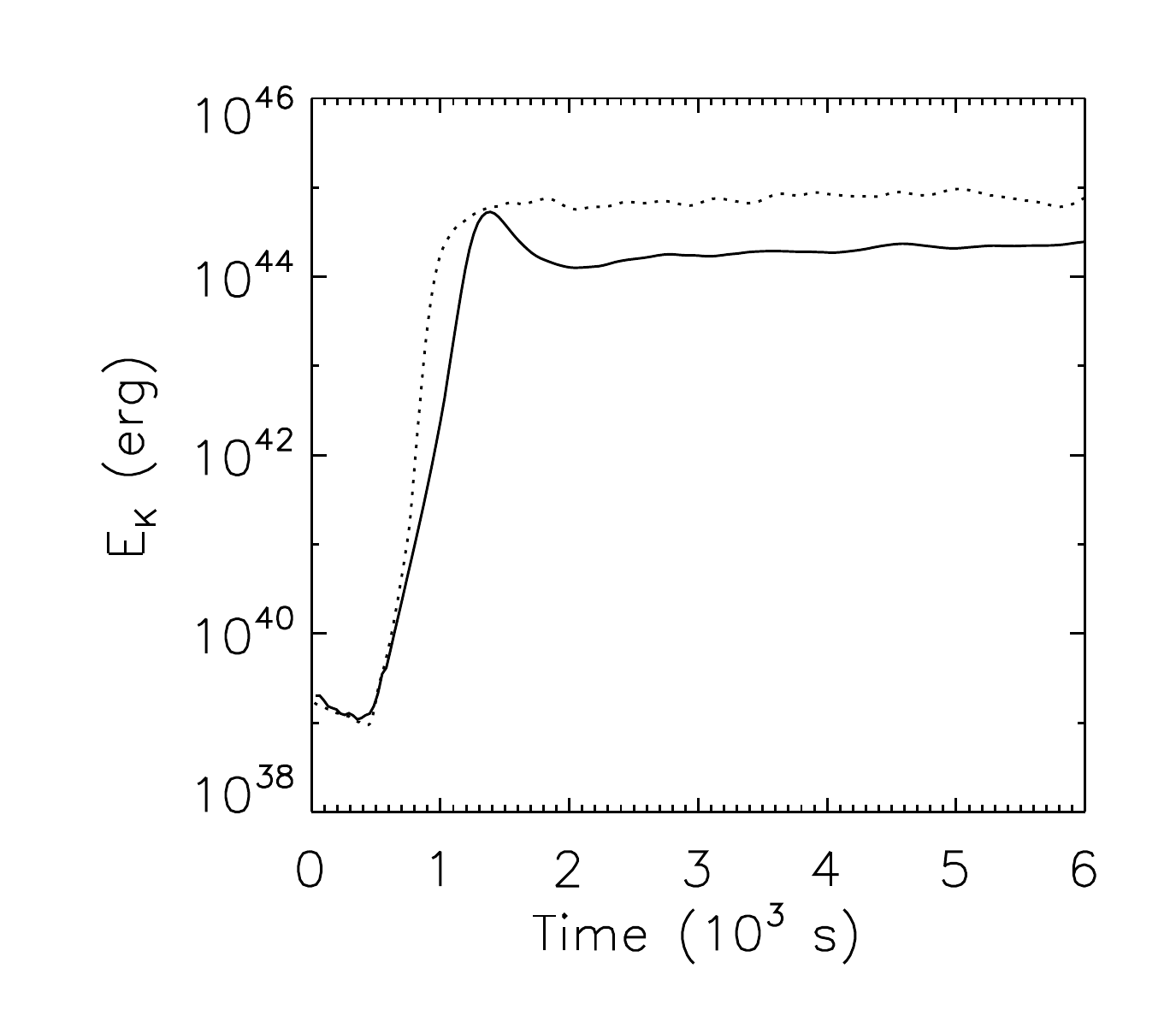} 
\caption{Temporal evolution of the total kinetic energy E$_K$ of
  models hefl.3d (solid) and hefl.2d.a (dotted), respectively.}
\label{fig4.3}
\end{figure}  

We are limited with our explicit hydrodynamics code by the CFL
condition which is most restrictive near the center of the spherical
grid. Therefore, we cut off the inner part of the grid at a radius of
$2\times 10^{8}$\,cm that is still sufficiently far away from the
radius of the temperature inversion at r$\,\sim 5 \times
10^{8}\,$cm. To trigger convection, we imposed a random flow field
with a maximum (absolute) velocity of $10 \cms$, and a random density
perturbation $\Delta \rho / \rho \le 10^{-2}$.  Imposing some
explicit non-radial perturbations is necessary, because a spherically
symmetric model evolved with Herakles on a grid in spherical
coordinates will remain spherically symmetric forever. This is
different from the study of \cite{Asida2000}, who did not perturb the
initial model by an artificial random flow, as instabilities were
growing from round-off errors. The different
perturbation techniques seem not to influence the final thermally
relaxed steady state \citep{MeakinArnett2007}.

Because thermal transport of energy by conduction and radiation is
roughly seven orders of magnitude smaller than the convective energy
flux, it has been neglected in our simulations. Most of the liberated
nuclear energy is carried away by convection.  All computed models 
are non-rotating, because rotation seems not to play an important role 
during the core helium flash \citep{LattanzioDearborn2006}.

%
\subsection{Temporal evolution}
The 3D model hefl.3d and the corresponding 2D model hefl.2d.a evolve
initially (t $<$ 1200 s) similarly. Convection sets in after roughly
1000~s (the thermal relaxation time), and hot rising bubbles appear in
the region where helium burns in a thin shell (r$\,\sim 5\times
10^{8}\,$K). After another $\sim 200\,$s, the bubbles cover the
complete height of the convection zone (R\,$\sim$ $4.8\times
10^{8}$\,cm). The flow eventually approaches a quasi-steady state consisting
of several upstreams (or plumes) of hot gas carrying the liberated
nuclear energy off the burning region, thereby inhibiting a
thermonuclear runaway.

The models exhibit a sandwich-like stratification: two stable layers
enclose the convection zone at the top and bottom, respectively
(Fig.\,\ref{fig4.1.2}).  The convection zone is characterized by a
large kinetic energy density, and the adjacent convectively stable
layers show waves induced by convection. The region of high kinetic
energy density appearing at $\sim 3000\,$s in the bottom layer in
model hefl.3d (see top panel of Fig.\,\ref{fig4.1.2}) is an
artifact caused by the proximity of the reflective inner radial 
boundary (at $2\times 10^{8}$\,cm) of the computational grid.

\begin{figure*}
\includegraphics[width=6.0cm]{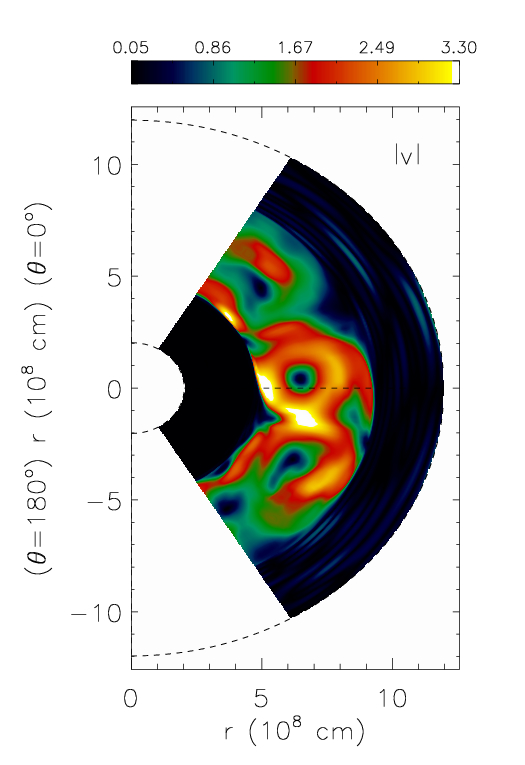}
\includegraphics[width=6.0cm]{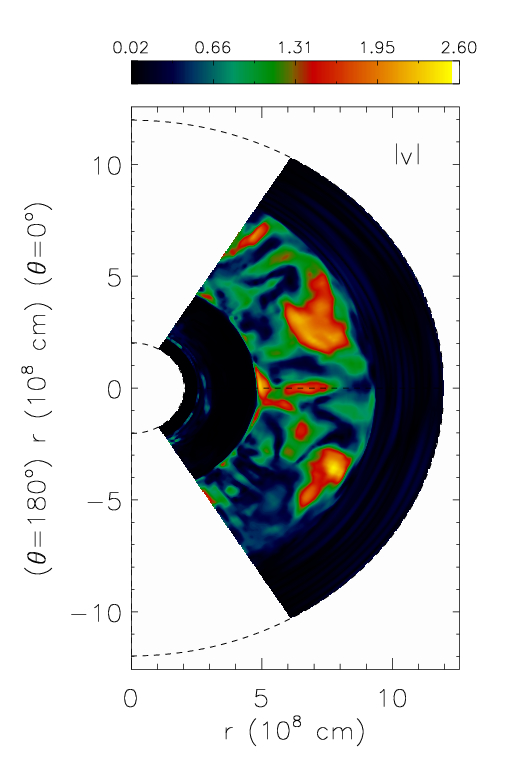} 
\includegraphics[width=6.0cm]{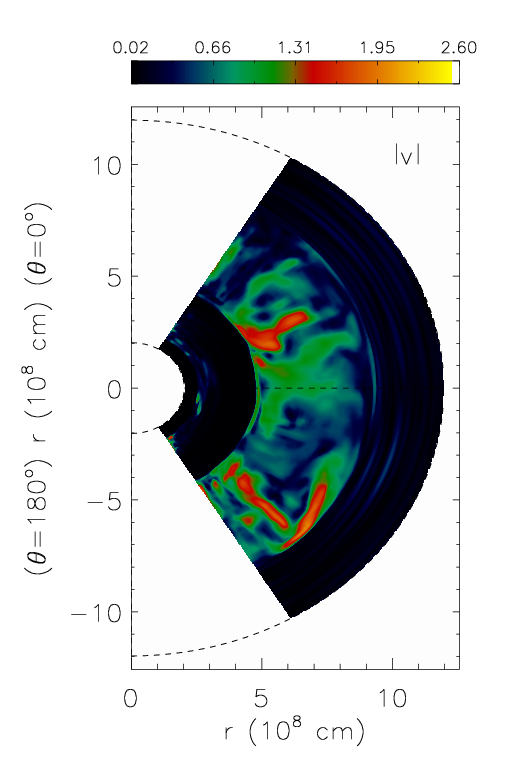} 
\includegraphics[width=0.49\hsize]{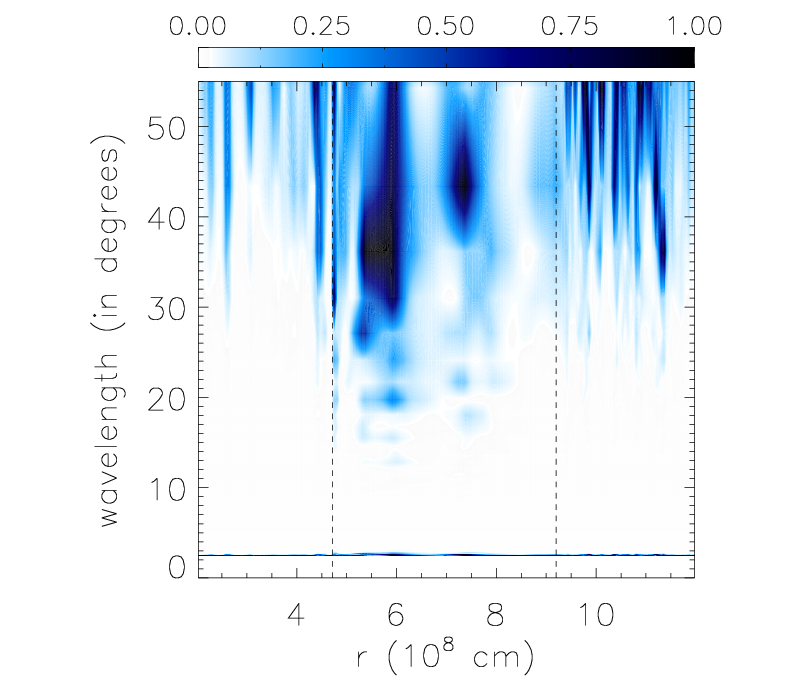}
\includegraphics[width=0.49\hsize]{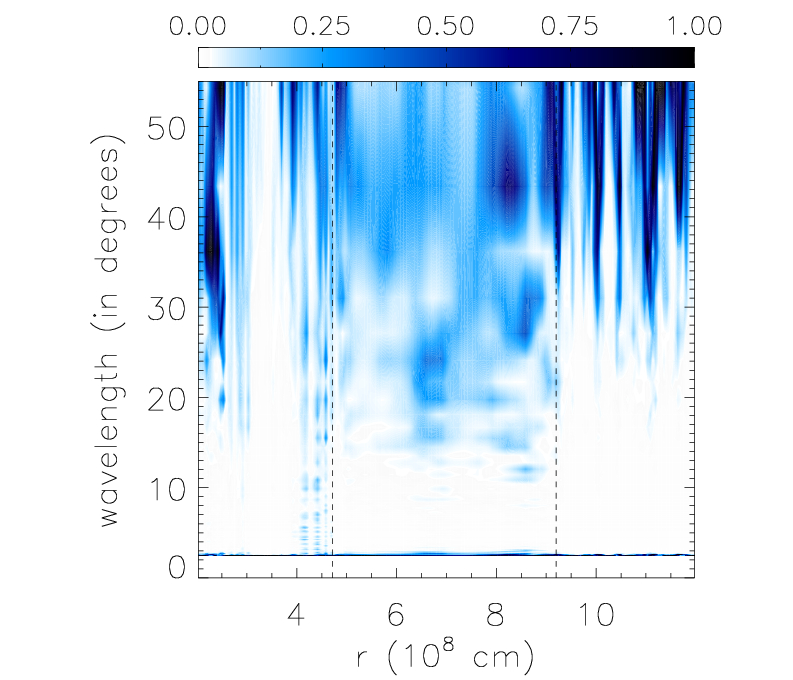}
\caption{
  {\it Upper panels:} Snapshots taken at $\sim 4754\,$s showing
  contour plots of the absolute value of the velocity (in units of
  $10^{6} \cms$) for the 2D model hefl.2d.a (left), and for the 3D
  model hefl.3d in a meridional plane of azimuth angle $\phi = 50\dgr$
  (middle) and $\phi = 70\dgr$ (right), respectively.
  {\it Lower panels:} Normalized power spectra of angular
  fluctuations in the absolute velocity as a function of radius for
  the 2D model helf.2d.a (left) averaged over time from 2100\,s to
  9500\,s, and for the 3D model hefl.3d (right) averaged over time
  from 2250\,s to 6000\,s, and azimuthal angle. The dashed vertical
  lines mark the edges of the convective zone of the initial model
  according to the Schwarzschild criterion.}
\label{fig4.c1}
\end{figure*} 

The models reach a steady state after roughly 2000\,s
(Figs.\,\ref{fig4.1.2} and \ref{fig4.3}). The steep increase of the
total kinetic energy from 10$^{39}\,$erg to $10^{45}\,$erg
(Fig.\,\ref{fig4.3}) marks the onset of convection. The kinetic energy
density shows small fluctuations in the fully evolved convection zone,
and is by an order of magnitude larger in model hefl.2d.a than in
model hefl.3d. This is in agreement with other studies, as it is well
know that two-dimensional turbulence is more intensive (see, \eg
\cite{Muthsam1995}). The total energy production is about 10\% higher
in the 3D model due to its slightly higher temperature and the strong
dependence of the triple-$\alpha$ reaction rate on temperature.  

We observe buoyancy driven gravity waves in the convectively stable
layers (Fig.\,\ref{fig4.1.2},~Fig.\,\ref{fig4.14.15}) \citep{Zahn1991, 
HurlburtToomre1986,
HurlburtToomre1994, MeakinArnett2007}, but we do not discuss these
waves any further here, because their properties are likely biased by
the reflective boundaries \citep{Asida2000}.  We only point out that 
the differences in
amplitude and frequency seen in Fig.\,\ref{fig4.1.2} are physical, \ie
gravity waves have a lower frequency and amplitude in 3D than in 2D,
although the energy carried by them is likely similar
\citep{Kiraga1999}.

%
\subsection{Structure of the convective flow}
Fully evolved convection ($t > 2000\,$s) in the 3D model hefl.3d
differs significantly from that in the corresponding 2D model
hefl.2d.a.  The convective flow is dominated in the 2D model by
vortices having (angular) diameters ranging from 30\degr to 50\degr
(Fig.\ref{fig4.c1}), and an aspect ratio of close to one.  The
vortices are qualitatively similar to those found in other
two-dimensional simulations \citep{HurlburtToomre1986,
  HurlburtToomre1994, PorterWoodward1994, Bazan1998}.  This vortex
structure of 2D turbulence is quite typical, and arises from the
self-organization of the flow \citep{Fornberg1977, McWilliams1984}.

The convective flow in the 3D model hefl.3d consists of column-shaped
plumes (Fig.\,\ref{fig4.c1}, and \ref{fig4.14.15}), and contrary to
the 2D model hefl.2d.a, it does not show any dominant angular mode.
The typical angular size of turbulent features ranges from 10\degr\,
to\, 30\degr\, (Fig.\ref{fig4.c1}). The power spectra of angular
velocity fluctuations show that turbulent elements have an almost
time-independent characteristic angular size of 30\degr\,-\,50\degr in
case of the 2D model, while the spectra computed for the 3D model
change with time and exhibit no dominant angular mode.

We find turbulent flow features across the whole convection zone
resulting from the interaction of convective up and downflows.  Close
to the edges of the convective zone we observe the smallest turbulent
flow features that form when compact turbulent plumes are decelerated
and break-up \citep{Brummell2002}. Shear instabilities likely play an
important role in the development of the turbulent flow as well
\citep{Cattaneo1991}.

\begin{table}
\caption{Root mean square fluctuation amplitudes of various variables
  within the convection zone ({\it{cnvz}}: $5\times10^{8}\,\mathrm{cm}
  \le r \le 9.2\times 10^{8}\,$cm) averaged over a time period of
  about $2500\,$s: temperature $T^{'} / \langle T \rangle$, density
  $\rho^{'} / \langle \rho \rangle$, helium abundance $^{4}$He$^{'} /
  \langle ^{4}$He$\rangle$, and carbon abundance $^{12}$C$^{'} /
  \langle ^{12}$C$\rangle$.}
\centering 
\begin{tabular}{l|c|cccc} 
\hline\hline 
pos & run 
& $T^{'} / \langle T \rangle$ 
& $\rho^{'} / \langle \rho \rangle$ 
& $^{4}$He$^{'} / \langle ^{4}$He$\rangle$
& $^{12}$C$^{'} / \langle ^{12}$C$\rangle$
\\ [0.3ex]
\hline 
&hefl.3d & 0.00058 & 0.00015 & 0.00009 &  0.01433  \\[-0.2ex]
\raisebox{1.5ex}{cnvz} \raisebox{1.5ex}
&hefl.2d.a & 0.00074 & 0.00021 & 0.00007  & 0.01272  \\[0.5ex]
\hline
\end{tabular}
\label{tab:mflcnvz}
\end{table}

Tables\,\ref{tab:mflcnvz} and \ref{tab:mlftbnry} provide time-averaged
root mean square fluctuation amplitudes of various variables of the
convective flow inside the convection zone and near its edges,
respectively for models hefl.3d and hefl.2d.a. The temperature and
density fluctuations are 30-40\,\% larger in the 2D model than in the
3D one. This is expected, because vortices are stable in 2D flows but
decay in 3D ones (Fig.\ref{fig4.c1}). The fluctuations in the 
composition ($^{4}$He and $^{12}$C) are larger by 10-30\% in the 3D
model which is a result of more ``broken'' and hence more non-uniform mixing 
of chemical elements than in the 2D one.

The temperature and density fluctuation amplitudes are a
factor of 2-3 larger near the inner edge than near the outer edge of
the convective layer.  At both edges are the fluctuation amplitudes by a
factor of 2-4 larger in 2D than in 3D models.  The fluctuations in the
composition ($^{4}$He and $^{12}$C) in both models differ close to 
the convective boundaries as well, by up to a factor of two.

\begin{table}[!ht]
\caption{Root mean square fluctuation amplitudes of various variables
  at the {\it{inner}} ($r = 5\times 10^{8}\,$cm) and {\it{outer}} ($r
  = 9.2\times 10^{8}\,$cm) edge of the convection zone averaged over a
  time period of about $2500\,$s: temperature $T^{'} / \langle T
  \rangle$, density $\rho^{'} / \langle \rho \rangle$, helium
  abundance $^{4}$He$^{'} / \langle ^{4}$He$\rangle$, and carbon
  abundance $^{12}$C$^{'} / \langle ^{12}$C$\rangle$.}
\centering 
\begin{tabular}{l|c|cccc} 
\hline\hline 
pos & run 
& $T^{'} / \langle T \rangle$ 
& $\rho^{'} / \langle \rho \rangle$ 
& $^{4}$He$^{'} / \langle ^{4}$He$\rangle$
& $^{12}$C$^{'} / \langle ^{12}$C$\rangle$
\\ [0.3ex]
\hline 
&hefl.3d & 0.00643 & 0.00144 & 0.00045 &  0.11497  \\[-0.2ex]
\raisebox{1.5ex}{inner} \raisebox{1.5ex}
&hefl.2d.a & 0.02027 & 0.00441 & 0.00024 & 0.08958  \\[0.5ex]
\hline
&hefl.3d & 0.00420 & 0.00117 & 0.00089 & 0.45105 \\[-0.2ex]
\raisebox{1.5ex}{outer} \raisebox{1.5ex}
& hefl.2d.a & 0.00626 & 0.00177 & 0.00141 & 0.62193  \\[0.5ex]
\hline 
\end{tabular}
\label{tab:mlftbnry}
\end{table}

%
\subsection{Stability of flow structures}
\begin{figure} 
\includegraphics[width=0.99\hsize]{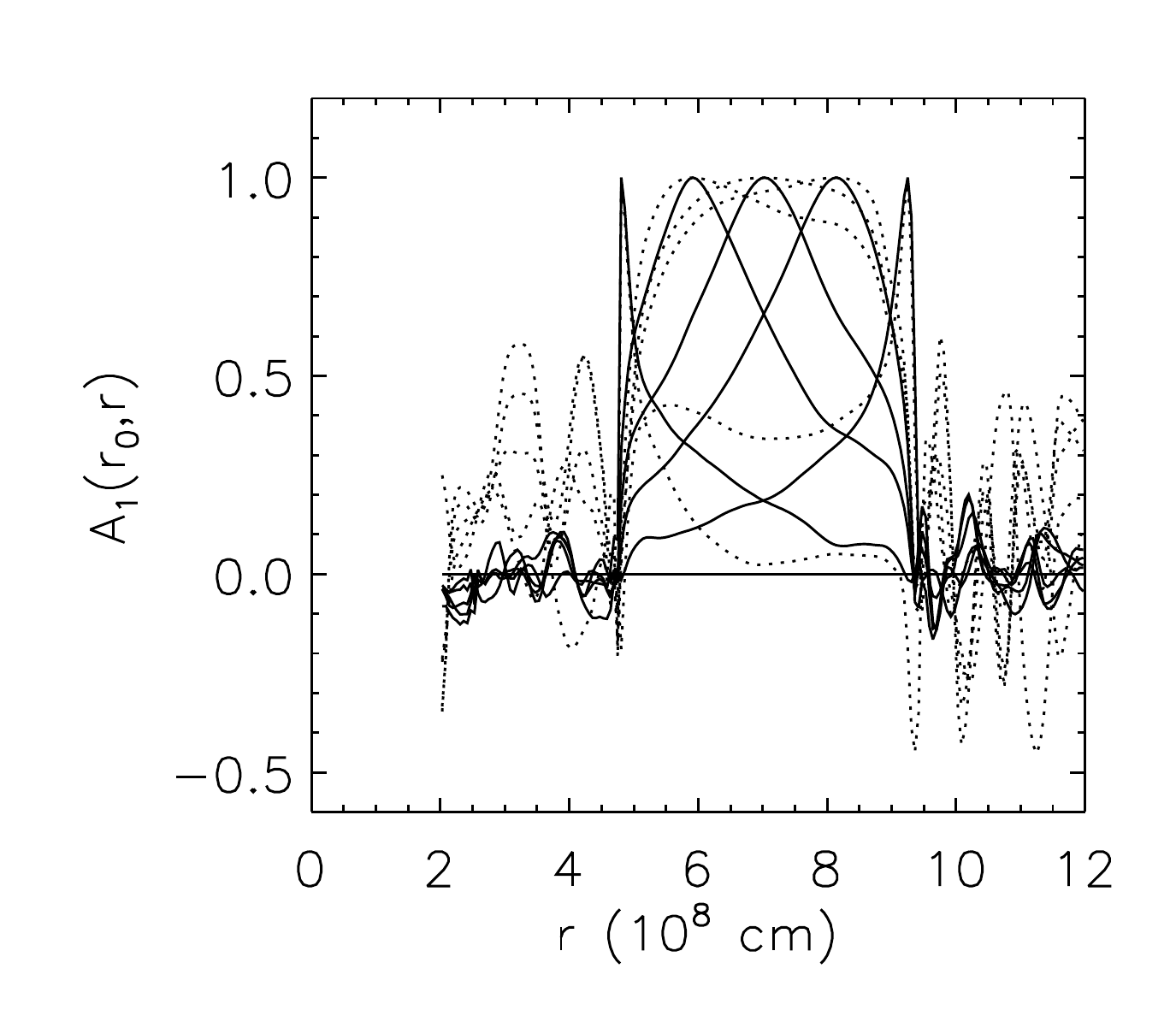}
\caption{Auto-correlation function $A_{1}(r_{0};r)$ measuring the
  radial extent of flow patterns (see Eq.\,\ref{eq.auto1}) at
  different radii $r_{0}$ ($4.8\times 10^{8}\,$cm, $5.9\times
  10^{8}\,$cm, $7\times 10^{8}\,$cm, $8.1\times 10^{8}\,$cm, and
  $9.25\times 10^{8}\,$cm) and $t \sim 4000\,$s measuring the radial
  extent of flow patterns for the 2D model hefl.2d.a (dotted lines)
  and the 3D hefl.3d (solid lines), respectively. }
\label{fig4.4}
\end{figure} 

\begin{figure} 
\includegraphics[width=0.99\hsize]{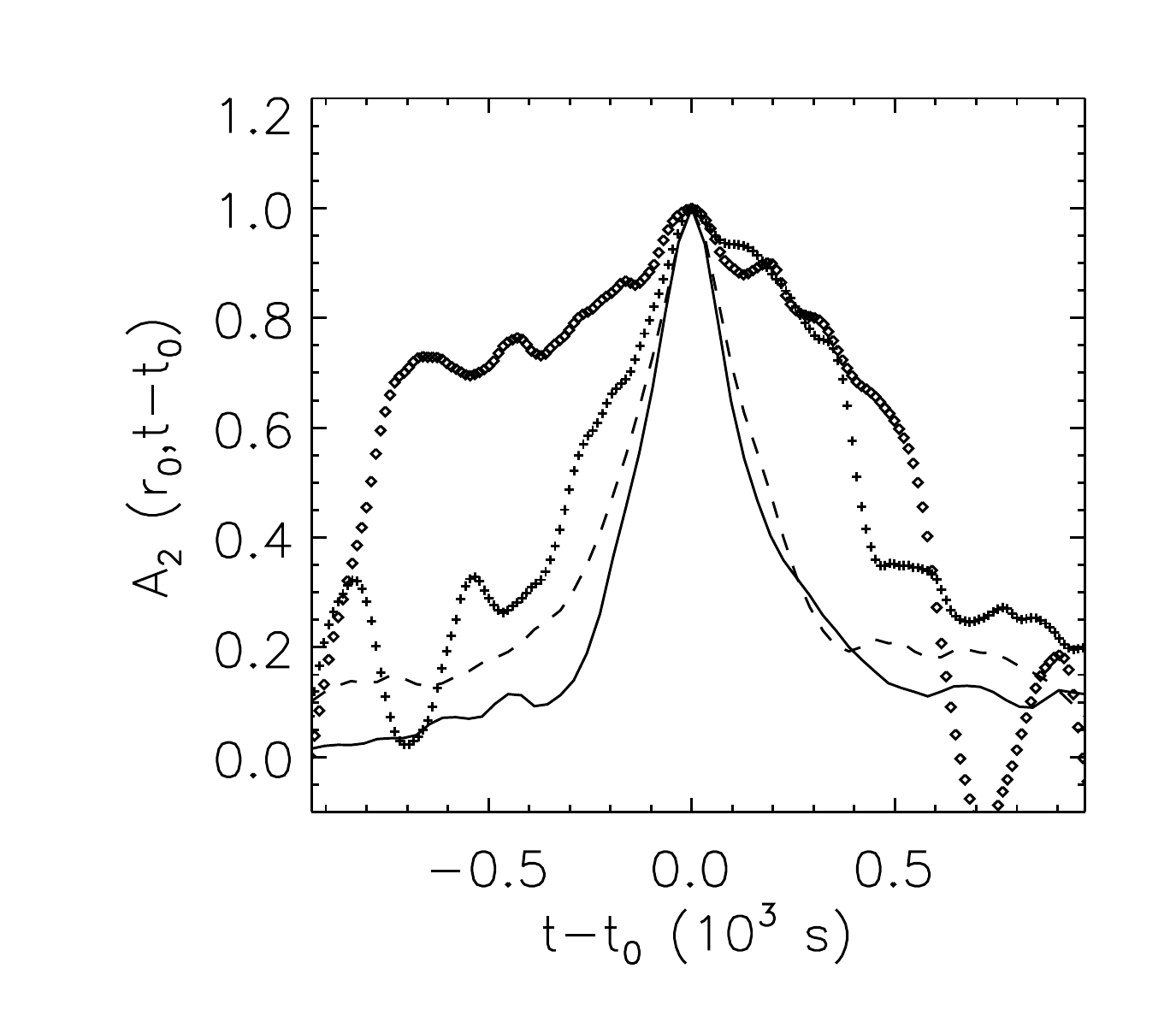}
\caption{Auto-correlation function $A_{2}(r_{0};t-t_{0})$ measuring
  the lifetime of flow patterns (see Eq.\,\ref{eq.auto2}) at two
  different epochs for the 3D model hefl.3d (solid: $t_{0} \sim
  2260\,$s, dashed: $t_{0} \sim 2900\,$s), and for the 2D model
  hefl.2d.a (crosses, diamonds), respectively. The radius $r_{0}$ is
  $7.6\times 10^{8}\,$cm. }
\label{fig4.5}
\end{figure}

To analyze the size and the stability of the vortices we introduce an
auto-correlation function of the radial velocity
\begin{equation}
A_{1}(r_{0};r) = \frac{\langle v_{r}(r_{0})\, v_{r}(r)\rangle_{\Omega},t}{
                    \langle v_{r}(r_{0})^{2}\rangle^{1/2}_{\Omega,t}\, 
                    \langle v_{r}(r    )^{2}\rangle^{1/2}_{\Omega,t}}
\label{eq.auto1}
\end{equation}
that measures the radial extent of flow patterns at a given radius
$r_{0}$, or equivalently the radial size of vortices. The notation
$\langle \rangle_{\Omega,t}$ indicates averaging over angles and
time. A second auto-correlation function
\begin{equation}
A_{2}(r_{0};t-t_{0}) = \frac{\langle v_{r}(r_{0},t_{0})\,
                                    v_{r}(r_{0},t)\rangle_{\Omega}}{
                            \langle v_{r}(r_{0},t_{0})^{2}
                            \rangle^{1/2}_{\Omega} \, 
                            \langle v_{r}(r_{0},t)^{2} \rangle^{1/2}_{\Omega}}
\label{eq.auto2}
\end{equation}
provides a measure of the lifetime of flow patterns at radius $r_{0}$,
beyond time $t_{0}$. Here, we average only over angles. Both
correlation functions have the properties $-1 \le A_{1,2} \le 1$, and
$A_{1}(r_{0};r_{0}) = 1$ and $A_{2}(r_{0};t_{0}-t_{0}) = 1$,
respectively. They are similar to the autocorrelation function used by
\citet{ChanSofia1982, ChanSofia1986}.

Figure \ref{fig4.4} displays the radial auto-correlation for models 
hefl.2d.a and hefl.3d.a and confirms the extension of the convective flow 
across the whole convective region as determined by the Schwarzschild 
criterion in the initial stellar model.  
The broad plateaus with $A_{1} \approx 1$ corresponding to the
2D model hefl.2d.a bear eveidence of the axial symetry 
imposed in the two-dimensional case which leads to 
pronouced circular vortices. In the three-dimensional case, the 
distributions of $A_{1}$ tend to differ from unity at nearly 
all radii. 

Figure\,\ref{fig4.5} shows the temporal auto-correlations with 
two typical results for different $t_{0}$. The three-dimensional
model always shows the typical behaviour of a decrease of the
function value to 0.5 within 200-250 s. From this we conclude that
the flow pattern fluctuates always in the same way. This is different
in the two-dimensional model, where $A_{2}$ can keep high values for a 
much longer time implying rather persistent structure (the vortices) 
of the convective flow.


%
\subsection{Turbulent entrainment and the width of the convection zone}
Convection may induce mixing in convectively stable layers adjacent to
convectively unstable regions. Following \citet{MeakinArnett2007}, we
prefer to call this process turbulent entrainment (or mixing), a term 
also known in oceanography, see \eg \citet{Fernando1991}. The commonly used term
(convective) overshooting accounts only for localized ascending or
descending plumes crossing the edge of the convection zone. If the
filling factor of these plumes or their crossing frequency is high,
they can change the entropy in convectively stable regions surrounding
convection zones, a process that is known as penetration
\citep{Brummell2002}. Turbulent entrainment accounts for both
overshooting and penetration.

\begin{figure} 
\includegraphics[width=0.99\hsize]{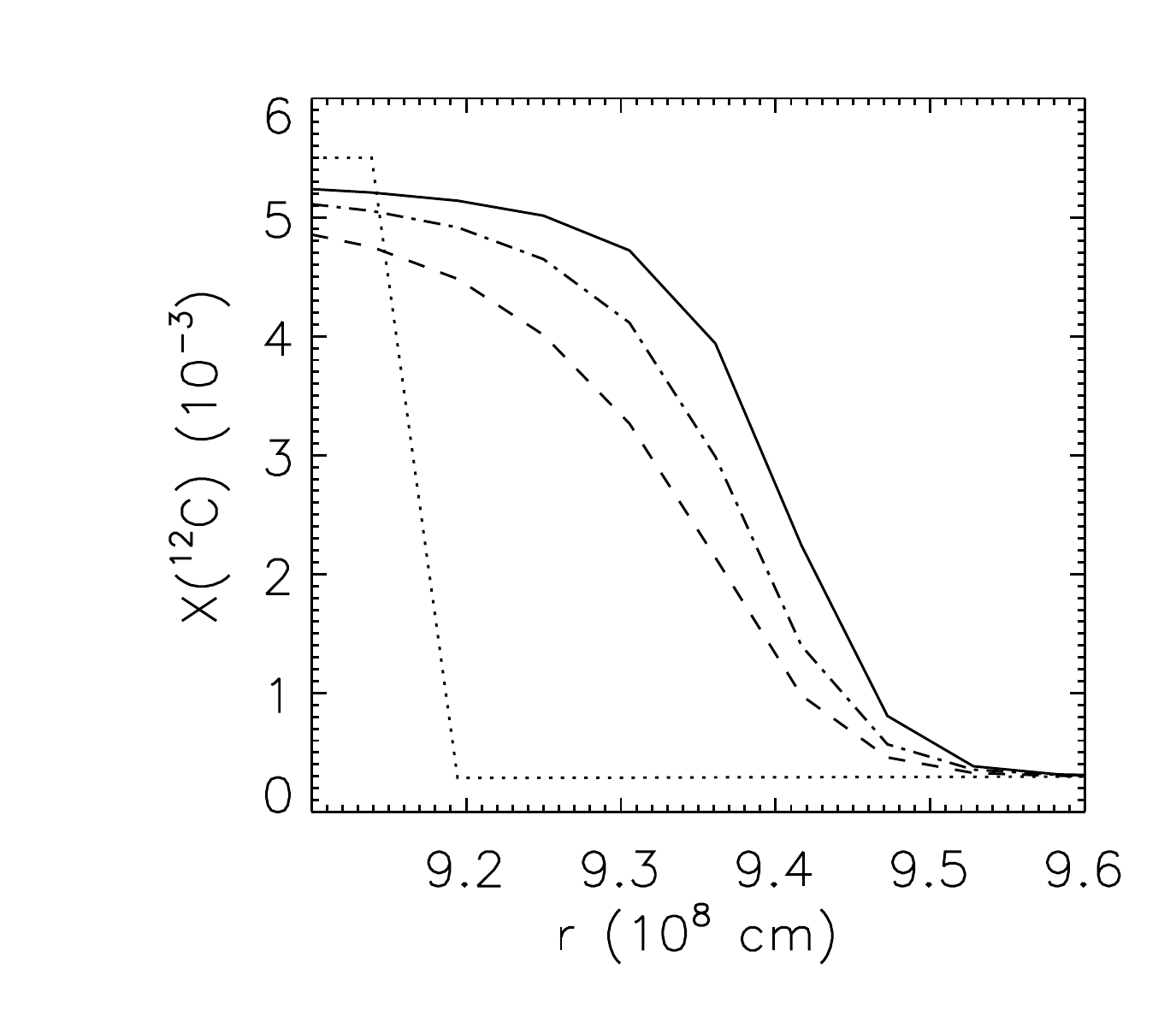}
\includegraphics[width=0.99\hsize]{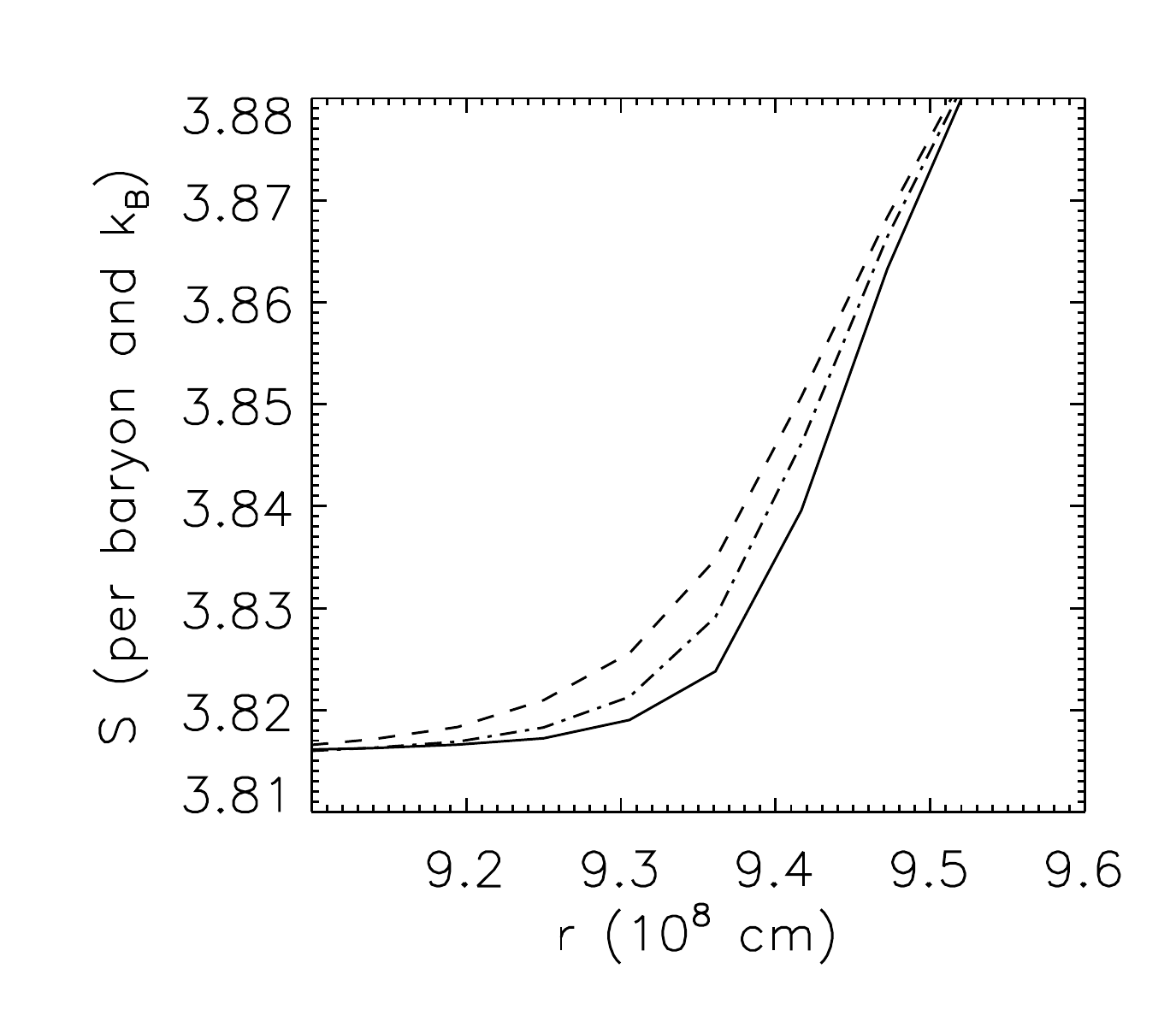}
\includegraphics[width=0.99\hsize]{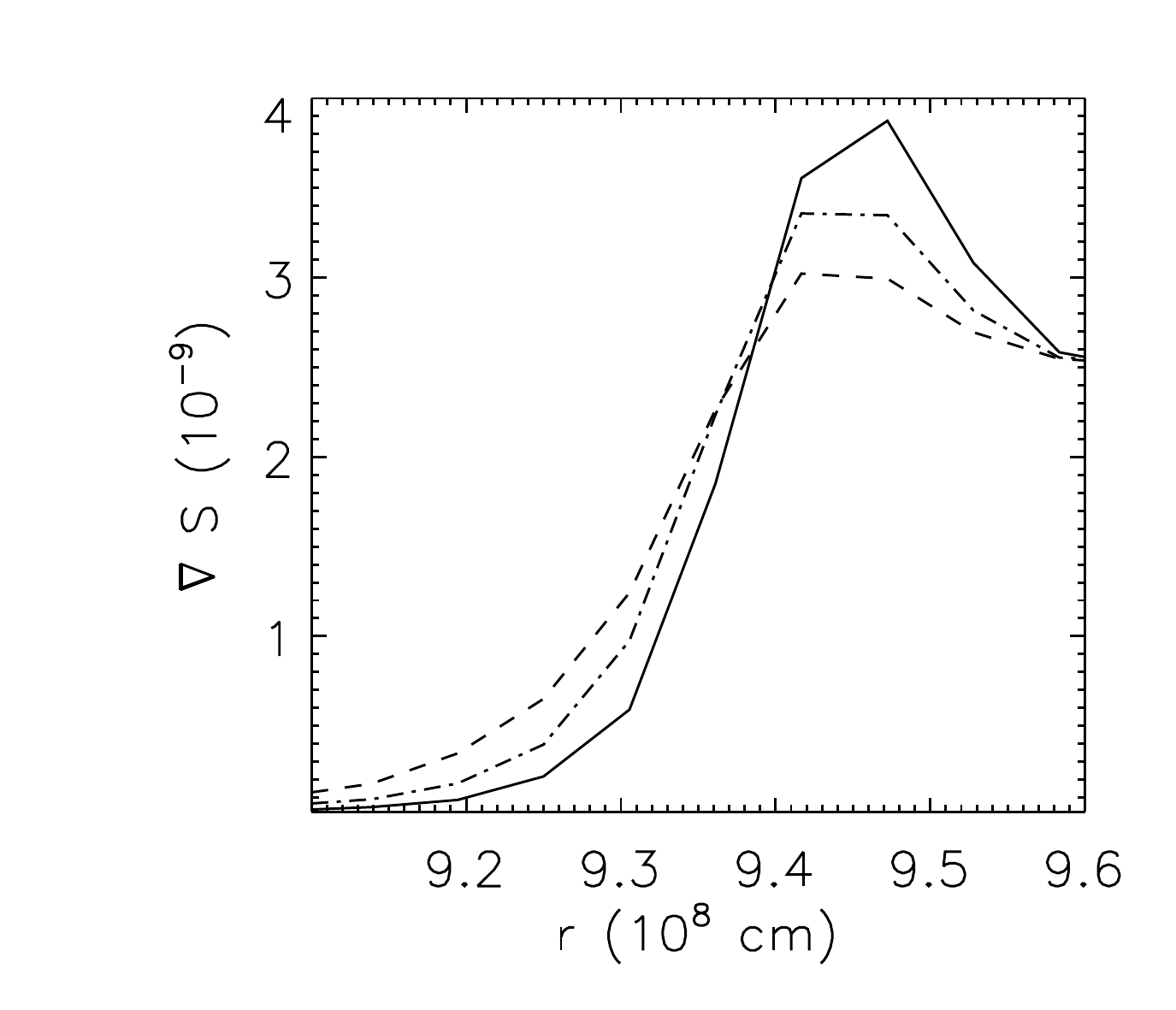}
\caption{Carbon mass fraction X($^{12}$C) (top), entropy S (middle),
  and entropy gradient $\nabla S$ (bottom) as a function of radius
  near the outer edge of the convection zone of model hefl.3d at three
  different epochs: $t_1 = 2000\,$s (dashed), $t_1 = 4000 \,$s
  (dash-dotted), and $t_1 = 6000 \,$ (solid). In addition, the initial
  X($^{12}$C) profile is shown in the top panel (dotted). }
\label{fig4.6.7.8}
\end{figure} 

Contrary to \citet{HurlburtToomre1994} we determine the depth of
the entrainment neither by the radius where the kinetic energy carried
into the stable layers is zero, nor by the radius where the kinetic
energy has dropped to a certain fraction of its maximum value
\citep{Brummell2002}. We find both conditions insufficient, because
the kinetic energy flux becomes zero much faster than the convective
flux, which is another possible indicator of the depth of the
entrainment (see next subsection). Instead, we rather use the $^{12}$C
mass fraction, as it is low outside the convection zone during the
flash (X($^{12}$C) $< 2\times 10^{-3}$), and as it can rise there only
due to turbulent entrainment.

In this study, we use the condition X($^{12}$C) = $2\times 10^{-3}$ to
define the edges of the convection zone. Due to the turbulent
entrainment, these edges are pushed towards the stellar center and
surface (Fig.\,\ref{fig4.6.7.8}). Hence, the width of the convection
zone increases on the dynamic timescale, which is in contradiction
with the predictions of one-dimensional hydrostatic stellar modeling,
where the width of the convection zone is determined by the local
Schwarzschild or Ledoux criterion.  However, the criterion for the
width of the convection zone cannot be a local one due to turbulent
entrainment caused by convection.

The speed, at which the radius of the outer edge of the convection zone
increases with time due to entrainment, is estimated for models
hefl.2d.a and hefl.3d to be at most 14\,$\mes$.  The radius of the
inner edge of the convection zone changes at a much smaller rate
\citep{Bazan1998,MeakinArnett2007}, as the region interior to the
convection zone is more stable against convection and has a higher
density than the region exterior to the convection zone
\citep{Singh1995}.  The entrainment at the bottom of the convection
zone also leads to a heating of the cool interior layers
\citep{DeupreeCole1983, ColeDemDeupree1985}. This seems to be a robust
feature of convection zones driven by nuclear burning, and is observed
in other studies too (\eg \citet{Asida2000}).
   
The region just interior to the convection zone shows less entrainment
\citep{Bazan1998,MeakinArnett2007}, as the square of the
Brunt-V\"ais\"al\"a frequency is almost ten times larger there than
that in the region just outside the outer edge of the convection zone.
The Brunt-V\"ais\"al\"a frequency is a good stability indicator since
it is related to the behavior of convective elements within a
convection zone, a fact also pointed out by
\citet{HurlburtToomre1994}. The Brunt-V\"ais\"al\"a frequency can be
written as \citep{MeakinArnett2007}:
\begin{equation}
  N^2 = -g \left( \frac{\partial \ln \rho}{\partial \, r} - 
                  \frac{\partial \ln \rho}{\partial \, r} 
           \bigg|_{s}  \right) \, ,
\label{eq:3}
\end{equation}
where $g$, $\rho$ ,and $r$ are the gravitational acceleration, the
density and the radius, respectively. A layer is convectively stable,
if $N^2 > 0$, and unstable otherwise
\footnote{The Brunt-V\"ais\"al\"a frequency is related to the bulk
  Richardson number known from oceanography, which is a more direct
  measure of the stability of the edges of a convection zone in the
  presence of a turbulent flow \citep{MeakinArnett2007}.}
(see, e.g., \cite{KipWeigert1990}). 

\begin{figure} 
\includegraphics[width=0.99\hsize]{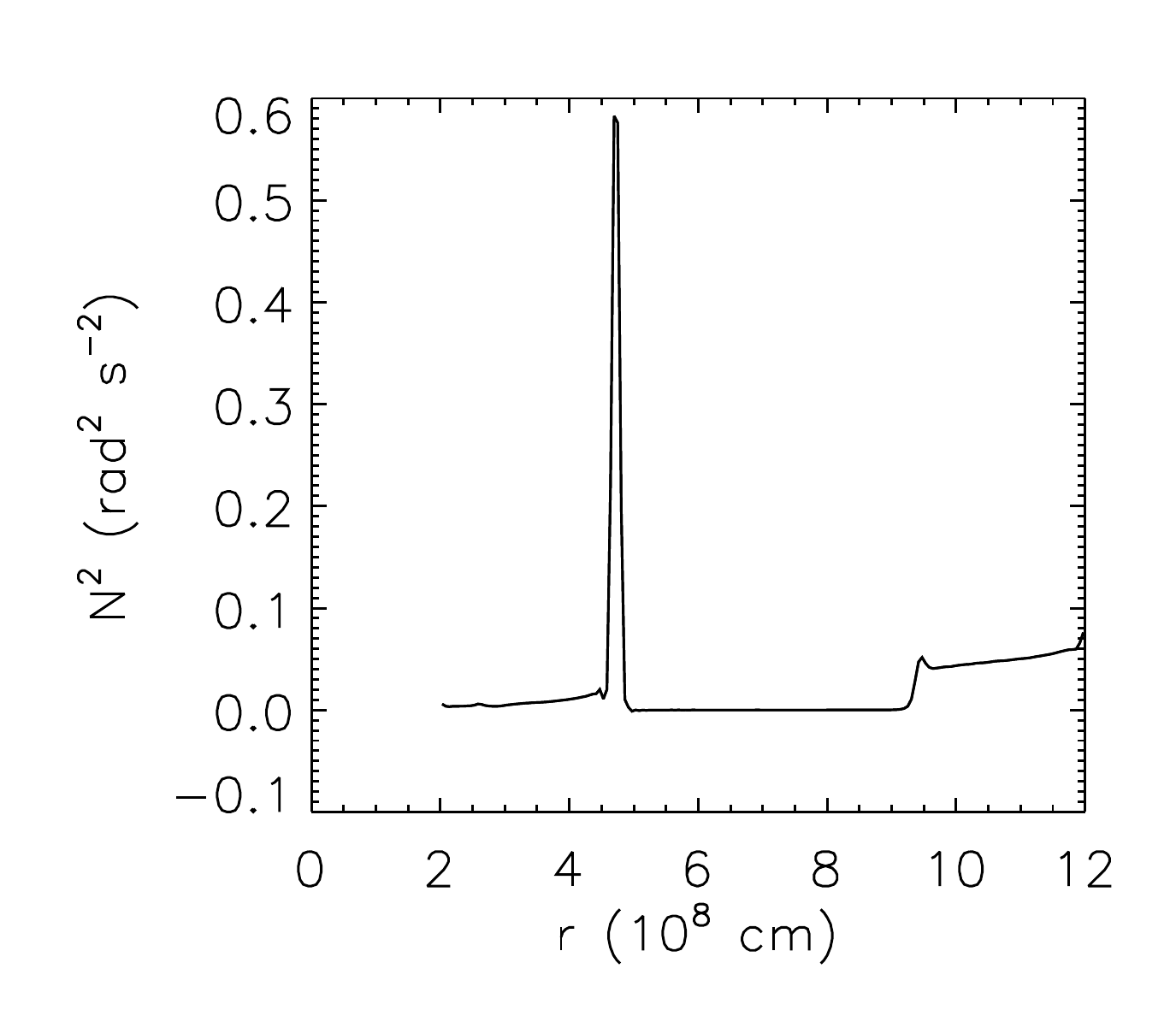}
\caption{Square of the Brunt-V\"ais\"al\"a buoyancy frequency as a
  function of radius for the 3D model hefl.3d at $t = 4638\,$s.}
\label{fig4.9}
\end{figure} 

The Brunt-V\"ais\"al\"a frequency differs in the 2D and 3D simulations
slightly, and has on average a very small negative value inside the
convection zone: N$^2$ = $-4.4\times 10^{-6}$ \radcm in the 2D model
hefl.2d.a, and N$^2$ = $-1.2\times 10^{-6}$ \radcm in the 3D model
hefl.3d. Just outside (inside) the inner (outer) edge of the
convection zone of the 2D model hefl.2d.a we find N$^2$ = 0.580 \radcm
(0.053 \radcm), and N$^2$ = 0.583 \radcm (0.052 \radcm) for the 3D
model hefl.3d, respectively (Fig.\,\ref{fig4.9}).  In the 
2D model hefl.2d.b, these frequencies are higher by about a factor of 
two.

Due to its high stability, the radius of the bottom edge of the
convection zone did not change during the time covered by our
simulations, except for an initial jump over one radial grid zone 
from $r= 4.69\times 10^{8}$\,cm to $r= 4.63\times 10^{8}$\,cm, 
when it was touched by convective downflows 
for the first time. However, entrainment may move the edge further towards the
stellar center later in the evolution (see Sect.\,\ref{sect:6}).  The
entrainment rate (\ie the velocity at which the convective boundary
moves) is lower by a factor of $\sim$ 5-6 at the bottom of the
convection zone in our 2D simulations of the core helium flash as
compared to that at the outer boundary ($\sim 14\,\mes$; see
above). This behavior was also observed in 3D simulations of oxygen
burning shell \citep{MeakinArnett2007}.  This implies that the
entrainment rate at the inner edge of the convection zone is $\sim
2.5\,\mes$ in our core helium flash simulations. The corresponding change
in radius is only $1.5\times 10^6\,$cm or about a quarter of the width
of a radial zone during the time covered by the simulations, and hence
too small to be seen.  As these estimates are resolution dependent, 
the values presented should be considered as order of magnitude
estimates, only.

The entrainment is more efficient in the 2D model hefl.2d.a than in
the 3D model hefl.3da.  In 2D, the observed 
convective flow structures are large and fast rotating vortices that 
due to the imposed axisymmetry are
actually tori \citep{Bazan1998}. They have a high filling factor near
the edge of the convection zone where they overshoot or penetrate
\citep{Brummell2002}. 3D structures crossing the edge of the
convection zones are smaller (localized) plumes with a lower filling
factor and smaller velocities.

\begin{figure*}
\includegraphics[width=8.5cm]{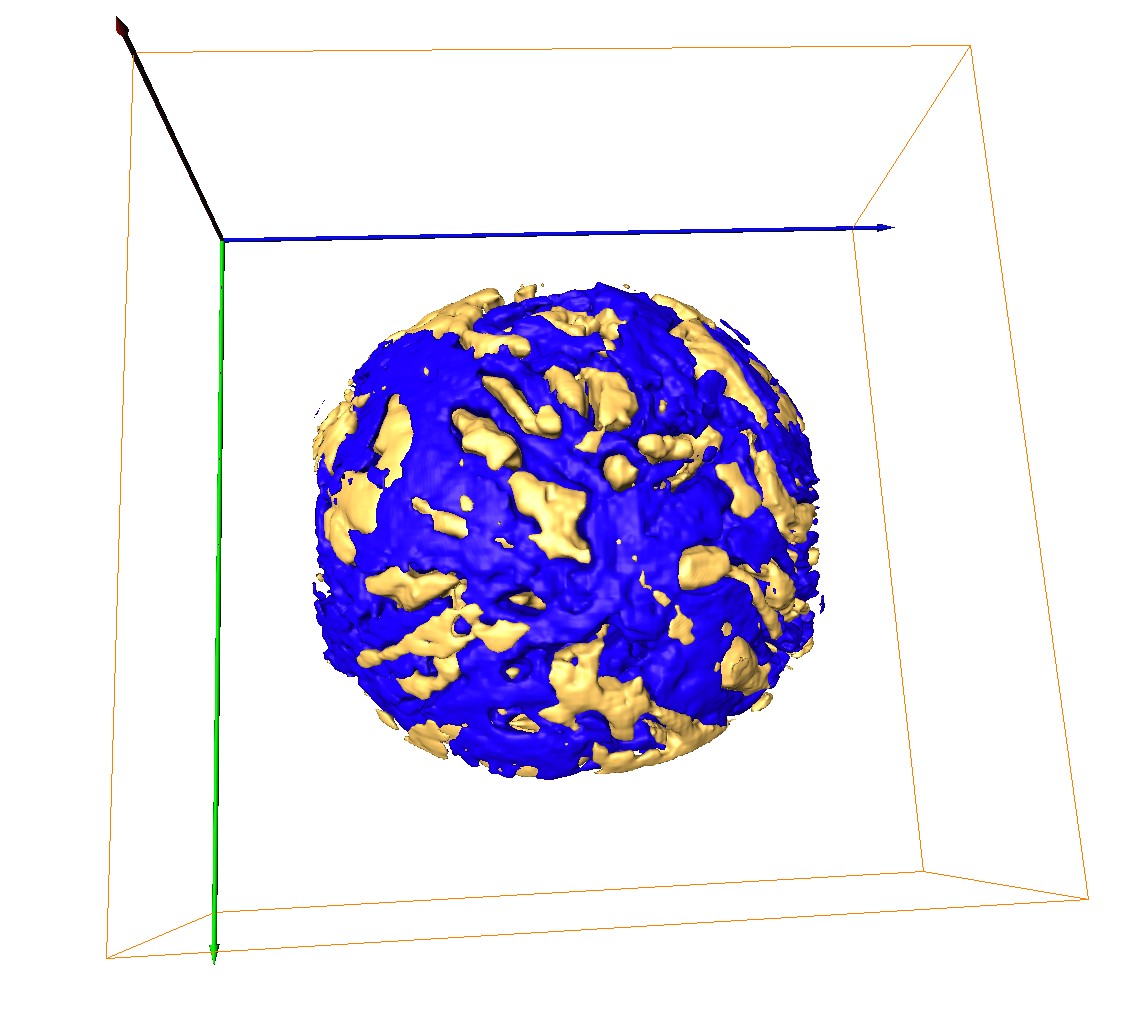} 
\includegraphics[width=8.5cm, angle=90.]{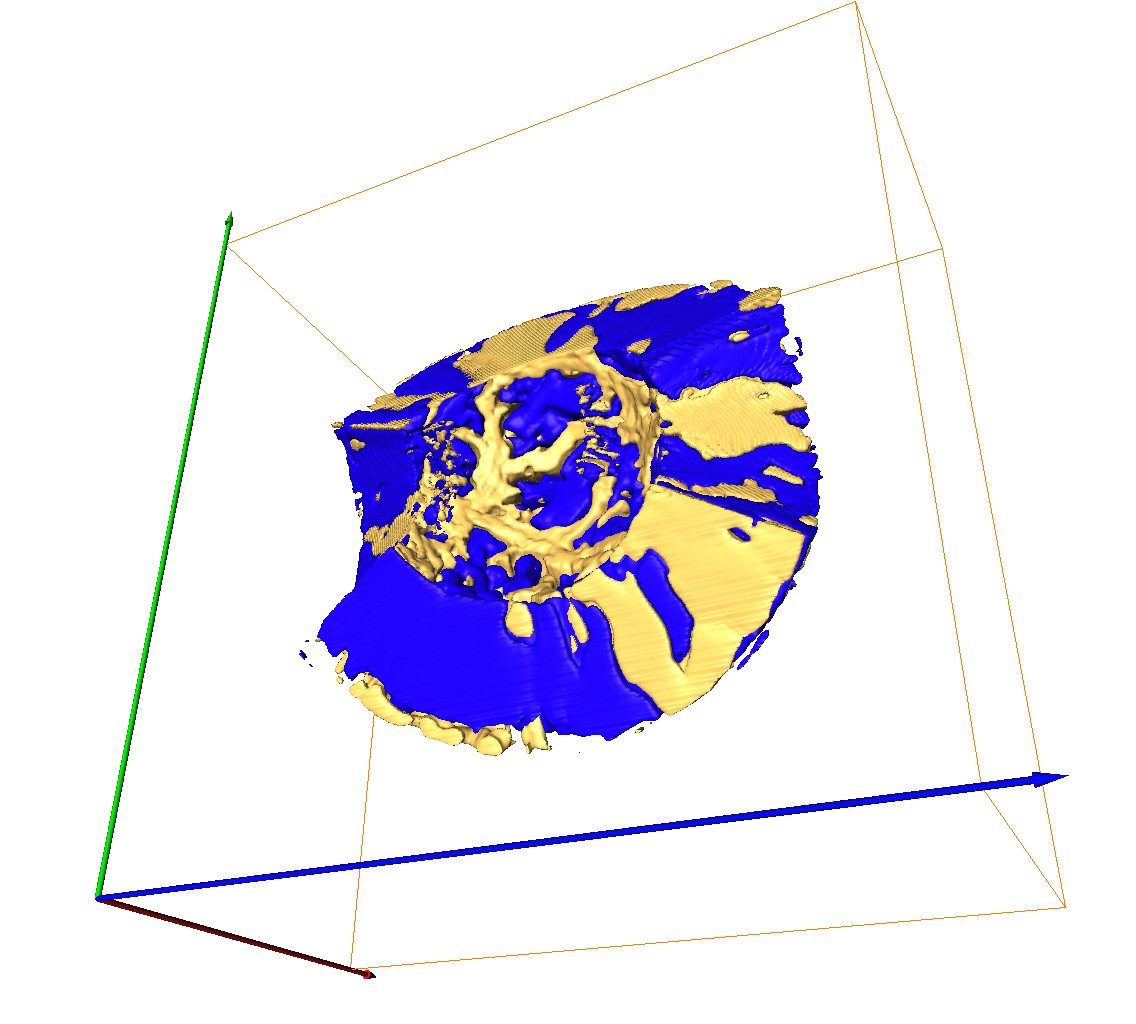} 
\includegraphics[width=8.5cm]{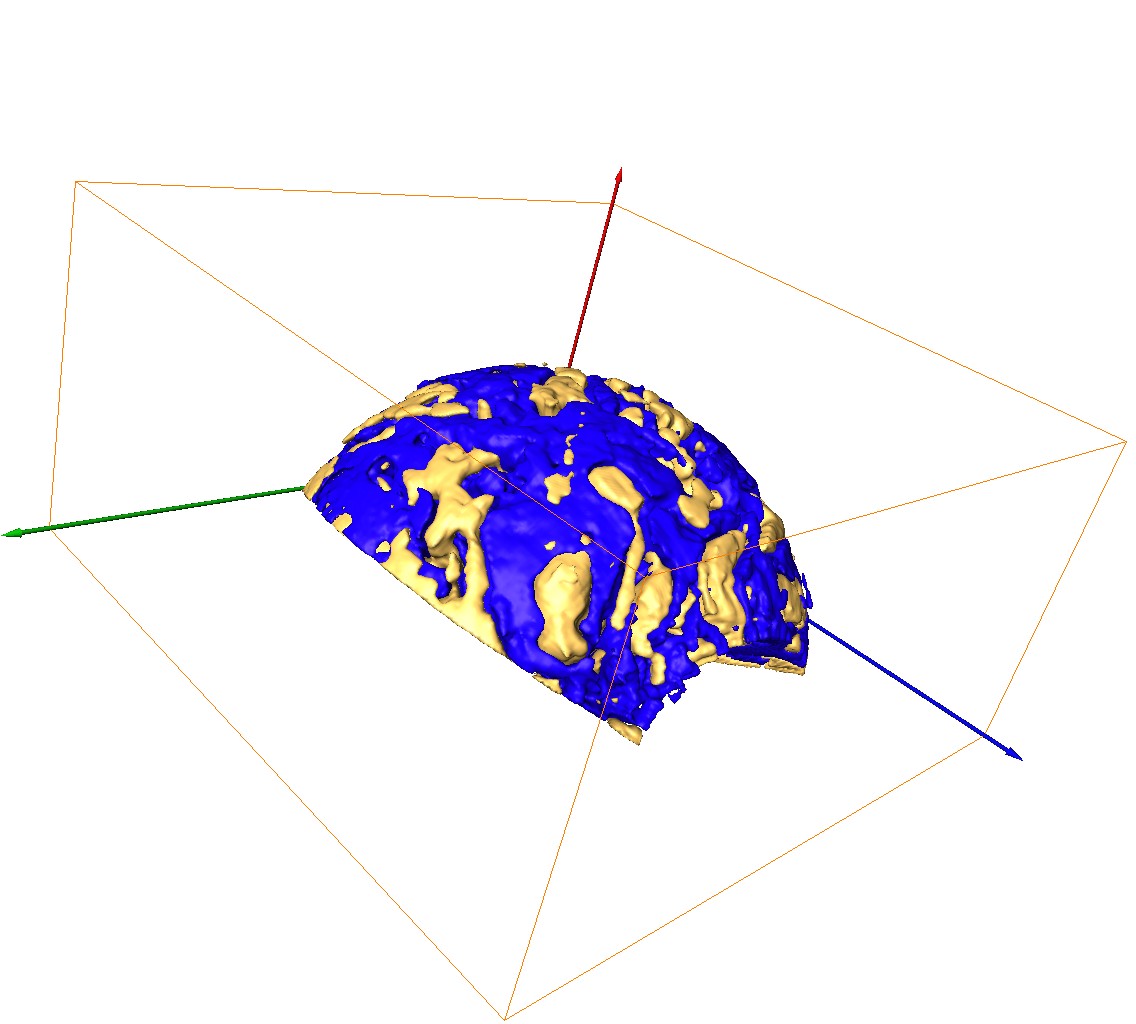} 
\hspace{2.5cm}
\includegraphics[width=6.cm]{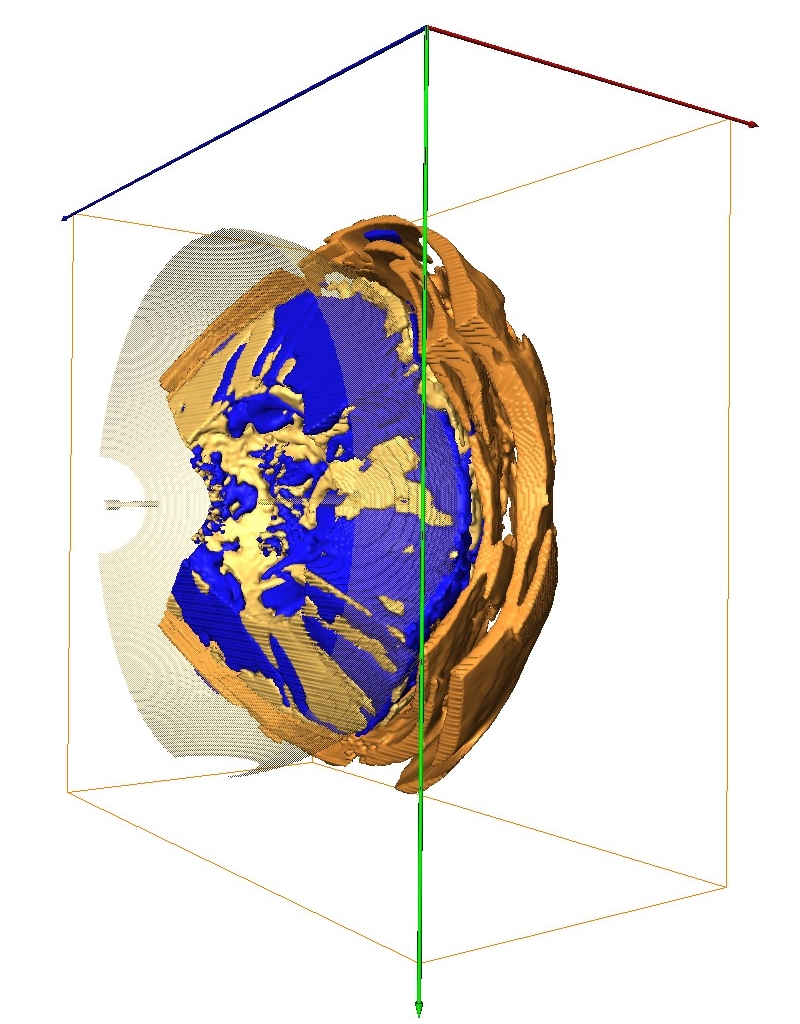}
\caption{Different views of isosurfaces of the velocity field for the
  3D model hefl.3d at $t = 3000\,$s. The blue isosurface corresponds
  to radial downflows with $v_{r} = -6\times 10^{5}\cms$, and the
  yellow and brown isosurfaces show radial upflows with $v_{r} =
  6\times 10^{5}\cms$, and $1\times 10^{4}\cms$ (gravity waves), 
  respectively. The edge sizes of the box are $1.2 \times 10^{9}\,$cm 
  and $2.4\times 10^{9}\,$cm,
  respectively. The yellow-greenish sphere in the bottom right panel marks
  the top of the convection zone according to the Schwarzschild criterion.}
\label{fig4.14.15}
\end{figure*}

\begin{figure*} 
\includegraphics[width=0.49\hsize]{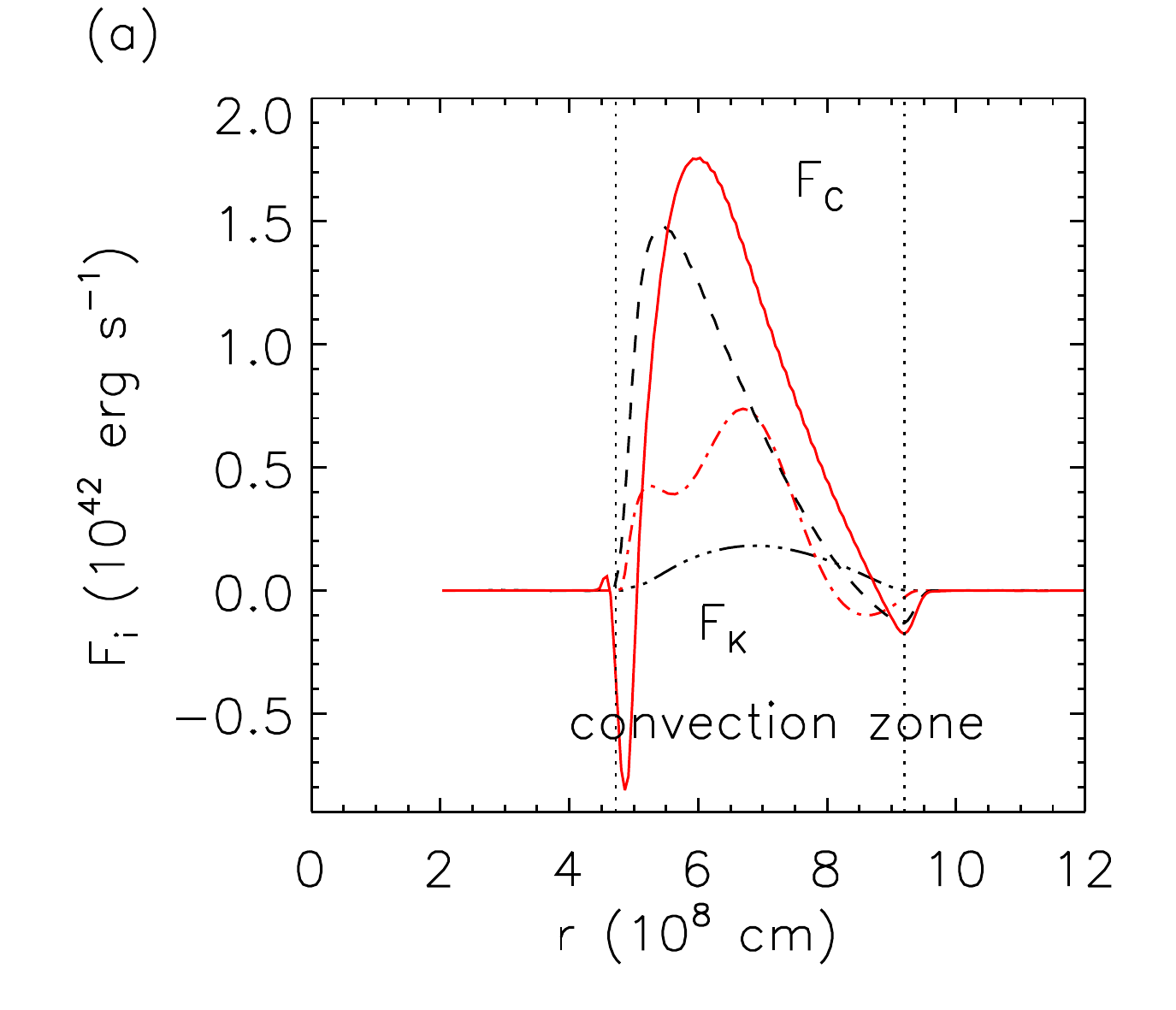}
\includegraphics[width=0.49\hsize]{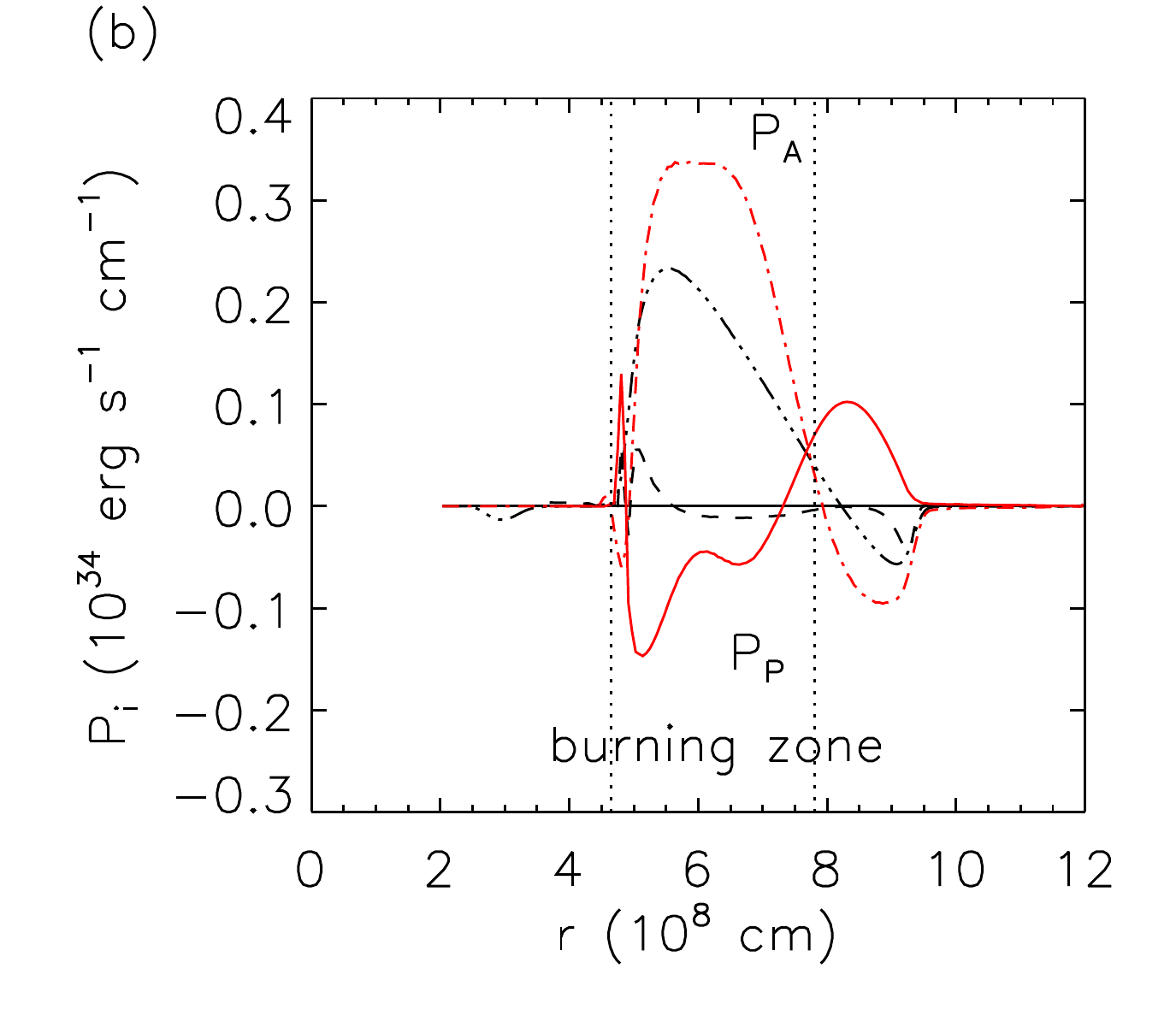}
\includegraphics[width=0.49\hsize]{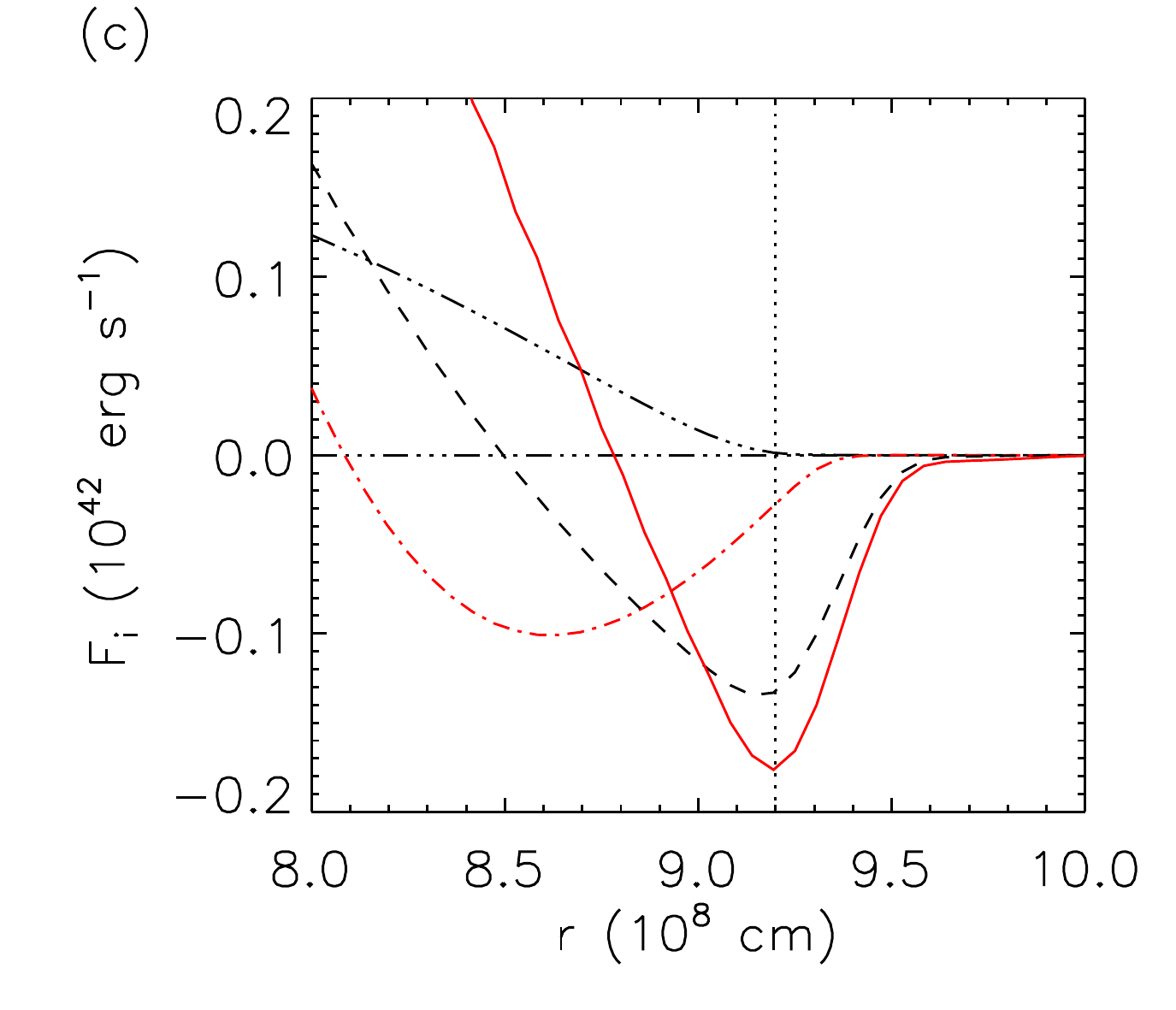}
\includegraphics[width=0.49\hsize]{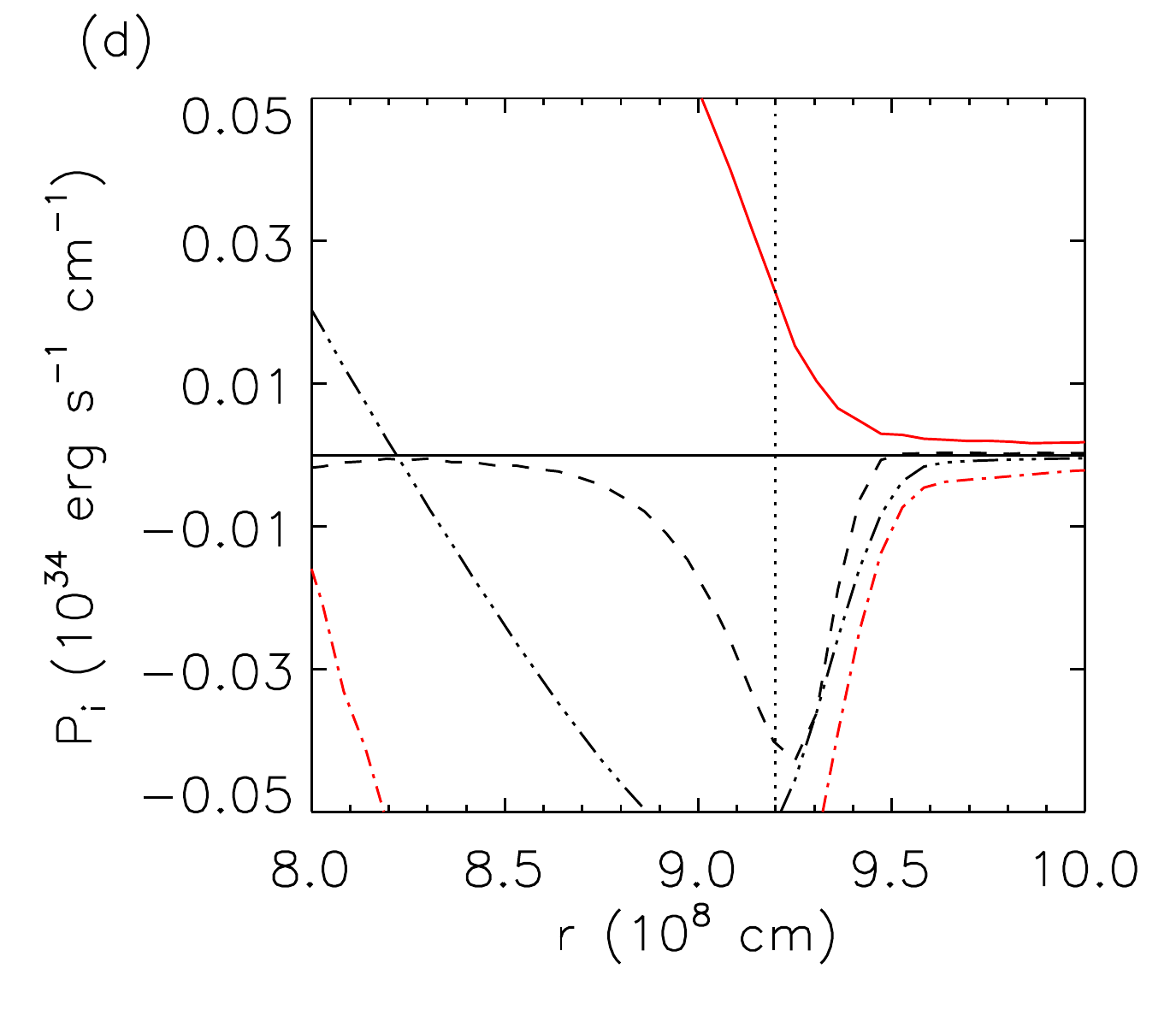}
\caption{Various energy fluxes and source terms as a function of
  radius averaged (from 2000\,s to 6000\,s) over about four convective
  turnover times. 
  {\it Panel (a)} shows the convective flux $F_C$ of the 2D model
  hefl.2d.a (solid-red) and the 3D model hefl.3d (dashed-black)
  together with the kinetic energy flux $F_K$ in the 2D
  (dash-dotted-red) and 3D (dash-dot-dotted-black) model,
  respectively. The dotted vertical lines mark the edges of the
  convection zone in the initial model according to the Schwarzschild
  criterion. 
  {\it Panel (b)} gives the source terms due to the work done by
  buoyancy forces $P_{A}$ (dash-dot-dotted black) and due to volume changes
  $P_P$ (dashed black) in the 3D model hefl.3d, and in the 2D model
  hefl.2d.a (dash-dotted-red, solid red) respectively.  The vertical
  lines enclose the nuclear burning zone (T$\,> 10^{8}$ K).  
  {\it Panels (c)} and (d) show an enlarged view of the energy
  fluxes and source terms displayed in panels (a) and (b) near the
  outer edge of the convection zone. }
%
\label{fig4.10.11.12.13}
\end{figure*} 

We studied convective stability in detail only for the layer above the
convection zone, where the gas is only weakly degenerate, and we are
thus able to compare our results with those of similar systems
simulated in 3D at high Reynolds numbers ($\sim\,10^{4}$) by
\citet{Brummell2002}. According to these authors, a stable layer
allows for more ``overshooting'', if its entropy gradient is smaller.
We found that the turbulent entrainment lowers the entropy gradient
at the outer edge of the convection zone in our simulations
(Fig.\,\ref{fig4.6.7.8}). Hence, the stability of the exterior stable
layer decreases with time, allowing for more turbulent entrainment.

Contrary to \citet{Brummell2002}, who studied only single fluid flow,
our simulations involve a mixture of different fluids of different
composition. This complicates the above argumentation, as in a
multi-fluid flow a shallow entropy gradient does not necessarily imply
that the layer is less stable against turbulent entrainment. We plan
to address this issue in more detail elsewhere.  We have not analyzed
the stability of the region below the inner edge of the convection
zone, because it is highly degenerate and appears to be very stable.
No significant entrainment of $^{12}$C was observed there during the
entire simulation of models hefl.2d.a and hefl.3d, respectively.
Simulations covering longer periods of time are therefore needed (see
Sect.\,\ref{sect:6}).  The analysis of the energy fluxes inside the
convection zone and near the outer edge of the convection zone, which
will be discussed in the next subsection, will provide more
information about the phenomenon of the turbulent entrainment.

%
\subsection{Energy fluxes}
\label{sect:4.4}
Energy fluxes are a useful tool for understanding convective flows
\citep{ChanSofia1986, HurlburtToomre1986, Muthsam1995,
  HurlburtToomre1994, Brummell2002, MeakinArnett2007}. They allow one
to discriminate the energy transport due to different processes and
mechanisms (\eg due to convection, heat conduction, etc).  We thus
analyzed various energy fluxes and source terms that are defined in
Appendix\,\ref{app:fluxes}.  All energy fluxes and source terms (see, 
Fig.\,\ref{fig4.10.11.12.13}) are averaged over several convective
turnover times, as they change considerably with time due to the
appearance of plumes \citep{Muthsam1995}.


The convective energy flux is mainly positive as heat is mostly
transported upwards by convection. It is larger in the 2D model
hefl.2d.a, as in 2D the convective flow structures are more laminar
and ordered, and thus experience less dissipation. The smaller value
of $F_{C}$ in the 3D model hefl.3d is a result of a less ordered flow
throughout the whole convection zone. This is in agreement with
Table\,\ref{tab:mflcnvz} which shows that the fluctuations in
temperature and density are smaller in the 3D model.

A kinetic energy flux arises from deviations from the mean convective
flow (\ie mainly from the upflow-downflow asymmetry). Typically it is
largest in the most turbulent regions of the flow, where on  
top of the kinetic flux due to the up- and down-flow asymmetry,
there is also a significant contribution due to the asymmetry of 
localized turbulent (convective) elements. Directly connected to 
this is the 
offset between the maxima of the kinetic and the convective flux (see
Fig.\,\ref{fig4.10.11.12.13}), which reflects the fact that the convective
flow decays more efficiently when the flow becomes strongly turbulent.
  
The work done by buoyancy forces is positive in the whole region of
dominant nuclear burning (see Fig.\,\ref{fig4.10.11.12.13}) indicating
either less dense and hot gas moving upwards or more dense and cooler
gas moving downwards.  A negative value of the work done by buoyancy
forces implies the opposite situation, \ie less dense and hot gas
flowing downward or denser and cooler gas flowing upward. The latter
is known as buoyancy breaking leading to a deceleration of the flow
motion \citep{Brummell2002}, and to the unusual situation that hot
matter tends to sink, and cool matter is likely to rise.

The gas should expand while rising up through the convection zone, \ie
the work done by buoyancy ($P_A$) and by volume changes ($P_P$) should
always be anti-correlated. The anti-correlation is clearly seen only in
the 2D model hefl.2d.a, whereas in the 3D model hefl.3d the quantities
are on average anti-correlated only in the central region of the
convection zone.  At the inner edge of the convection zone, buoyancy
drives gas upwards which is on average simultaneously compressed,
probably by the broad downflows. At the outer edge where buoyancy
braking occurs, the gas on average expands. Hence, it must expand
faster than the upflows cool and are being compressed (a situation
observed for the 2D model hefl.2d.a).

All the fluxes discussed here agree qualitatively well with those of
our previous high-resolution 2D simulations \citep{Mocak2008}.  They
are also qualitatively very similar to those of the high-resolution 3D
simulations of \citet{Brummell2002} who investigated a stratified
model with a convectively stable region located on top of a
convectively unstable region both consisting of an ideal gas with a
very high Reynolds number ($\sim 10^{4}$).  The angular and time
averaged radial distributions of the kinetic and convective fluxes
seem to be robust in the convectively unstable region, as our 3D
results are qualitatively similar to those obtained in several other
3D studies \citep{HurlburtToomre1994, Brummell2002, MeakinArnett2007}.

The outer part of the convection zone, where buoyancy breaking occurs,
resembles the overshooting region due to active penetration of plumes
described by \citet{HurlburtToomre1986, HurlburtToomre1994} and
\citet{Brummell2002}. Note, however, that this region is convectively
stable at the beginning of their simulations, \ie buoyancy breaking
takes already place inside the convection zone in our models.

\begin{figure*}
\includegraphics[width=0.49\hsize]{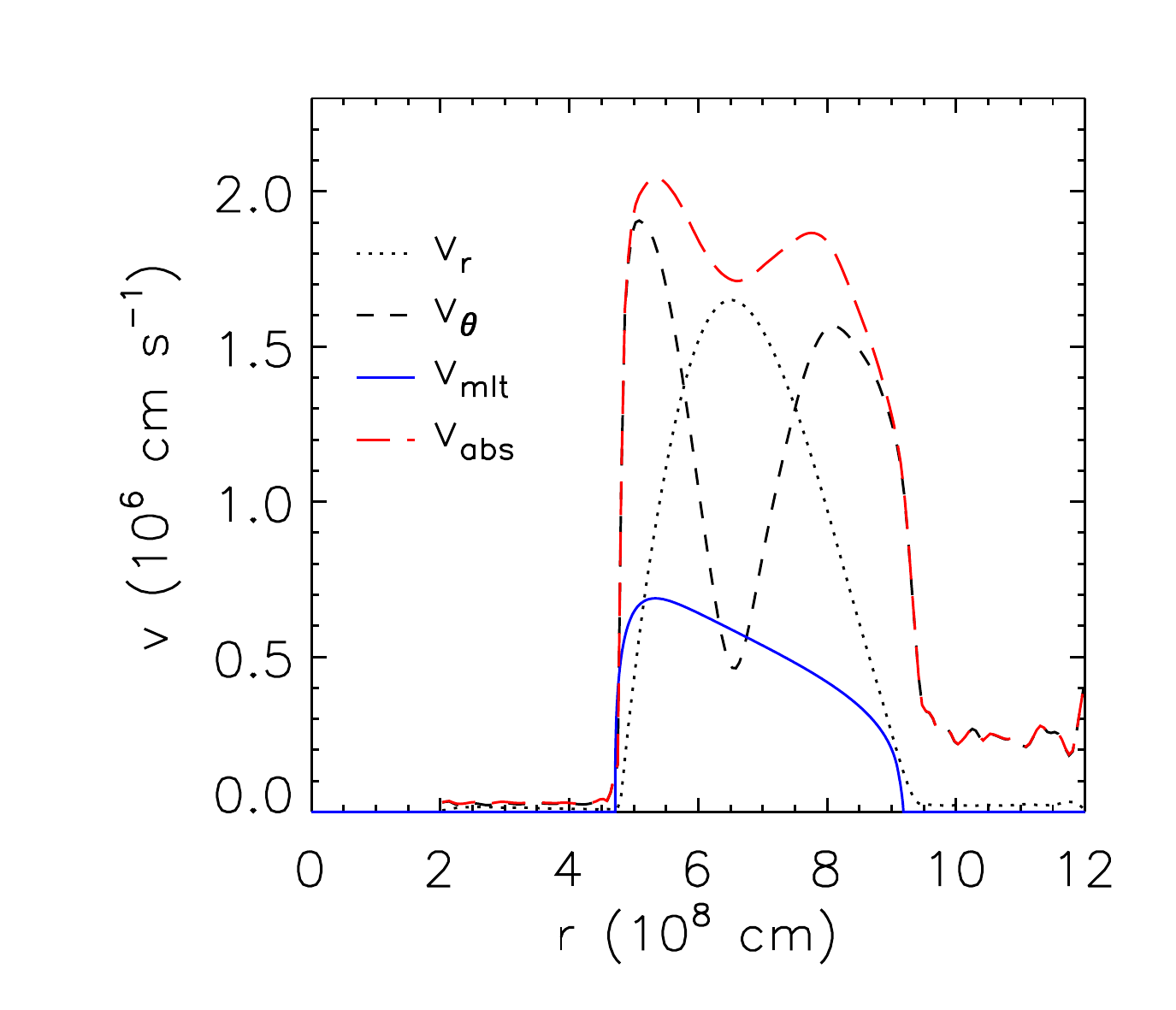}
\includegraphics[width=0.49\hsize]{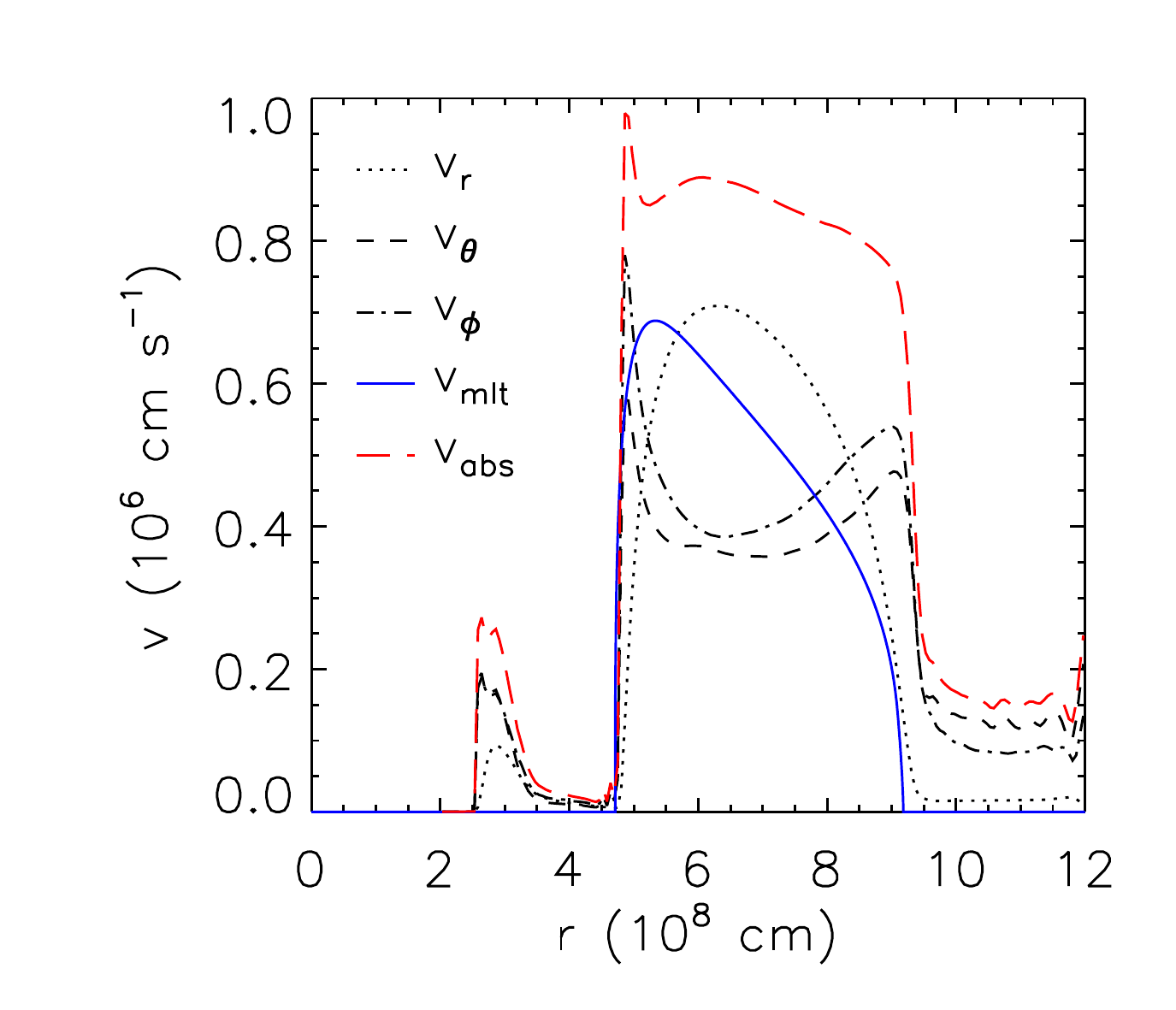}
\caption{
  Radial distributions of the time (from 2000\,s
  to 4000\,s) and angle-averaged velocity components ($v_r$,
  $v_\theta$, $v_\phi$) and velocity modulus ($v_{abs}$) for the 2D
  model hefl.2d.a (left), and the 3D model hefl.3d (right),
  respectively.  The panels also show the velocity predicted by the
  mixing-length theory ($v_{mlt}$) .}
\label{fig4.18.19}
\end{figure*} 

The distribution of the kinetic flux (Fig.\,\ref{fig4.10.11.12.13})
exhibits structural differences in the convection zone between 2D and
3D flows. The typical evolved 2D flow contains well defined vortices
(Fig.\,\ref{fig4.c1}) whose central parts never interact with the
region of dominant nuclear burning.  This results in a reduced kinetic
energy flux at r $\sim 5.5\times 10^{8}$\,cm, as
this region corresponds to the central region of the vortices, which
do not experience any strong radial motion.  On the other hand, the
distribution of the kinetic energy flux in the 3D model hefl.3da is
rather smooth as a result of the column-shaped flow structures
(Fig.\,\ref{fig4.14.15}).

The convective flux changes sign in stable layers since the downflows
or upflows penetrating into the stable zones are suddenly too hot or
cold compared to the surrounding gas. The penetration continues until
the momentum of the convective elements is used up or diffusion
smooths out the perturbations, and the convective flux approaches
zero \citep{Brummell2002}. This fingerprint of turbulent entrainment
is clearly present at both convective boundaries of our models
(Fig.\,\ref{fig4.10.11.12.13}).  The convective energy flux is
relatively strong even in regions where the kinetic energy flux is
almost zero. Therefore, the ``zero'' kinetic flux criterion
\citep{HurlburtToomre1994, Brummell2002} seems to be a bad indicator
of entrainment which may extend well beyond the location where the
kinetic flux becomes small.  

In fact, what is happening at the
convective boundaries is an exchange between the potential energy of
the stratification given by the buoyancy jump $db = N^2 ~dr$ and the
kinetic energy of the turbulence. Turbulence looses its kinetic energy
by doing work against gravity, which leads to a reduction of the
buoyancy jump, and hence to stability weakening of the 
stable boundary layer to the effects of the turbulent entrainment. 
The buoyancy jump
$db$ is a direct measure of the stability of the boundary layer. To
mix gas into the boundary layer the buoyancy must be reduced. This is
accomplished through the buoyancy flux $q = P_A / \rho$, which is
related to the temporal variation of the buoyancy jump by $db/dt = -
div(q)$, where $P_A$ and $\rho$ are the sink/source term of the
kinetic energy due to buoyancy forces and the density, respectively.

The convective flux can directly be related to the buoyancy flux that
is a function of $P_A$ by a linear relation $F_C = F_C (q)$ described
in more detail by \citet{MeakinArnett2007}.
Figure\,\ref{fig4.10.11.12.13} shows that this agrees well with what
we observe in our simulations. It also supports our previous
conclusion that entrainment is well indicated by the convective flux
and the related buoyancy flux which via the equation $db/dt = -
div(q)$ leads to the decrease of the buoyancy jump in the stable
layer.  This in turn reduces the convective stability of that layer.

The properties of the entrainment at the outer convective boundary differs
in models hefl.2d.a and hefl.3d (see Fig.\,\ref{fig4.10.11.12.13}).
In the 2D model entrainment is reaching deeper into the stable layer
due to a more active convection zone with higher typical velocities
(Fig.\,\ref{fig4.18.19}) than in the 3D model. The radial distribution
of the work done by buoyancy $P_A$ is qualitatively similar in both
the 2D and 3D models, \ie it is negative indicating buoyancy
breaking. Nevertheless, the work done by gas compression or expansion
$P_P$ is different. In the 2D model hefl.2d.a the gas on average is
compressed ($P_P$ is positive), while in the 3D model hefl.3d the gas
is expanding at the boundary. This again confirms that 2D and 3D
convective flows are qualitatively different.

\begin{figure}
\includegraphics[width=0.99\hsize]{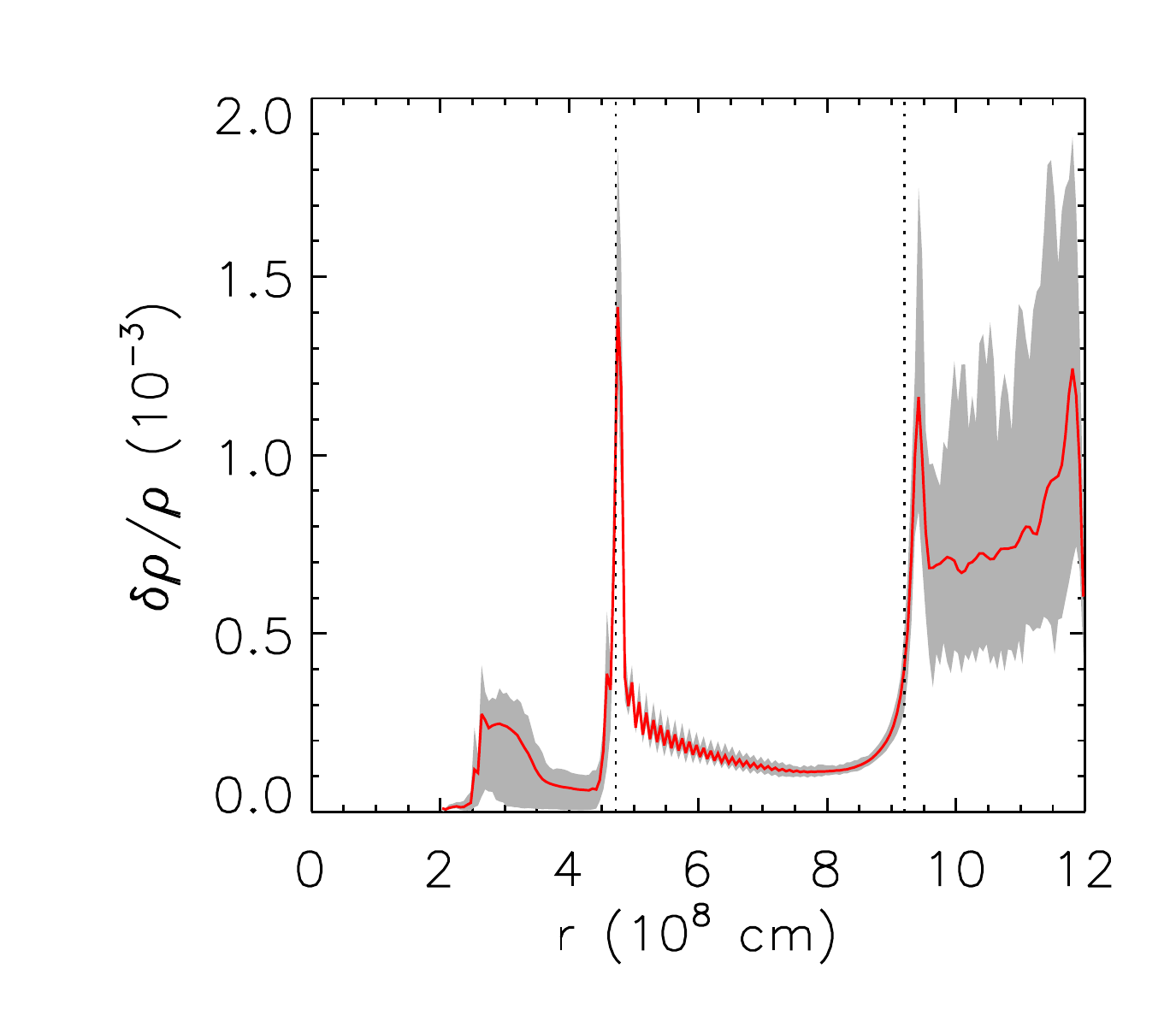}

\includegraphics[width=0.99\hsize]{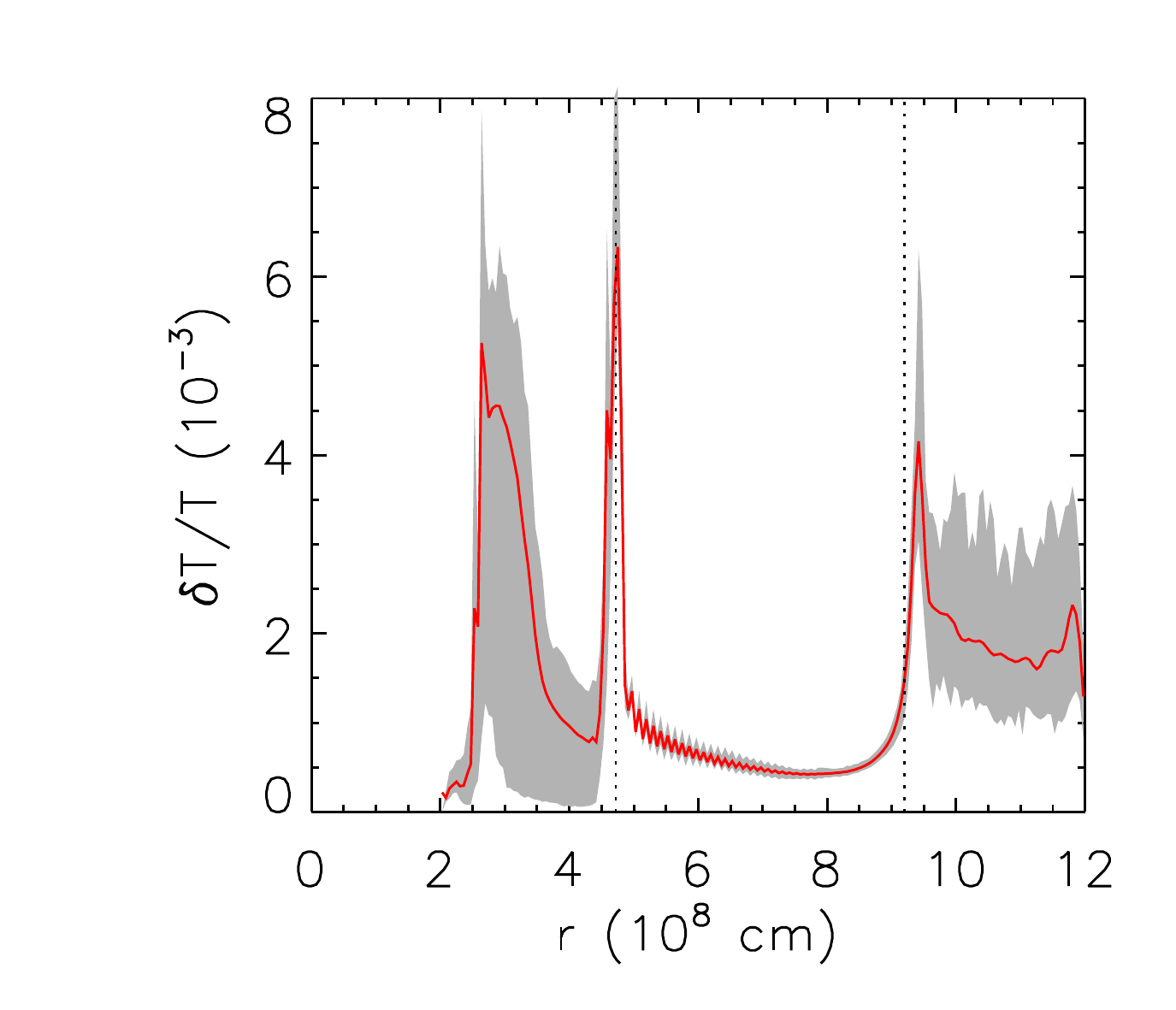}
\caption{
  Radial distributions of the time (from 2800\,s
  to 4800\,s) and angle-averaged density (upper) and temperature
  (lower) fluctuations in the 3D model hefl.3d (solid red
  line). The panels also show the angular variation of the respective
  quantity at a given radius (gray shaded region). The dotted vertical 
  lines mark the edges of the convection zone as determined by the 
  Schwarzschild criterion.}
\label{fig4.16.17}
\end{figure} 

\begin{figure} 
\includegraphics[width=0.99\hsize]{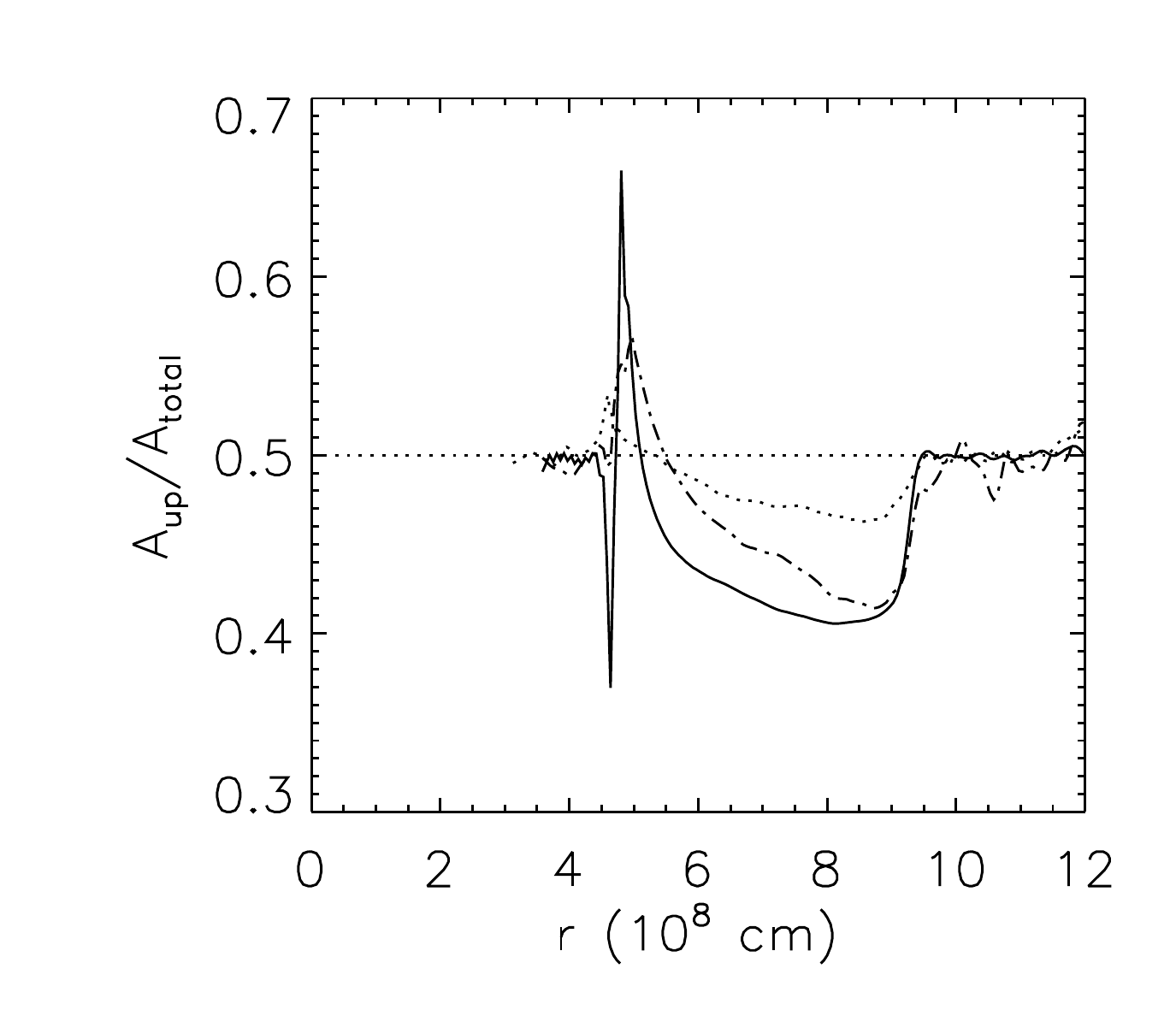} 
\caption{Fractional volume occupied by upflow and downflow streams
  averaged over $\sim$2000\,s as a function of radius for the 3D model
  hefl.3d (solid), the 2D low-resolution model hefl.2d.a
  (dashed-dotted), and the 2D high-resolution model hefl.2d.b
  (dotted), respectively.}
\label{fig4.20}
\end{figure}

%
\subsection{The flow within the convection zone}
The amount of energy ($F_C + F_K$) which has to be transported by 
convection in order to prevent a thermonuclear runaway during the flash is
similar in models hefl.2d.a and hefl.3d. Since the convective flux is
almost the same in both models, the resulting typical convective
velocities are higher in the 2D model than in the 3D one
(Fig.\,\ref{fig4.18.19}).  

The velocities in the 3D model hefl.3d match those predicted by
mixing-length theory better than in the 2D model hefl.2d.a, where they
are about a factor of 2 larger.  This behavior was also observed in
other hydrodynamic simulations of convective flows; see \eg
\citet{Muthsam1995, MeakinArnett2007}.  The radial velocities in the
regions above and below the convection zone are smaller than the
angular velocities in both models, which is a typical feature of
gravity waves \citep{Asida2000}. 

In the 3D model hefl.3d, the flow in the convectively stable layer
beneath the convection zone exhibits some numerical artefact's due to
the proximity of the inner grid boundary (see Fig.\,\ref{fig4.18.19}).
The radial distributions of the time and angle-averaged components of
the velocity field and of the density and temperature fluctuations
show pronounced maxima at $\sim 3\times 10^{8}$\,cm and a sharp
cut-off at $r = 2.5\times 10^{8}$\,cm (Fig.\,\ref{fig4.18.19},~
Fig.\,\ref{fig4.16.17}). The sharp cut-off is caused by
the artificial damping we had to apply to the velocity field in the
innermost grid zones to prevent numerical instabilities from
spreading limitless to the convection zone.
    
Although, the flow velocities in the 3D model match those predicted 
by mixing-length theory very well, one should keep in mind that with
increasing resolution the flow velocities will likely increase due to
the reduced numerical viscosity.  This trend is confirmed by the
velocities obtained for the high-resolution 2D model hefl.2d.b that
are a factor of two higher than in the low-resolution model hefl.2d.a.

Near both edges of the convective zone there are large narrow peaks
visible in the radial distributions of the time and angle-averaged
density and temperature fluctuations (Fig.\,\ref{fig4.16.17}). These
peaks are not caused by compression or expansion, but they are a
result of the density and temperature discontinuities at the edges of
the convection zone \citep{Meakin2007ane,Arnett2007}, because any
angle-dependent radial perturbation will cause large angular
variations of density and temperature at these discontinuities. 

The temperature fluctuations within the convection zone are rather
uniformly distributed, but they are more intense near the outer edge
of the convection zone, where they are only weakly correlated with the
radial velocity (Fig.\,\ref{fig4.c2}).  At the top of the convection zone 
the emerging rising plumes are embedded in an environment which sinks down
(Fig.\,\ref{fig4.c2}, bottom panels). This
situation is similar to the sinking down-drafts with upwelling centers
found in simulations by \citet{NordlundDravins1990} and \citet{Cattaneo1991}.

%
\subsection{Upflow-downflow asymmetry}
The 2D and 3D simulations share the common property of an
upflow-downflow asymmetry \citep{HurlburtToomre1994, Muthsam1995,
  Brummell2002}. The downflows cover a much larger volume fraction of
the convection zone than the upflows (Fig.\,\ref{fig4.20}). The
filling factor of the downflows increases with decreasing depth
\citep{Rast1993, MeakinArnett2007} across most of the convection zone,
and the downflows are more dominant in the 3D model hefl.3d than in
the 2D models. Contrary to the simulations of oxygen burning shell of
\citet{Asida2000} we find that the absolute velocities are about 40\% higher in
the upflows than in the downflows. Hence, the downflows in the
convection zone of a star at the peak of the core helium flash are
slower and broader than the faster and narrower upflows.

%
\subsection{Mixing}

\begin{figure*}
\includegraphics[width=6.3cm]{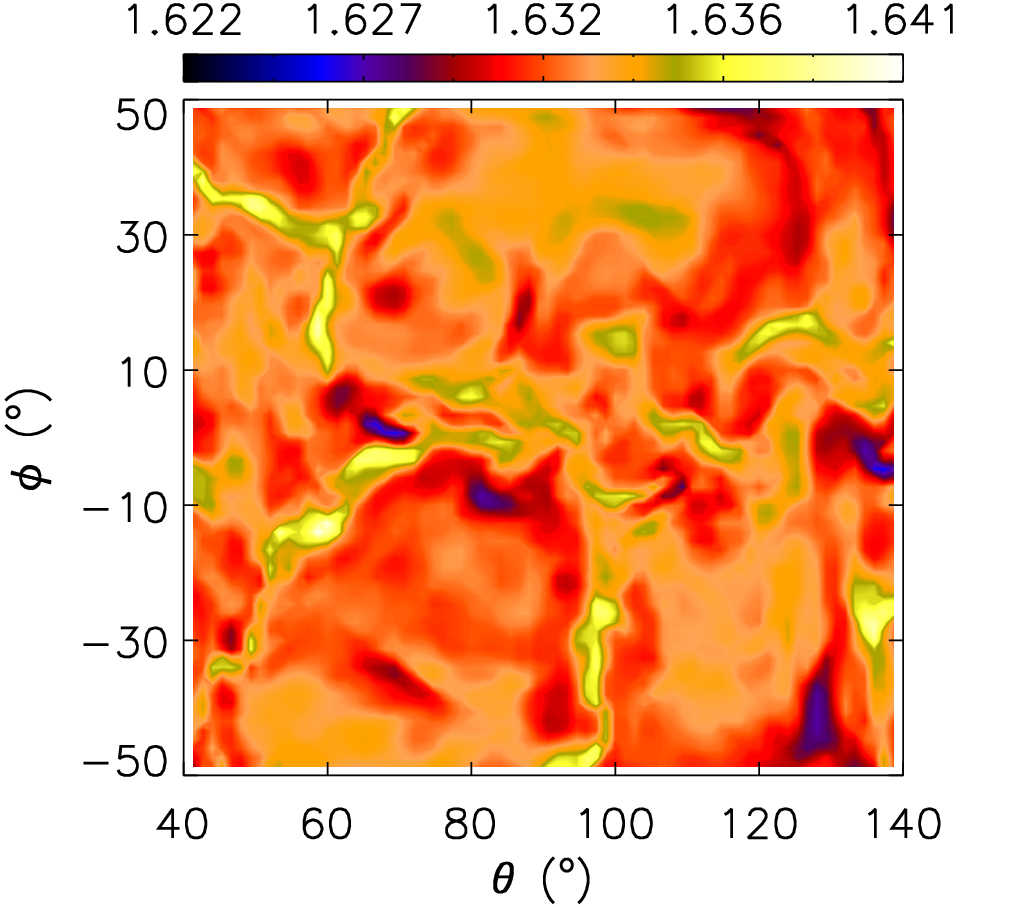}
\includegraphics[width=6.3cm]{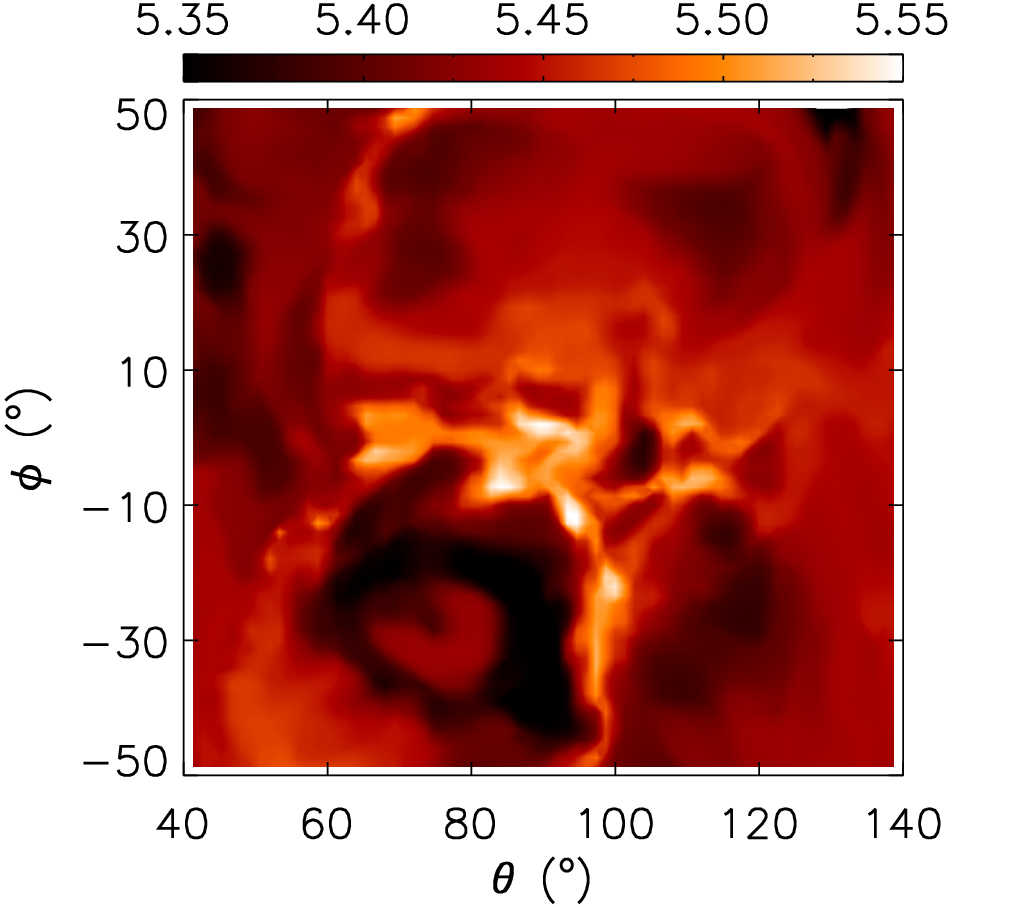} 
\includegraphics[width=6.3cm]{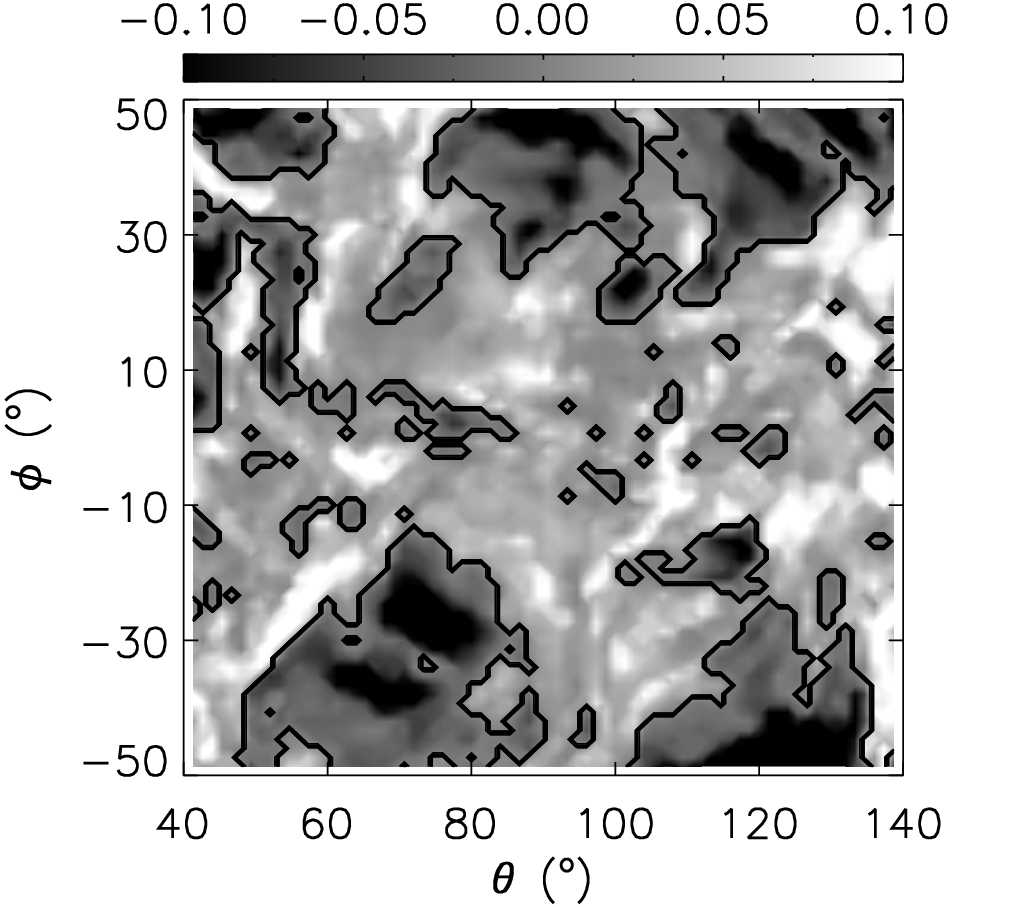}
\hspace{0.1cm}
\includegraphics[width=6.3cm]{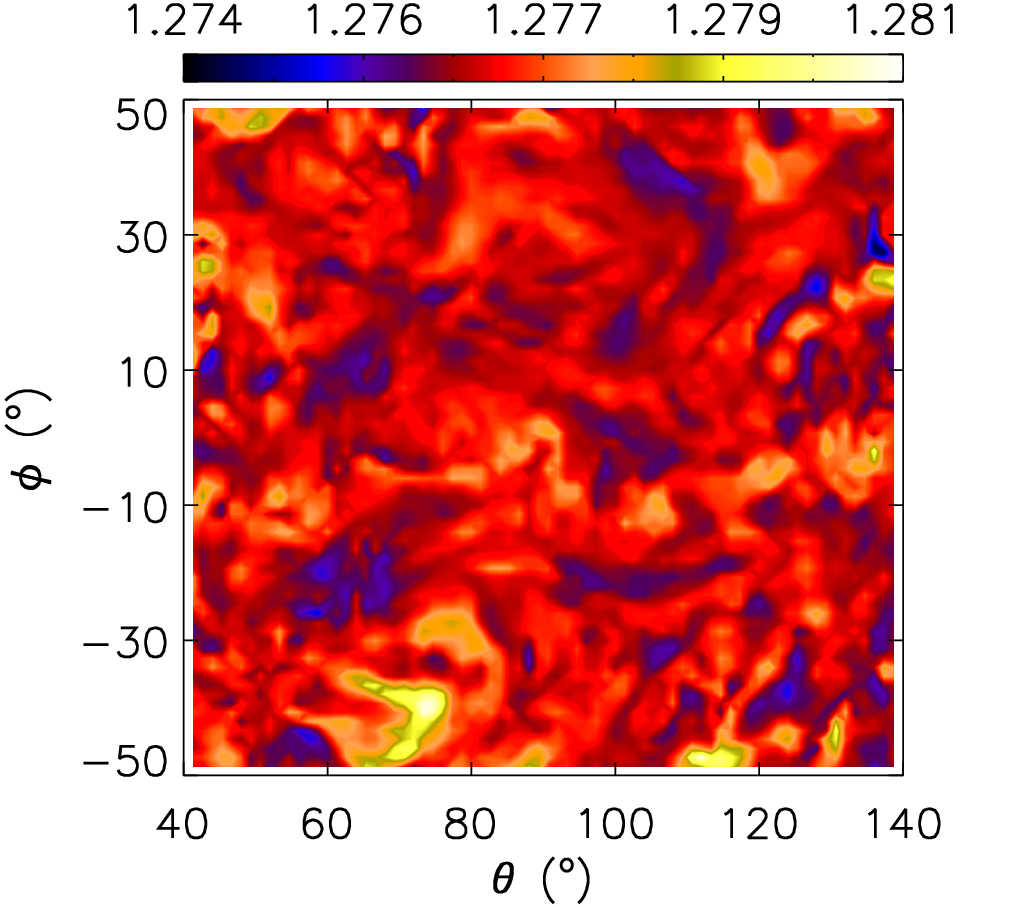}
\includegraphics[width=6.3cm]{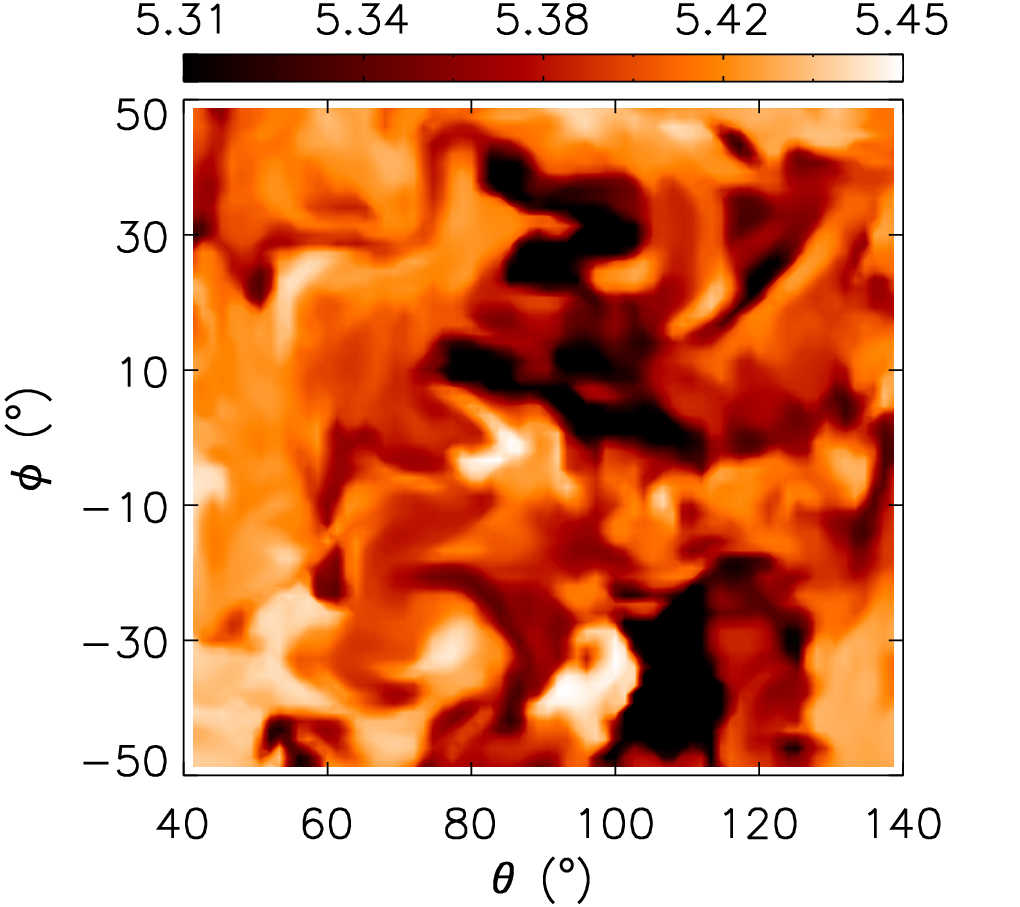} 
\includegraphics[width=6.3cm]{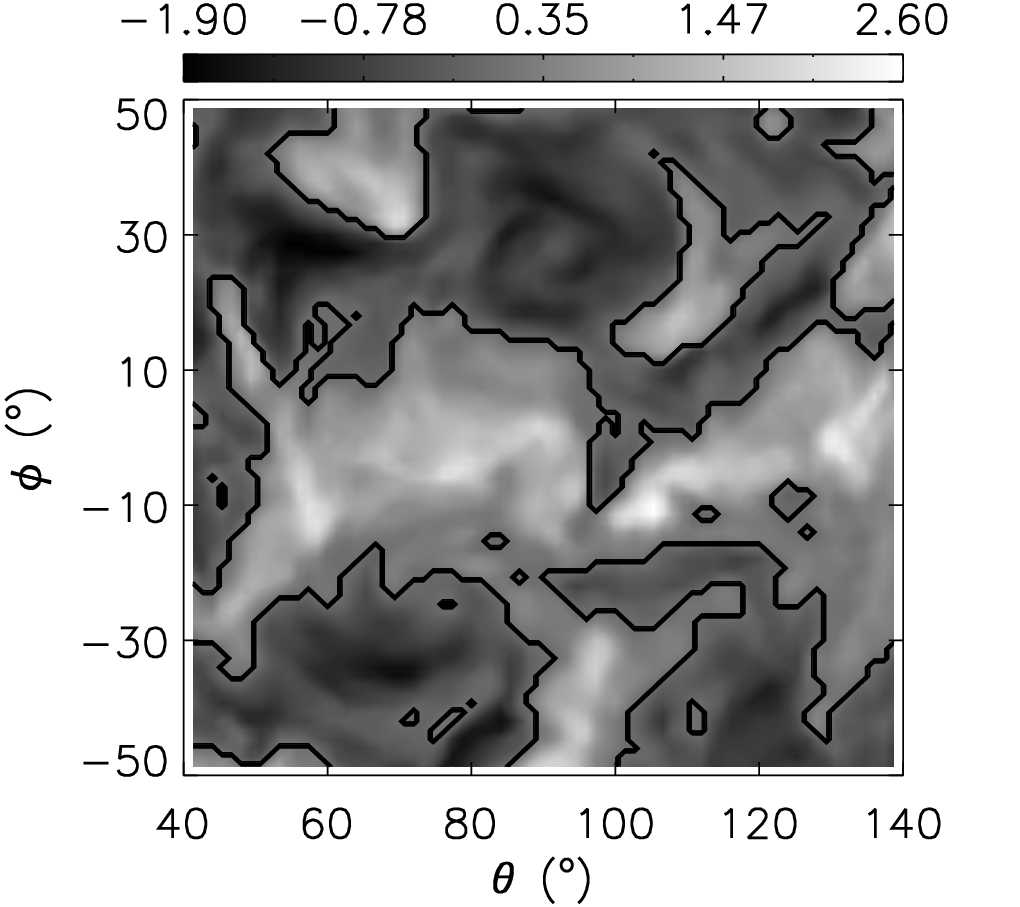} 
\hspace{0.1cm}
\includegraphics[width=6.3cm]{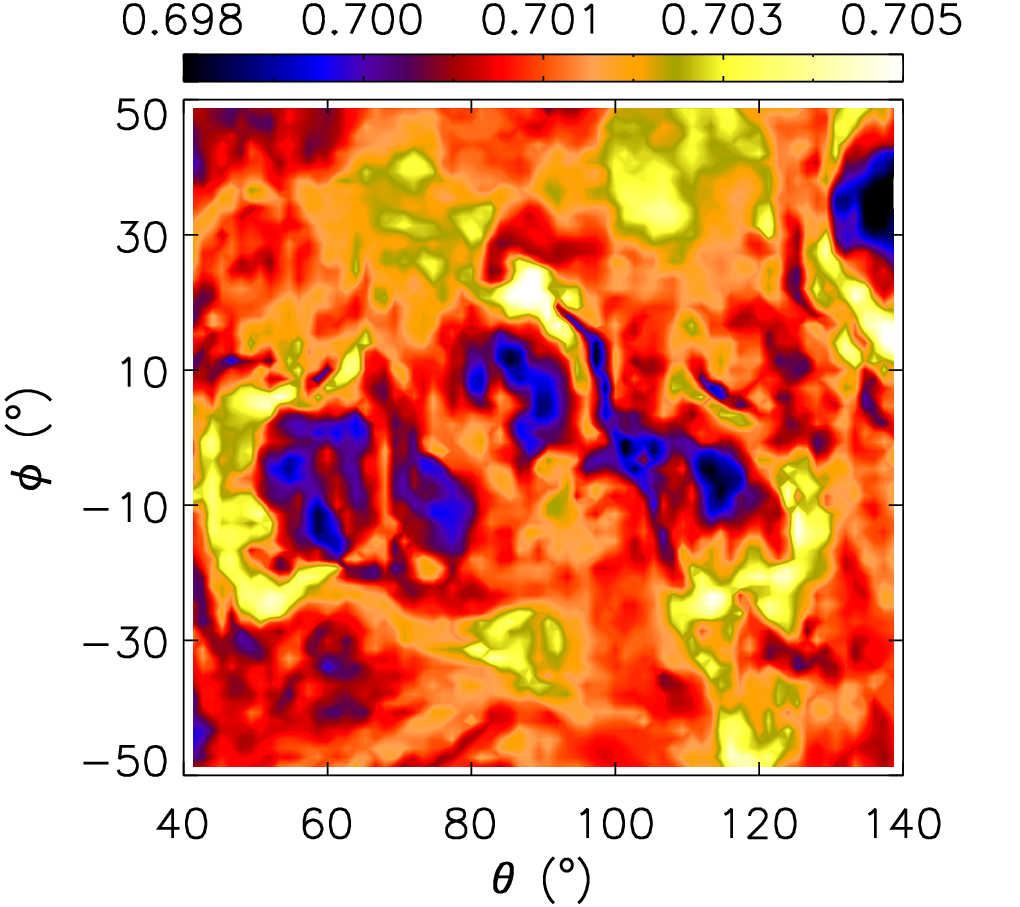}
\includegraphics[width=6.3cm]{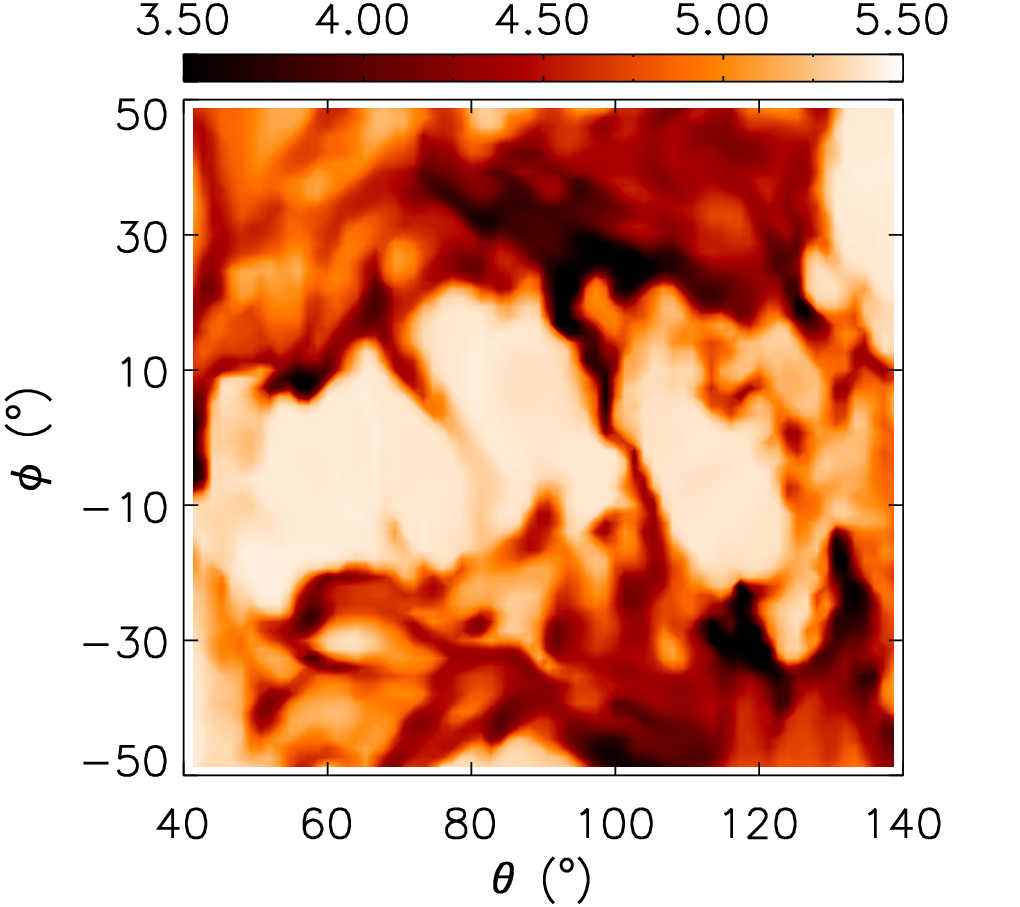} 
\includegraphics[width=6.3cm]{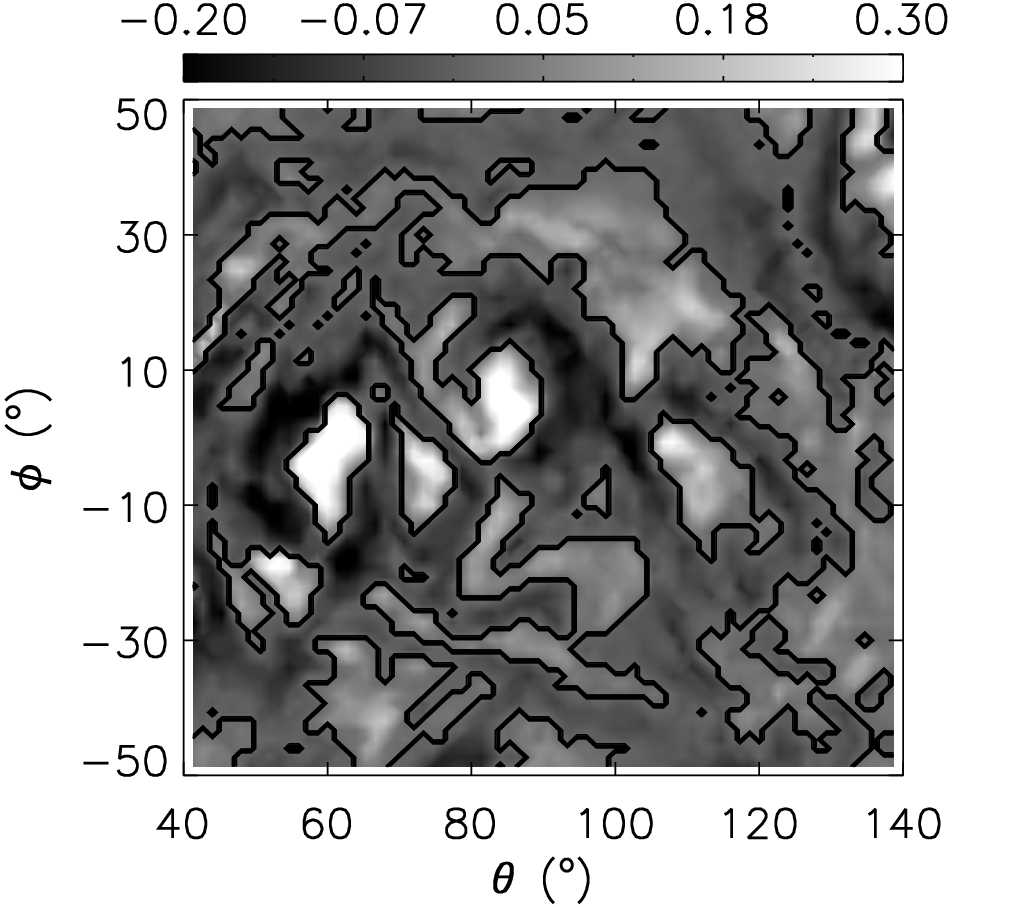} 
\caption{ Cuts through the 3D model hefl.3d at $t = 4815\,$s showing
  the angular variation of temperature (in units of $10^{8}\,$K; left
  panels), $^{12}$C mass fraction (in units of $10^{-3}$; middle
  panels), and radial velocity (in units of $10^{5} \cms$; right
  panels), respectively, at three different radii: $r_{1}$ =
  $4.8\times 10^{8}$ cm (temperature maximum; top), $r_{2}$ =
  $6.5\times 10^{8}$cm (center of the convection zone; middle), and
  $r_{3}$ = $9.3\times 10^{8}$cm (top of convection zone, bottom). The
  black lines in the right panels mark the boundaries between positive
  and negative radial velocities.}
\label{fig4.c2}
\end{figure*} 

Cuts through the 3D model hefl.3d at $t = 4815\,$s showing the angular
variation of temperature, $^{12}$C mass fraction, and radial velocity
at three different radii (Fig.\,\ref{fig4.c2}) demonstrate that the
helium core at the peak of the core helium flash is a very turbulent
environment at all heights of the convection zone.

The bottom of the convection zone contains hot filaments of gas where
the temperature exceeds that of the environment by about 1\%. The
filaments contain ashes from helium burning, \ie $^{12}$C and
$^{16}$O, and they move across the whole bottom of the convection zone
in a random way. The filaments are correlated with upflows, as the hot
gas of burned matter is forced by buoyancy to rise towards the top of
the convection zone.  

The apparent turbulent nature of the convective
flow indicated by our simulations implies that the treatment of mixing
in stars as a diffusive process may lead to inaccurate or even
incorrect results.  Convective flows are rather advective, as
suggested by \citet{WoodwardHerwig2008}.

\begin{figure} 
\includegraphics[width=0.9\hsize]{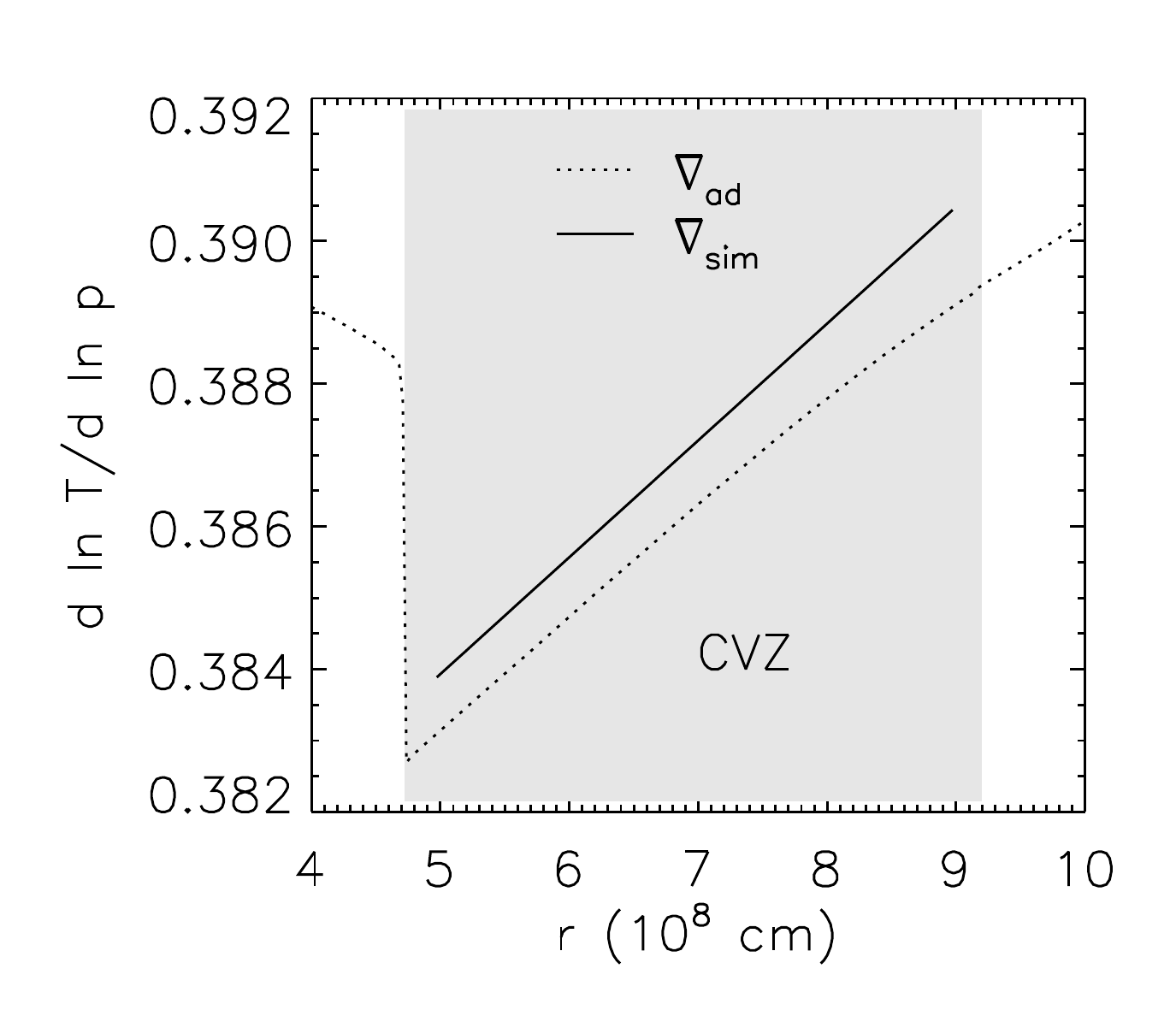}
\includegraphics[width=0.9\hsize]{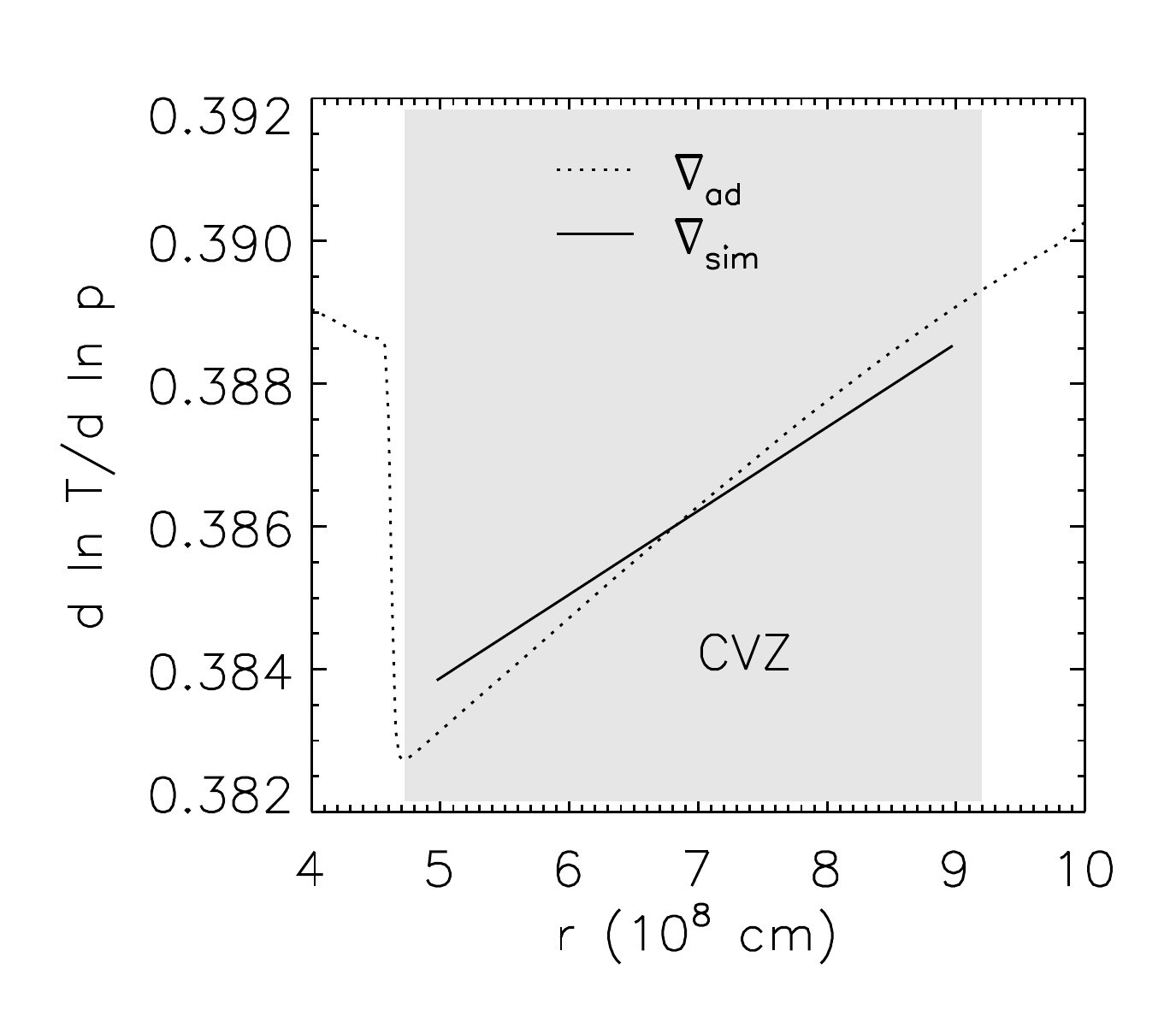}
\caption{
  {\it Upper panel:} Radial distributions of the adiabatic temperature
  gradient $\nabla_{ad}$ (dotted) and of the temperature gradients of
  model hefl.3d (solid), respectively. The latter distribution is a
  linear fit to the gradients averaged over angle and over the first
  200\,s of the evolution of the model. The gray shaded region marks
  the convection zone CVZ.
  {\it Lower panel:} Same as above, but showing the radial
  distributions in the evolved convection zone averaged over roughly
  3000\,s of evolutionary time. }
\vspace{-0.2cm}
\label{fig4.22.23}
\end{figure}

%
\section{Mixing length theory and simulations}
\label{sect:5}
Mixing length theory (or MLT) commonly used for treating convection in
stellar evolutionary calculations relies on assumptions and parameters
that are often chosen based on convenient ad-hoc arguments about the
convective flow, like \eg the value of the mixing length, the amount
of upflow-downflow symmetry or the position where, within the
convection zone, convective elements start to rise
\citep{KipWeigert1990, CoxGiuli2008}.

MLT assumes that the temperature of a convective element (blob)
is the same as that of the ambient medium surrounding it when it
starts to rise. However, as a blob will not rise until it is hotter
than the surroundings, this MLT assumption is contradictory. MLT
further assumes that once the blobs begin to rise they carry their
surplus of heat lossless over a distance given by the mixing length
before they release it to the surrounding gas instantaneously at the
end of their path. These assumptions are also not fulfilled in
general.
Our simulations show that convective elements typically start their
rise deep inside the star from the region of dominant nuclear burning
where they are accelerated by buoyant forces. The assumptions of MLT
that convective blobs form and begin their motion at different depths
of the convection zone, and that the average convective blob
propagates a distance equal to half of the assumed mixing length
before dissolving with the surrounding gas \citep{KipWeigert1990},
therefore do not hold.  MLT finally also assumes a correlation between
the thermodynamic variables and the velocity of the flow in a
convection zone. However, the results of our simulations falsify this
assumption (Fig.\,\ref{fig4.c2}).

According to MLT, the temperature fluctuations in a convection zone
are directly proportional to the mixing length and to the deviation of
the temperature gradient of the model $\nabla_{\mbox{sim}} =
(d\,\ln T / d\,\ln p)_{\mbox{sim}}$ from the adiabatic one
$\nabla_{\mbox{ad}} = (d\,\ln T/d\,\ln p)_{\mbox{ad}} $:
\begin{equation}  
  \frac{T^{'}}{T} = (\nabla_{\mbox{sim}} - 
                     \nabla_{\mbox{ad}}) \frac{1}{H_{p}} \frac{\Lambda}{2}
\label{eq.11}
\end{equation}
where $\Lambda$ is the mixing length, $H_p$ the pressure scale height,
$T^{'}$ the absolute value of the temperature deviation from the mean
(horizontally averaged) temperature T, and p the pressure,
respectively.  

Since $T{'}/T$ and $\nabla_{\mbox{sim}} - \nabla_{\mbox{ad}}$ can
directly be obtained from our simulations, we attempted to test MLT in
a qualitative manner.  Our simulations show that in the outer part of
the convection zone, \ie in the region where the buoyancy force is
getting smaller (Fig.\,\ref{fig4.10.11.12.13}, panel b), the temperature 
gradient of the
models $\nabla_{\mbox{sim}}$ becomes sub-adiabatic (see lower panel of
Fig.\,\ref{fig4.22.23}).  Equation (\ref{eq.11}) which was derived for the
adiabatic rise of convective bubbles would then imply that the
temperature of convective elements should be lower than that of the
surrounding gas (hence no convection) in the outer part of the 
convection zone, or that the
value of $\Lambda$ should be negative.  However, convective elements
do not rise adiabatically in our hydrodynamic simulations and the
sub-adiabatic gradient means only that the convective elements start to
cool faster than their surroundings. It does not imply necessarily that the
elements are already cooler than the surrounding gas which would prevent
the gas from being convectively active. Note that initially, 
the temperature gradient is super-adiabatic in the whole convection zone 
(see upper panel of Fig.\,\ref{fig4.22.23}), because the stellar evolutionary
model used as initial input for our simulations is computed under the
assumptions of the MLT.

%
\section{Long-term evolution}
\label{sect:6}
Two-dimensional simulations are biased due to the imposed symmetry restriction
which tend to overestimate the activity in the convection zone, but
qualitative similarities with geometrically unconstrained 3D models
exists.  Moreover, many phenomena that we observe in 2D models happen
also in the 3D ones, just slower.  Hence, we think it is justified to
explore the long-term evolution (\ie covering a few tens of hours
instead of a few hours) of our initial core helium flash models by
performing cost-effective 2D instead of very costly 3D simulations.

\begin{figure*}
\includegraphics[width=6.3cm]{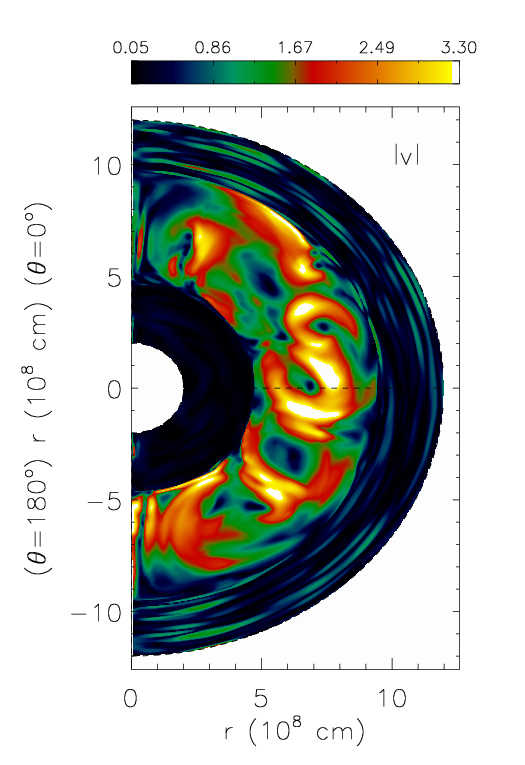}
\includegraphics[width=6.3cm]{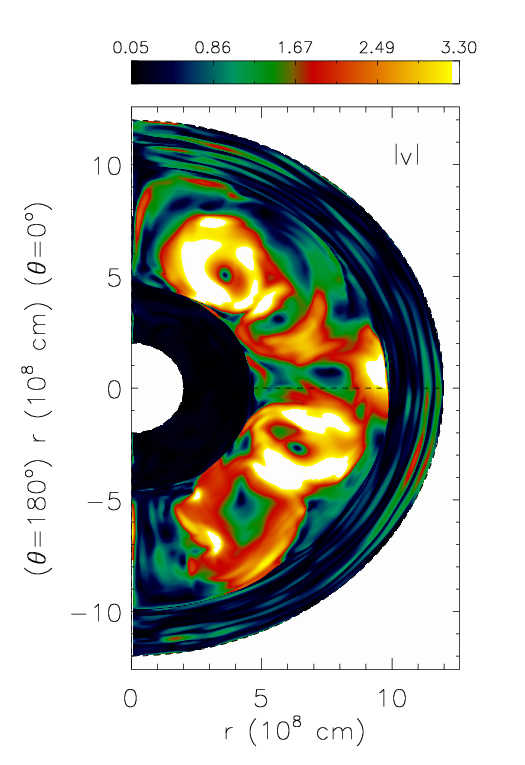} 
\includegraphics[width=6.3cm]{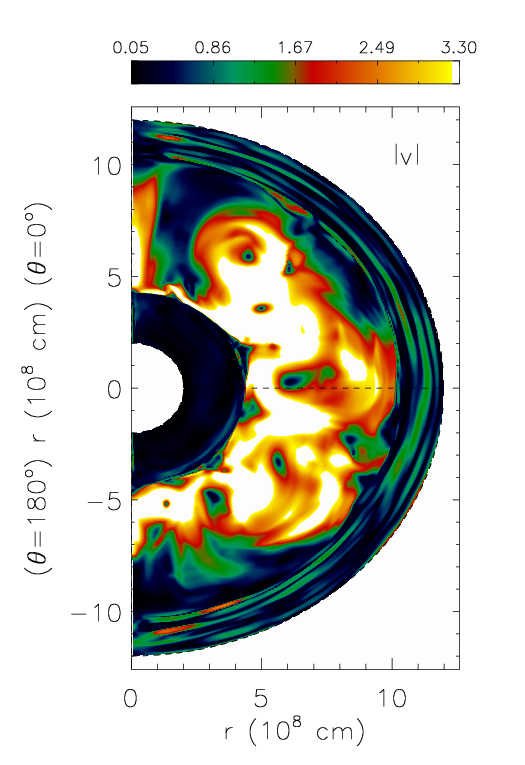} 
\caption{Snapshots of the spatial distribution of the velocity modulus
  $|v|$ (in units of $10^{6} \cms$) for the 2D model hefl.2d.b at
  24\,000\,s (left), 60\,000\,s (middle), and 120\,000\,s (right),
  respectively.}
\label{fig6.c1}
\end{figure*} 

\begin{figure} 
\includegraphics[width=0.99\hsize]{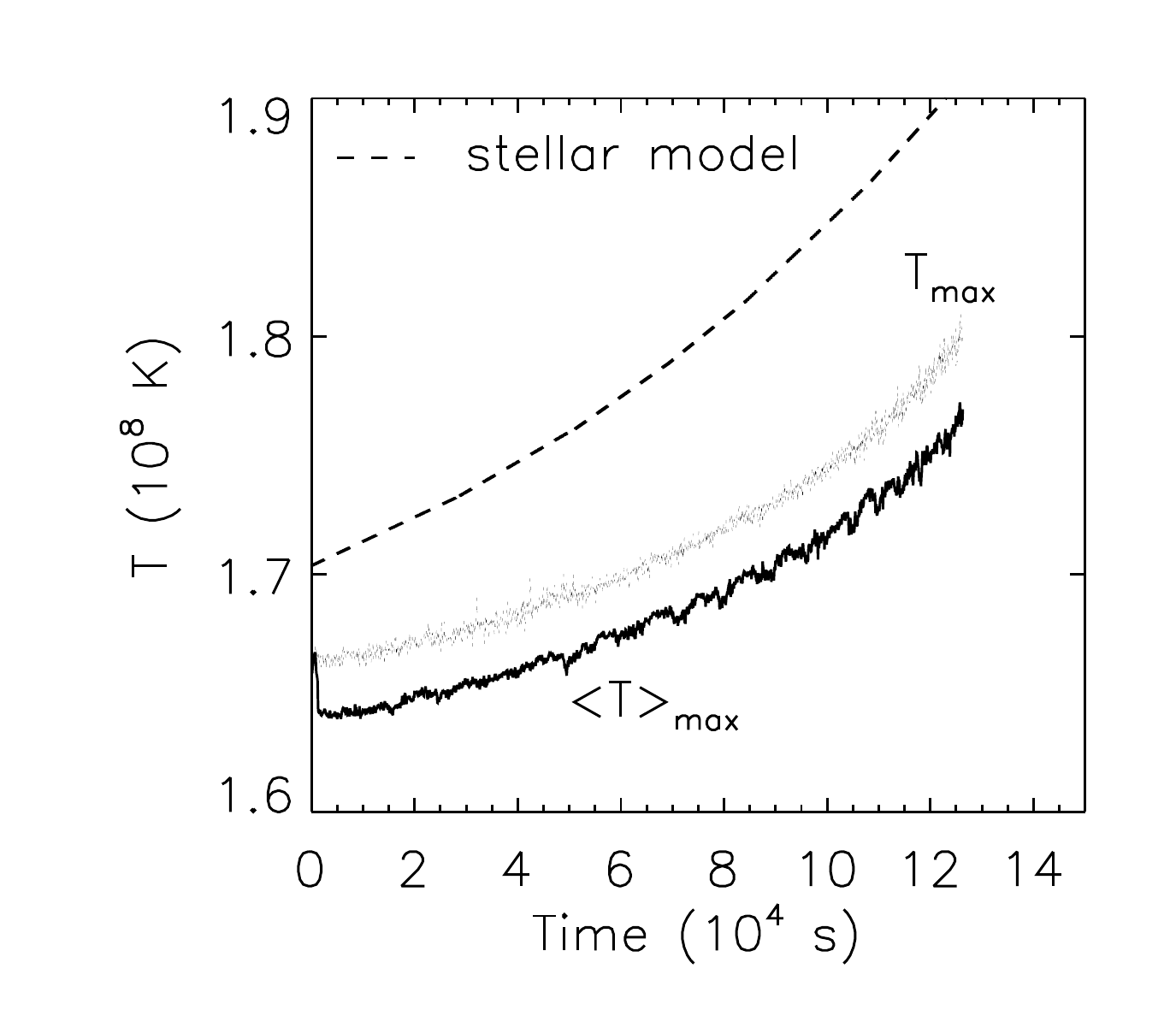}
\caption{Temporal evolution of the horizontally averaged temperature
  maximum $\langle T \rangle_{max}$ (solid), and of the global
  temperature maximum $T_{max}$ (solid thin) in the long-term 2D model
  hefl.2d.b. The dashed line correspond to the temporal
  evolution of the maximum temperature in the stellar evolutionary
  calculation. }
\label{fig6.1}
\end{figure}

In the following we describe the long-term evolution of a 2D model
whose early evolution, covering 8\,hrs, was discussed in detail by
\citet{Mocak2008}. The model is characterized by a very dynamic flow
involving typical convective velocities of $1.8\times 10^{6}\,\cms$.
Our long-term hydrodynamic simulation of this model covering 36\,hrs
(see Fig.\,\ref{fig6.1}) has revealed that the global and
angle-averaged maximum temperatures continue to rise at the initial
rate of $40\,\Ks$ that is 60\% lower than the rate predicted by
stellar evolutionary calculations. As a consequence, the typical
convective velocities increase by about 50\% and reach a level of
$2.8\times 10^{6}\,\cms$ at the end of the simulation
(Fig.\,\ref{fig6.c1}).

Hydrodynamic simulations of convection driven by nuclear burning
covering several convective turnover times show a rapid growth of the
convection zone due to the turbulent entrainment
\citep{MeakinArnett2007}.  An analysis of our simulations based on a
tracing of the radial position of the convective boundaries (defined
by the condition X($^{12}C$) = $2\times 10^{-3}$; see
Sect.\,\ref{sect:4.4}), shows a similar behavior.

\begin{figure*}
\includegraphics[width=0.49\hsize]{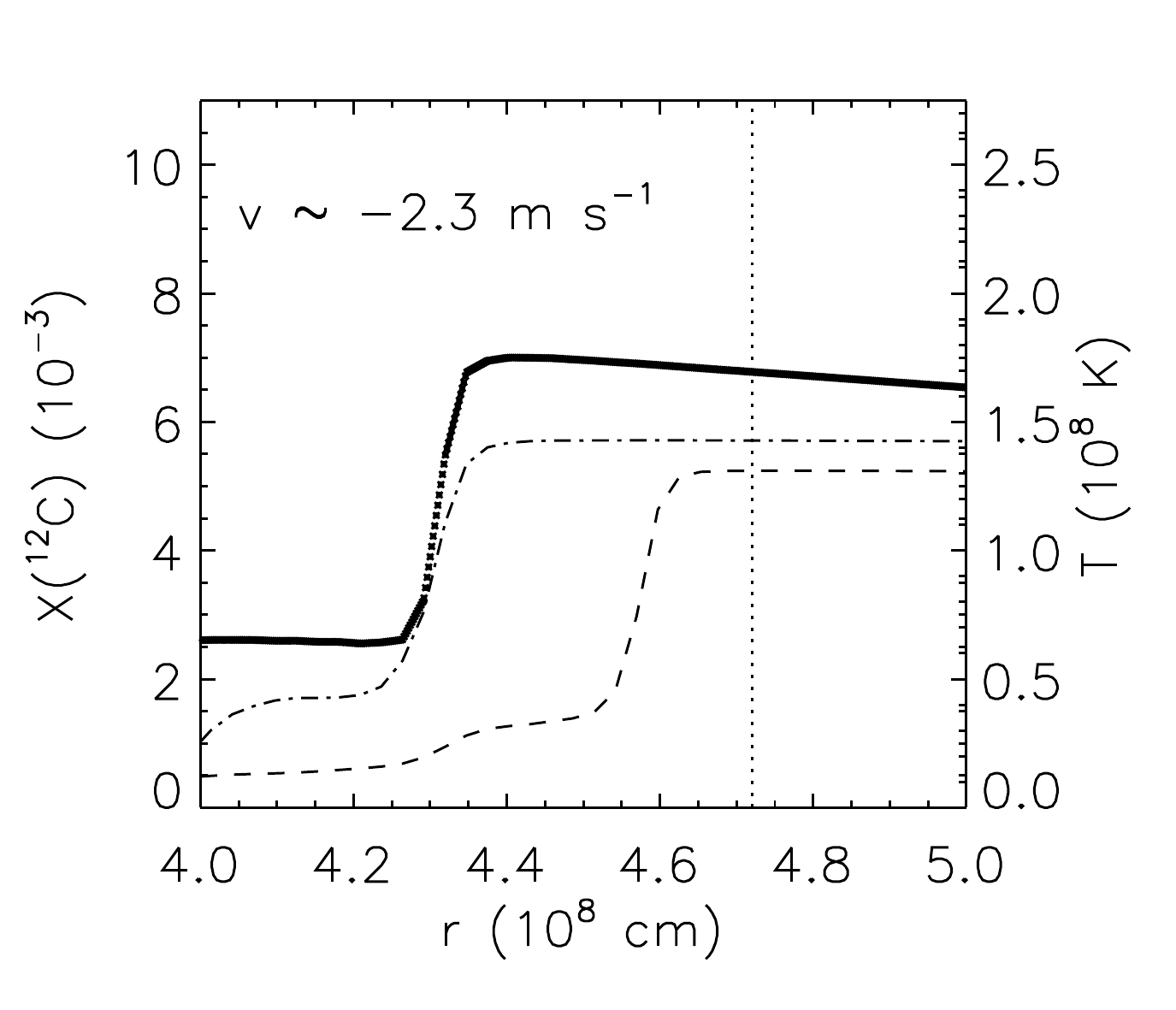} 
\includegraphics[width=0.49\hsize]{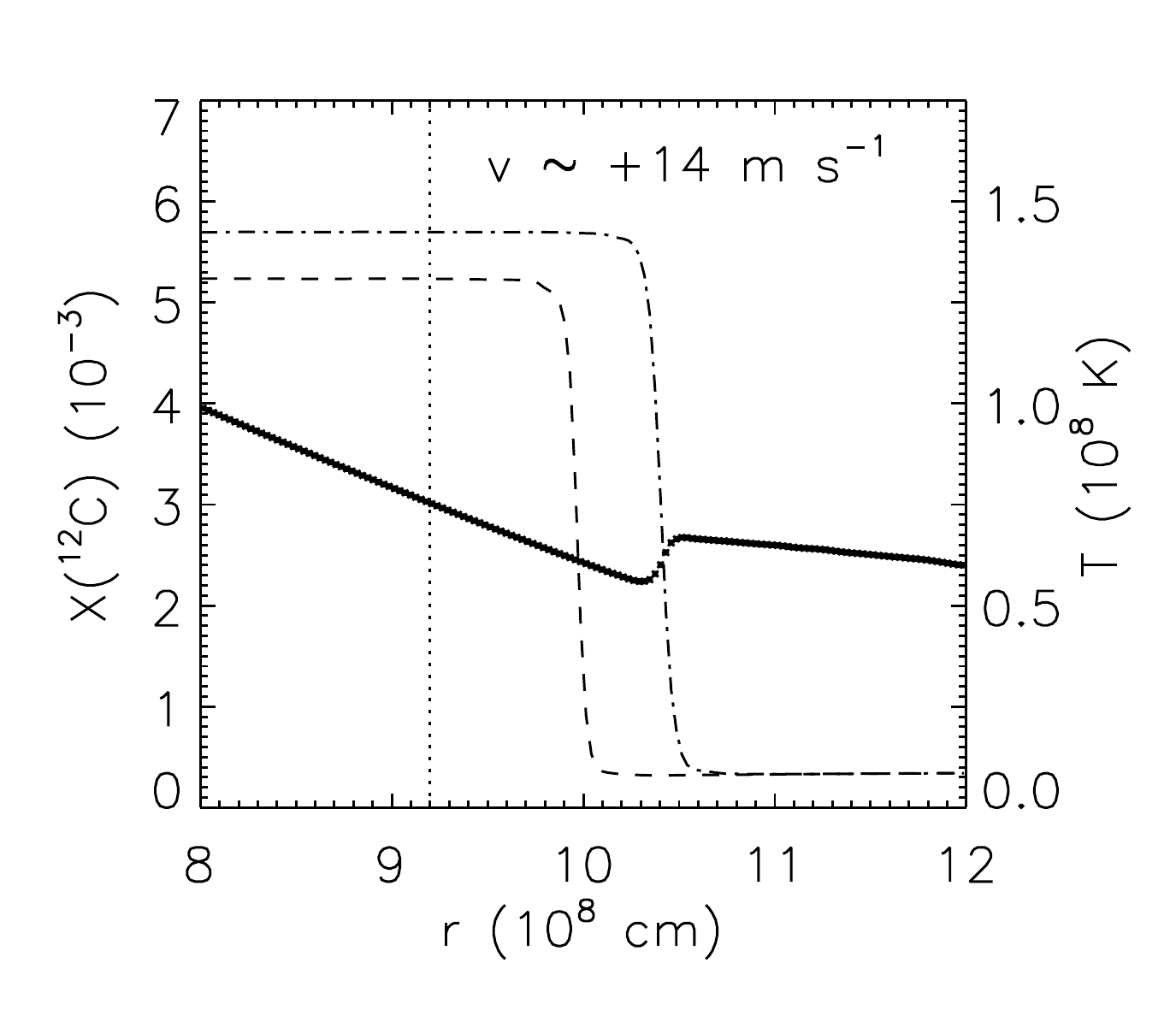} 

\includegraphics[width=0.49\hsize]{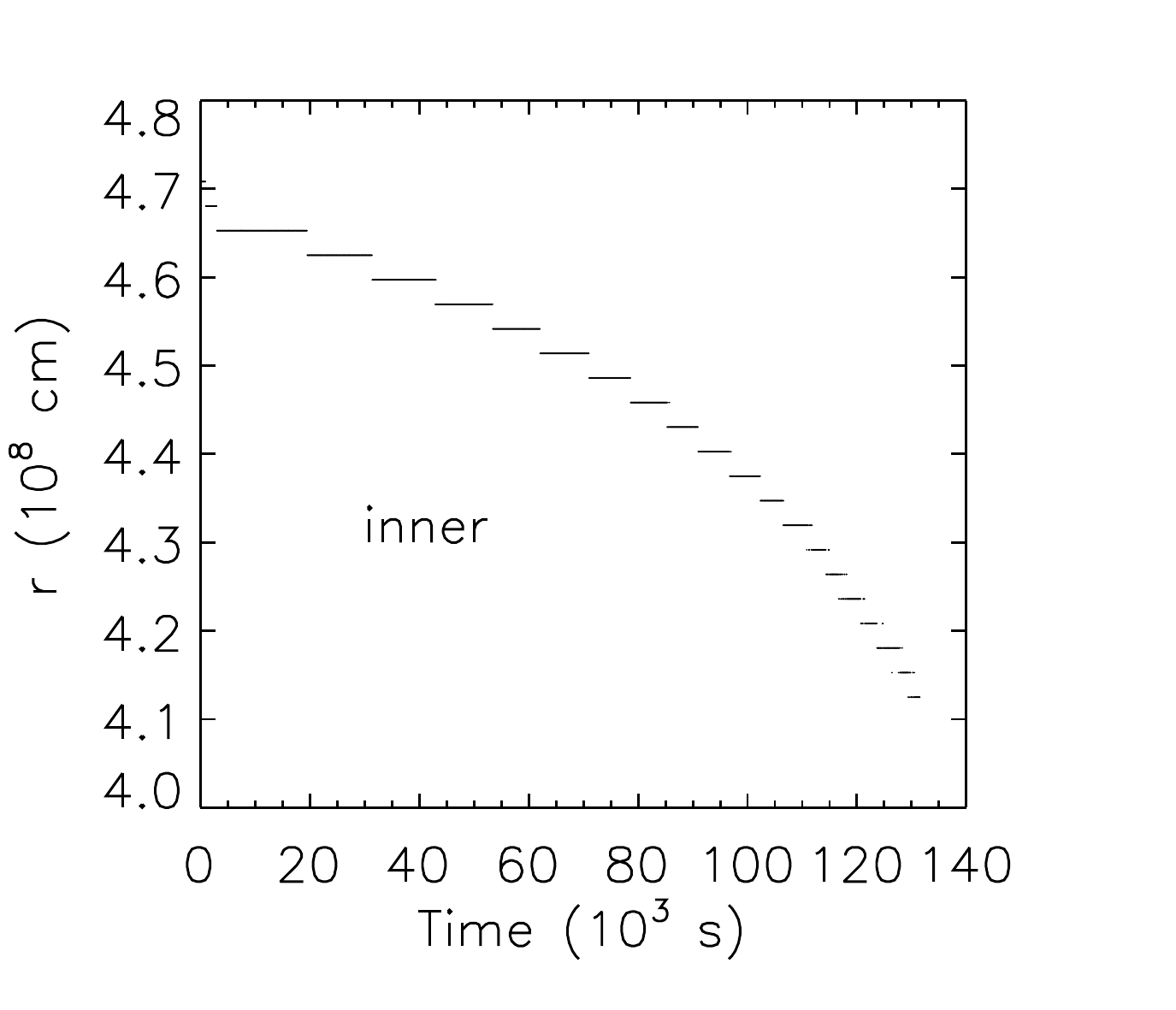} 
\includegraphics[width=0.49\hsize]{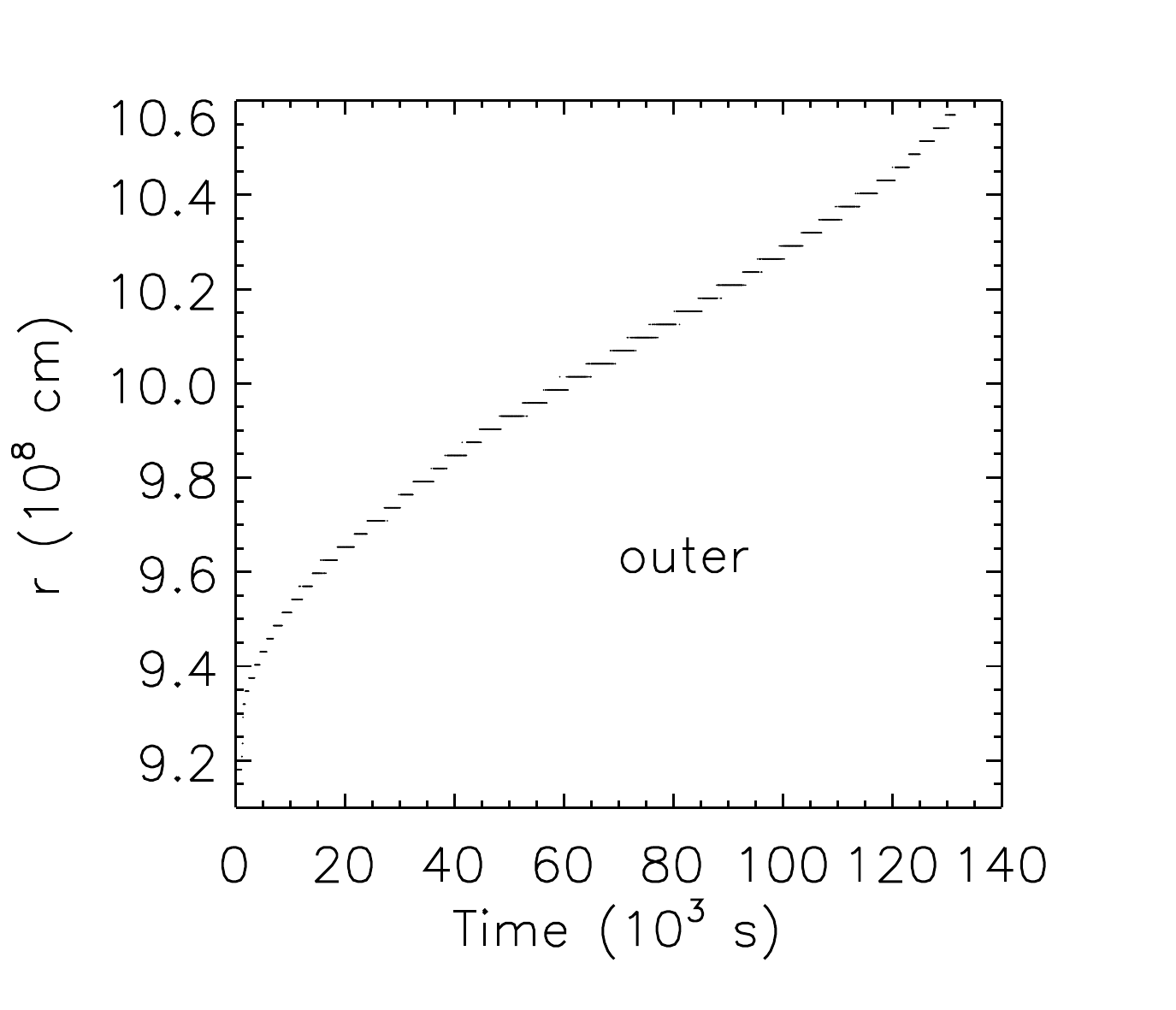} 
\caption{
  {\it Upper panels:} Angular averaged $^{12}C$ mass fraction (dashed)
  and temperature stratification (thick) as a function of radius near
  the inner (left) and outer edge (right) of the convection zone in
  the long-term 2D model hefl.2d.b at $t = 60\,000\s$ and $t =
  120\,000\s$, respectively.  The vertical dotted lines mark the
  boundaries of the convection zone at $t = 0\,$s.
  {\it Lower panels:} Temporal evolution of the position of the inner
  (left) and outer (right) edge of the convection zone in model
  hefl.2d.b, respectively. }
\label{fig6.2.3}
\end{figure*}

Turbulent motion near the upper edge of the convection zone pumps
material into the convectively stable layer at an entrainment rate of
14\,$\mes$ without any significant slowdown over the whole duration of
the simulation (Fig.\,\ref{fig6.2.3}) covering $\sim$ 130\,000\,s (or
more than 250 convective turnover times). The entrainment rate
at the inner convective boundary is about a factor of six smaller
(2.3\,$\mes$) slightly increasing during the second half of the
simulation ($t > 60000\,$; see Fig.\,\ref{fig6.2.3}).  These entrainment
rates have to be considered as upper limits because of the imposed
axi\-symmetry which leads to exaggerated convective velocities and large
filling factors for the penetrating plumes.  The turbulent entrainment
causes a growth of the convection zone on a dynamic timescale, in
agreement with the oxygen shell burning hydrodynamic models of
\citet{MeakinArnett2007}.

\begin{table} 
\caption{Approximate rates at which the mean mass fractions 
  of $^{4}$He, $^{12}$C,
  and $^{16}$O evolve in the long-term 2D model hefl.2d.b in the
  convection zone within the first $40\,000\,$s ($R_i$; in units of
  $10^{-9}$), and within the time interval $40\,000\,$s to
  $130\,000\,$s ($R_f$; in units of $10^{-9}$), respectively. The
  quantities $X_i$ and $X_f$ give the initial ($t = 0\,$s) and final
  ($t = 130\,000\,$s) mass fraction of $^{4}$He and mass fractions 
  of $^{12}$C, and $^{16}$O abundances (in units of 10$^{-3}$), 
  respectively.}
\begin{center}
\begin{tabular}{p{1.cm}|p{0.7cm}p{0.7cm}p{0.7cm}p{0.7cm}} 
\hline
\hline
element & $R_i$ & $R_f$ & $X_i$ & $X_f$ 
\\
\hline 
$^{4}$He &  +7.74 &  -7.85 &  0.975  & 0.974 \\
$^{12}$C &  -7.41 &  +7.71 &  5.502  & 5.918 \\
$^{16}$O &  -0.58 &  +1.13 &  0.927  & 1.008 \\
\hline
\end{tabular} 
\end{center}
\label{tab:elevol} 
\end{table}  

Both interfaces at the edges of the convection zone remain sharp
during the whole length of the simulation (Fig.\,\ref{fig6.2.3}). The
entrainment is correlated with a decrease of the temperature at the
outer edge of the convection zone due to the decreasing entropy
(Fig.\,\ref{fig4.6.7.8}). At the inner edge of the convection zone the
entrainment leads to heating of the cold region at $r < 5\times
10^{8}\,$cm that is cooled by neutrinos (Fig.\,\ref{fig6.2.3}).
Contrary to the finding of \citet{Asida2000}, the heating does not
penetrate deeper into the star than the mixing of $^{12}$C and the
other nuclear ashes.

Due to the growth of the convection zone and due to nuclear burning
the mean $^{12}$C, $^{4}$He, and $^{16}$O mass fractions change in the
convection zone at rates listed in
Table\,\ref{tab:elevol}. The mean value of the $^{12}$C mass fraction
in the convection zone decreases at a rate of $-7.4\times 10^{-9}
s^{-1}$ until $t = 40\,000\,$ to a value of X($^{12}$C) = $5.2\times
10^{-3}$.  Then it begins to increase again at roughly the same
(absolute) rate $+7.7\times 10^{-9} s^{-1}$ . The $^{12}$C mass
fraction decreases, because the volume of the convection zone grows
initially almost discontinuously due to the sudden start of the
entrainment. Hence, nuclear reactions are for a start unable to
produce enough carbon to compensate for the volume increase. At $t
\sim 110\,000\,$s the $^{12}$C mass fraction has risen again to its
initial value of X($^{12}$C)$ = 5.5\times 10^{-3}$, and at the end of
the simulation at $t = 130\,000\,$s the $^{12}$C mass fraction is
$5.9\times 10^{-3}$, a value that is 7\% higher than the initial one.

The mean $^{16}$O mass fraction shows a similar trend as that of
$^{12}$C as its production depends directly on the $^{12}$C mass
fraction.  The mass fraction of $^{4}$He rises within the first
$40\,000\,$s because convection is dredging up fresh $^{4}$He from the
convectively stable layers. Later in the evolution the mass fraction
of $^{4}$He decreases, as it is being constantly burned.

%
%

%
\section{Summary}
\label{sect:7}
We performed, analyzed and compared 2D axisymmetric and 3D
hydrodynamic simulations of the core helium flash. In agreement with
our previous study of 2D hydrodynamic models of the core helium flash
we find that the core helium flash in three dimensions neither rips
the star apart, nor that it significantly alters its structure.

The evolved convection of the 3D models differs qualitatively from
that of the axisymmetric ones. The typical convective structure in the
2D simulations is a vortex with a diameter roughly equal to the width
of convection zone, whereas the 3D structures are smaller in extent
and have a plume-like shape. The typical convective velocities are
much higher in the 2D models than in the 3D ones. In the latter models
the convective velocities tend to fit those predicted by the mixing
length theory better. Both 2D and 3D models are characterized by an
upflow-downflow asymmetry, where downflows dominate.

Our hydrodynamic simulations show the presence of turbulent
entrainment at both the inner and outer edge of the convection zone,
which results in a growth of the convection zone on a dynamic
timescale. While entrainment occurs at an almost constant speed at the
outer boundary of the convection zone, it tends to accelerate at the
inner one.  The entrainment rates are higher at the outer edge than at
the inner edge of the convection zone, as the latter is more stable
against entrainment.

The upper part of the evolved convection zone is characterized by a
sub-adiabatic temperature gradient, where buoyancy breaking takes
places, \ie rising convective plumes start to slow down in this region
and eventually descend back into deeper layers of the star. Convection
should not exist in that mass layer according to mixing length theory.

The fast growth of the convection zone due to entrainment has some
potentially interesting implications.  As entrainment is not
considered in canonical stellar evolutionary calculations, stars
evolving towards the core helium flash may never reach a state as the
one given by our initial stellar model. Hence, this may influence the
growth of the convection zone observed in our hydrodynamic
simulations, as the thermodynamic conditions at the edges of the
convection zone may differ.

If a rapid growth of the convection zone indeed occurs, the main core
helium flash studied here will never be followed by subsequent
mini-flashes, as convection will lift the electron degeneracy in the
helium core within 10 days. In addition, the helium core will likely
experience an injection of hydrogen from the stellar envelope within a
month and undergo a violent nuclear burning phase powered by the CNO
cycle. However, the growth of the convection zone within the 
core that is simulated in our models does not have
to continue until it will reach the outer convection zone extending up to the
surface of the star. Hence,
mixing of nuclear ashes to the stellar atmosphere does not necessarily
take place. But a fast dynamic growth of the inner convection zone
will lead to a change of the composition of the stellar core (less
carbon and oxygen), and consequently of the luminosities of low-mass
stars on the horizontal branch.

We found significant differences between the properties and the
evolution of 2D and 3D models having a convection zone powered by
(semi-)degenerate helium burning.  However, as our 3D models are
likely not yet fully converged and as they cover only a relatively
short period of evolutionary time ($< 6000\,$s), long-term 3D
simulations using a higher grid resolution are needed to obtain a
better and more reliable understanding of the hydrodynamics of the
core helium flash.

%
\begin{acknowledgements}
The simulations were performed at the Leibniz-Rechenzentrum of the
Bavarian Academy of Sciences \& Humanities on the SGI Altix 4700
system.  The authors want to thank Frank Timmes for some of his
publicly available Fortran subroutines which we used in the Herakles
code, and Kurt Achatz whose unpublished hydrodynamic simulations of
the core helium flash performed during his diploma work in 1995
inspired this work.  We are indebted to Casey Meakin for several
enlightening discussions and helpful comments.
\end{acknowledgements}

\bibliography{referenc}

%
\appendix

\onecolumn

%
\section{Hydrodynamic equations in spherical polar coordinates}
\label{app:euler_eqs}
The hydrodynamic equations of a non-viscous multi-component reactive
gas subject to a gravitational potential $\Phi$ and having a heat
conductivity $K$ are given in spherical polar coordinates ($r, \theta,
\phi$) by
\begin{equation}
 \partial_t \rho 
    + \frac{1}{r^2}         \partial_r      (r^2 \rho v_r) 
    + \frac{1}{r\sin\theta} \partial_{\theta}(\sin\theta\, \rho v_{\theta}) 
    + \frac{1}{r\sin\theta} \partial_{\phi}  (\rho v_{\phi})  
    =  0
\label{app:cons-m}
\end{equation}

\begin{subequations}
\begin{equation} 
 \partial_t (\rho v_r) 
    + \frac{1}{r^2}         \partial_r (r^2 \rho v_{r}^2) 
    + \frac{1}{r\sin\theta} \partial_{\theta}(\sin\theta\, \rho v_r v_{\theta}) 
    + \frac{1}{r\sin\theta} \partial_{\phi}  (\rho v_r v_{\phi}) 
    - \frac{\rho v_{\theta}^2}{r} 
    - \frac{\rho v_{\phi}^2}{r} 
    + \partial_r p 
    = 
    - \rho \partial_r \Phi  
\end{equation}
\begin{equation} 
  \partial_t (\rho v_{\theta}) 
     + \frac{1}{r^2}         \partial_r      (r^2\rho v_r^2) 
     + \frac{1}{r\sin\theta} \partial_{\theta}(\sin\theta\, \rho v_{\theta}^2) 
     + \frac{1}{r\sin\theta} \partial_{\phi}  (\rho v_{\theta} v_{\phi}) 
     + \frac{\rho v_{\theta} v_r}{r} 
     - \frac{\rho v_{\phi}^2 \cos\theta}{r \sin\theta} 
     + \frac{1}{r} \partial_{\theta} p 
     = 
     - \frac{\rho}{r} \partial_{\theta} \Phi  
\end{equation}
\begin{equation} 
  \partial_t (\rho v_{\phi}) 
     + \frac{1}{r^2}         \partial_r      (r^2 \rho v_r^2) 
     + \frac{1}{r\sin\theta} \partial_{\theta}(\sin\theta\, \rho
                                                v_{\theta} v_{\phi}) 
     + \frac{1}{r\sin\theta} \partial_{\phi}  (\rho v_{\phi}^2 ) 
     + \frac{\rho v_{\phi} v_r}{r} 
     + \frac{\rho v_{\phi} v_{\theta} \cos\theta}{r \sin\theta} 
     + \frac{1}{r \sin\theta} \partial_{\phi} p 
     = 
     - \frac{\rho}{r \sin\theta} \partial_{\phi} \Phi  
\label{app:cons-i}
\end{equation}
\end{subequations}

\begin{eqnarray} 
 \partial_t (\rho e) 
    + \frac{1}{r^2} \partial_{r} \left\{ r^2 [v_r 
                                    (\rho e + p) - K \partial_r T ] 
                                   \right\} 
    + \frac{1}{r \sin\theta} \partial_{\theta} 
          \left\{ \sin\theta\, [v_{\theta} (\rho e + p) - 
                               \frac{K}{r} \partial_{\theta} T]
          \right\} 
    + \frac{1}{r \sin\theta} \partial_{\phi} 
          \left\{ v_{\phi} (\rho e + p) - 
                 \frac{K}{r \sin\theta} \partial_{\phi} \Phi 
          \right\} 
    = & \\
    - \rho \left(  v_r\partial_{r} \Phi 
                 + \frac{v_{\theta}}{r} \partial_{\theta} \Phi 
                 + \frac{v_{\phi}}{r \sin\theta} \partial_{\phi} \Phi
           \right) 
    + \rho \dot{\varepsilon}  
      & 
\label{app:cons-e}
\end{eqnarray}

\begin{equation} 
 \partial_{t} (\rho X_{k}) 
    + \frac{1}{r^{2}} \partial_{r}(r^{2} \rho X_{k}  v_{r}) 
    + \frac{1}{r \sin\theta} \partial_{\theta} 
                             (\sin\theta\, \rho  X_{k} v_{\theta} ) 
    + \frac{1}{r \sin\theta} \partial_{\phi} (\rho  X_{k} v_{\phi} ) 
    = \rho \dot{X}_{k} 
    \,\, , \quad
    k = 1 \dots N_{nuc}
\end{equation}
where $\rho$, $v_r$, $v_{\theta}$, $v_{\phi}$, $p$, $e$, $T$,
$\dot{\varepsilon}$, $X_k$, and $\dot{X}_k$ are the density, the
radial velocity, the $\theta$-velocity, the rotation velocity, the
pressure, the total specific energy, the temperature, the energy
generation rate per mass due to reactions, the mass fraction of
species $k$, and the change of this mass fraction due to reactions,
respectively.  $N_{nuc}$ is the number of species the gas is composed
of.

%
\section{Energy fluxes} 
\label{app:fluxes}
The various contributions to the total energy flux
\citep{HurlburtToomre1986, Achatz1995} can be obtained by first
integrating the hydrodynamic energy equation given in
Appendix\,\ref{app:euler_eqs} over the angular coordinates $\theta$
and $\phi$. Then, one decomposes both the specific enthalpy
$\varepsilon + p/\rho$ (where $\varepsilon$ is the specific thermal
energy) and the specific kinetic energy $v_iv_i/2$ into a horizontal
mean and a perturbation, $f \equiv \overline f+f'$, and obtains
\footnote{Note that the equations given in \citet{Mocak2008} that
  correspond to Eqs.\,(\ref{app:term-e}), (\ref{app:term-fe}) and
  (\ref{app:term-ek}) contain some small typographical errors. }
\begin{equation} 
   \partial_tE + \partial_r(F_C + F_K + F_R + F_E)=0 
\label{aformequenE} 
\end{equation} 
where 
\newcommand{\dOm}{\,r^2\ad\Omega} 
\begin{eqnarray} 
   E   &=& \int_V \rho e~dV \label{app:term-e} \\ 
   F_C &=& \oint v_r\rho\cdot \left(\varepsilon+\frac{p}{\rho}\right)' 
           \dOm \\ 
   F_K &=& \oint v_r\rho\cdot \left(\frac12 v_iv_i\right)' \dOm 
           \,\, , \quad i=1,2,3\\ 
   F_R &=& -\oint K\partial_rT \dOm \\ 
%
%
   F_E &=& 4\pi r^2\overline{v_r\rho}\cdot 
           \left(\,\overline{\varepsilon+\frac p\rho}+\overline{\frac12v_iv_i} 
            +\Phi\,\right) \label{app:term-fe} \,. 
\label{app_fe}
\end{eqnarray} 
Here, the gravitational potential $\Phi$ is assumed to be constant for
simplicity. The sum of the various flux terms $F_i$ give the total
energy transported per unit time across a sphere of radius $r$ by
different physical processes. One has the convective (or enthalpy)
flux, $F_C$, the flux of kinetic energy, $F_K$, and the flux due to
heat conduction and radiation, $F_R$.
%
%
Finally, $F_E$, includes all terms causing a spherical mass flow, \ie
the model's expansion or contraction, while $F_C$ and $F_K$ rest on
deviations from this mean energy flow (vortices). The latter are the
major contributors to the heat transport by convection.
%
%

In a similar way one can also formulate a conservation equation for
the mean horizontal kinetic energy that provides further insight into
the effects of convective motions. Using the other hydrodynamic
equations (Eqs.\,\ref{app:cons-m} to \ref{app:cons-i}), and the
relation $\partial_t (\rho v_iv_i/2) = v_i \partial_t(\rho v_i) - v_i
v_i \partial_t\rho/2$, one finds
\begin{equation} 
   \partial_tE_K + \partial_r(F_K + F_P + F_{E,K}) 
        = P_A + P_P + P_{E,K} 
      \label{aformequenEK} 
\end{equation} 
With $F_K$ as introduced above, one obtains
\begin{eqnarray} 
   E_K &=& \int_V \frac\rho2 v_i v_i ~dV  \label{app:term-ek} \\ 
   F_P &=& -\oint v_r p' \dOm \\ 
   F_{E,K} &=& 4\pi r^2\overline{v_r\rho}\cdot 
              \left(\,\overline{\frac p\rho+\frac{v_i v_i}2}\,\right) \\ 
   P_A &=& -\oint v_r\rho'\partial_r\Phi \dOm \\ 
   P_P &=& \oint p'\partial_i v_i \dOm \\ 
%
%
   P_{E,K} &=& 4\pi r^2\cdot 
          \left(\,\overline p\,\overline{\partial_i v_i}- 
           \overline v_r\overline\rho\,\partial_r\Phi\,\right) 
    \,\, , \quad i=1,2,3
\end{eqnarray} 
where the $P_i$ are source or sink terms of the kinetic energy. They
are separated into the effect of buoyancy forces ($P_A$), 
%
%
and the work due to density fluctuations ($P_P$, volume changes). By
analyzing the various $P_i$ one can determine what causes the braking
or acceleration of the convective flow. The acoustic flux, $F_P$,
describes the vertical transport of density fluctuations. $F_{E,K}$
and $P_{E,K}$ describe the effect of expansion (volume work, and work
against the gravitational potential), similar to $F_E$ in
Eq.\,(\ref{app_fe}).

\end{document}